\documentclass[aps,twocolumn,rmp]{revtex4}
\def\lap{\hbox{~{\lower -2.5pt\hbox{$<$}}\hskip -8pt\raise
-3.5pt\hbox{$\sim$}}}
\def\gap{\hbox{~{\lower -2.5pt\hbox{$>$}}\hskip -8pt\raise
-3.5pt\hbox{$\sim$}}}
\def\apg{\hbox{{\raise -2.5pt\hbox{$>$}}\hskip -8pt\lower -
2.5pt\hbox{$\sim$}}}
\def\apl{\hbox{{\raise -2.5pt\hbox{$<$}}\hskip -8pt\lower -
2.5pt\hbox{$\sim$}}}
\begin{document}
\title {DECOHERENCE, EINSELECTION,\\
AND THE QUANTUM ORIGINS OF THE CLASSICAL}

\author{Wojciech Hubert Zurek}
\address{Theory Division, LANL, Mail Stop B288\\
Los Alamos, New Mexico 87545}
\maketitle

\bigskip
{\it Decoherence} is caused by the interaction with the environment
which in effect monitors certain observables of the system, destroying
coherence between the {\it pointer states} corresponding to their
eigenvalues. This leads to {\it environment-induced superselection}
or {\it einselection}, a quantum process associated with
selective loss of information. Einselected pointer states are stable.
They can retain
correlations with the rest of the Universe in spite of the environment.
Einselection enforces classicality by imposing an effective ban on
the vast majority of the Hilbert space, eliminating especially the flagrantly
non-local ``Schr\"odinger cat'' states. Classical structure of
phase space emerges from the quantum Hilbert space in the appropriate
macroscopic limit: Combination of einselection with dynamics
leads to the idealizations of a point and of a classical trajectory.
In measurements, einselection replaces quantum entanglement between
the apparatus and the measured system with the classical correlation.
Only the preferred pointer observable of the apparatus can store
information that has predictive power. When the measured quantum system
is microscopic and isolated, this restriction on the predictive utility
of its correlations with the macroscopic apparatus results in
the effective ``collapse of the wavepacket''.
{\it Existential interpretation} implied by einselection regards
observers as open quantum systems, distinguished
only by their ability to acquire,
store, and process information. Spreading of the correlations with the
effectively classical pointer states throughout the environment
allows one to understand `classical reality' as a property based on the
{\it relatively objective existence} of the einselected states:
They can be ``found out'' without being re-prepared, e.g, by intercepting
the information already present in the environment. The redundancy of 
the records of pointer states in the environment (which can be thought of 
as their `fitness' in the Darwinian sense) is a measure of their 
classicality. A new symmetry appears in this setting: Environment - 
assisted invariance or {\it envariance} sheds a new light on the nature 
of ignorance of the state of the system due to quantum correlations 
with the environment, and leads to Born's rules and to the reduced 
density matrices, ultimately justifying basic principles of the program 
of decoherence and einselection.

\bigskip

\tableofcontents

\section{INTRODUCTION}

The issue of interpretation is as old as quantum theory. It dates back
to the discussions of Niels Bohr, Werner Heisenberg, Erwin Schr\"odinger,
(Bohr, 1928; 1949; Heisenberg, 1927; Schr\"odinger, 1926; 1935a,b;
see also Jammer, 1974; Wheeler and Zurek, 1983). Perhaps the most incisive
critique of the (then new) theory was due to Albert Einstein, who, searching
for inconsistencies, distilled the essence of the conceptual difficulties
of quantum mechanics through ingenious ``gedankenexperiments''.
We owe him and Bohr clarification of the significance of the quantum
indeterminacy in course of the Solvay congress debates (see Bohr, 1949)
and elucidation of the nature of quantum entanglement (Einstein, Podolsky,
and Rosen, 1935; Bohr, 1935, Schr\"odinger, 1935a,b). Issues identified then
are still a part of the subject.

Within the past two decades the focus of the research on the fundamental
aspects of quantum theory has shifted from esoteric and philosophical to
more ``down to earth'' as a result of three developments.  To begin with,
many of the old {\it gedankenexperiments} (such as the EPR ``paradox'') 
became compelling demonstrations of quantum physics. More or less 
simultaneously the role of decoherence begun to be appreciated and  
einselection was recognized as key in the emergence of classicality. 
Last not least, various developments have led to a new view of the role of
information in physics. This paper reviews progress with a focus on 
decoherence, einselection and the emergence of classicality, but also 
attempts a ``preview'' of the future of this exciting and fundamental area.

\subsection{The problem: Hilbert space is big}

The interpretation problem stems from the vastness of the Hilbert space,
which, by the principle of superposition, admits arbitrary linear combinations
of any states as a possible quantum state. This law, thoroughly tested in
the microscopic domain, bears consequences that defy classical
intuition: It appears to imply that the familiar classical states should be
an exceedingly rare exception. And, naively, one may guess that superposition
principle should always apply literally: Everything is ultimately made
out of quantum ``stuff''. Therefore, there is no {\it a priori} reason
for macroscopic objects to have definite position or momentum.
As Einstein noted\footnote{In a letter dated 1954, Albert Einstein wrote
to Max Born ``Let $\psi_1$ and $\psi_2$ be solutions of
the same Schr\"odinger equation.$\dots$. When the system is a macrosystem
and when $\psi_1$ and $\psi_2$ are `narrow' with respect to
the macrocoordinates, then in by far the greater  number of cases this
is no longer true for $\psi = \psi_1 + \psi_2$. Narrowness with respect
to macrocoordinates is not only independent of the principles of quantum
mechanics, but, moreover, incompatible with them." (The translation from
Born (1969) quoted here is due to Joos (1986).)}
localization with respect to macrocoordinates
is not just {\it independent}, but {\it incompatible} with quantum
theory. How can one then establish correspondence between
the quantum and the familiar classical reality?

\subsubsection{Copenhagen Interpretation}

Bohr's solution was to draw a border between the quantum and the classical
and to keep certain objects -- especially measuring devices and
observers -- on the classical side (Bohr, 1928; 1949).
The principle of superposition was suspended ``by decree'' in
the classical domain. The exact location of this
border was difficult to pinpoint, but measurements ``brought to a close''
quantum events. Indeed, in Bohr's view the classical domain was more
fundamental: Its laws were self-contained (they could be confirmed from
within) and established the framework necessary to define the quantum.

The first breach in the quantum-classical border appeared early:
In the famous Bohr -- Einstein double-slit debate, quantum Heisenberg
uncertainty was invoked by Bohr at the macroscopic level to preserve
wave-particle duality. Indeed, as the ultimate components of classical
objects are quantum, Bohr emphasized that the boundary must be
moveable, so that even the human nervous system could be
regarded as quantum provided that suitable classical
devices to detect its quantum features were available.
In the words of John Archibald Wheeler (1978; 1983) who has
elucidated Bohr's position
and decisively contributed to the revival of interest in these matters,
``No [quantum] phenomenon is a phenomenon until it is a recorded (observed)
phenomenon''.

This is a pithy summary of a point of view -- known as
the Copenhagen Interpretation (CI) -- that has kept many
a physicist out of despair. On the other hand, as long as a compelling reason
for the quantum-classical border could not be found, the CI Universe would be
governed by two sets of laws, with poorly defined domains of
jurisdiction. This fact has kept many a student,
not to mention their teachers, in despair (Mermin 1990a;~b; 1994).

\subsubsection{Many Worlds Interpretation}

The approach proposed by Hugh Everett (1957a,~b) and elucidated by
Wheeler (1957), Bryce DeWitt (1970) and others (see DeWitt and Graham,
1973; Zeh, 1970; 1973; Geroch, 1984; Deutsch, 1985, 1997, 2001) was to enlarge
the quantum domain.  Everything is now represented by a unitarily
evolving state vector, a gigantic superposition splitting to accommodate
all the alternatives consistent with the initial conditions.
This is the essence of the Many Worlds Interpretation (MWI).
It does not suffer from the dual nature of CI. However, it also
does not explain the emergence of classical reality.

The difficulty many have in accepting MWI stems from its violation of
the intuitively obvious ``conservation law'' -- that there is just
one Universe, the one we perceive. But even after this question is dealt
with, , many a convert from CI (which claims allegiance of a majority
of physicists) to MWI (which has steadily gained popularity;
see Tegmark and Wheeler, 2001, for an assessment) eventually realizes that
the original MWI does not address the ``preferred basis question''
posed by Einstein$^1$ (see Wheeler, 1983; Stein, 1984; Bell 1981, 1987;
Kent, 1990; for critical assessments of MWI). And as long as it is unclear
what singles out preferred states, perception of a unique outcome of
a measurement and, hence, of a single Universe cannot be explained
either\footnote{DeWitt, in the Many Worlds re-analysis
of quantum measurements makes this clear: in
DeWitt and Graham (1973), last paragraph of p. 189 he writes about
the key `remaining problem': ``Why is it so easy to find apparata
in states [with a well defined value of the pointer observable]? In the case
of macroscopic apparata it is well known that a small value for the mean
square deviation of a macroscopic observable is a fairly stable property
of the apparatus. But how does the mean square deviation become so small
in the  first place? Why is a large value of the mean square deviation
of a macroscopic observable virtually never, in fact, encountered in
practice? \dots a proof of this does not yet exist. It remains a program
for the future.''}.

In essence, Many Worlds Interpretation does not address but only postpones
the key question. The quantum - classical boundary 
is pushed all the way towards the observer, right against the border
between the material Universe and the ``consciousness'', leaving it
at a very uncomfortable place to do physics. MWI is incomplete:
It does not explain what is effectively classical and why. Nevertheless,
it was a crucial conceptual breakthrough: Everett reinstated quantum 
mechanics as a basic tool in the search for its interpretation.

\subsection{Decoherence and einselection}

Decoherence and einselection are two complementary views of
the consequences of the same process of environmental monitoring.
Decoherence is the destruction of quantum coherence between
preferred states associated with the observables monitored
by the environment. Einselection is its consequence -- the {\it de facto}
exclusion of all but a small set, a {\it classical domain} consisting
of pointer states -- from within a much larger Hilbert space.
Einselected states are
distinguished by their resilience -- stability in spite of the monitoring
environment.

The idea that ``openness'' of quantum systems may have anything
to do with the transition from quantum to classical was resolutely
ignored for a very long time, probably because in classical physics
problems of fundamental importance were always settled in isolated
systems. In the context of measurements Gottfried (1967) anticipated
some of the latter developments. The fragility of energy levels of
quantum systems was emphasized by seminal papers of
Dieter Zeh (1970; 1973), who argued
(inspired by remarks relevant to what would be called today ``deterministic
chaos'' (Borel, 1914)) that macroscopic quantum systems are in effect
impossible to isolate.

The understanding of how the environment distills the classical essence
from quantum systems is more recent (Zurek, 1981; 1982; 1993a). It combines
two observations: (1) In quantum physics ``reality'' can be attributed
to the measured states. (2) Information transfer usually associated with
measurements is a common result of almost
any interaction with the environment of a system.

Some quantum states are resilient to decoherence. This is the basis 
of einselection. Using Darwinian analogy, one might say that
pointer states are most `fit'. They survive monitoring by the environment
to leave `descendants' that inherit their properties.
Classical domain of pointer states offers a static
summary of the result of quantum decoherence.
Save for classical dynamics, (almost) nothing happens to
these einselected states, even though they are immersed in
the environment.

It is difficult to catch einselection ``in action'': Environment has little
effect on the pointer states, as they are already classical. Therefore,
it was easy to miss decoherence - driven dynamics of einselection
by taking for granted its result -- existence of the classical domain,
and a ban on arbitrary quantum superpositions.
Macroscopic superpositions of einselected states disappear
rapidly. Einselection creates effective
superselection rules (Wick, Wightman and Wigner, 1952; 1970; Wightman,
1995). However, in the microscopic, decoherence can be slow
in comparison with the dynamics.

Einselection is a quantum phenomenon. Its essence cannot
be even motivated classically: In classical physics arbitrarily accurate
measurements (also by the environment) can be in principle carried out without
disturbing the system. Only in quantum mechanics acquisition of information
inevitably alters the state of the system -- the fact that becomes apparent 
in double-slit and related experiments (Wootters and Zurek, 1979; 
Zurek, 1983).  

Quantum nature of
decoherence and the absence of classical analogues are a source of
misconceptions. For instance, decoherence is sometimes equated with relaxation
or classical noise that can be also introduced by the environment.
Indeed, {\it all} of these effects often appear together and as a consequence
of the ``openness''. The distinction between them can be briefly summed up:
{\it Relaxation and noise are caused by the environment perturbing
the system}, while {\it decoherence and einselection are caused by
the system perturbing the environment}.

Within the past few years decoherence and einselection became
familiar to many. This does not mean that their implications are
universally accepted (see comments in the April 1993 issue  of
{\it Physics Today}; d'Espagnat, 1989 and 1995; Bub, 1997; Leggett, 1998
and 2002; Stapp, 2001; exchange of views between Anderson, 2001,
and Adler, 2001).  In a field where controversy reigned for so long 
this resistance to a new paradigm is no surprise.

\subsection{The nature of the resolution and the role of envariance}

Our aim is to explain why does the quantum Universe appear classical.
This question can be motivated only in the context of the Universe
divided into systems, and must be phrased in the language of the
correlations between them. In the absence of systems Schr\"odinger equation
dictates deterministic evolution;
$$ |\Psi (t) \rangle  \ = \ \exp (-i H t /\hbar ) ~ |\Psi(0)\rangle  \ ,  \eqno(1.1)$$
and the problem of interpretation seems to disappear.

There is no need for ``collapse'' in a Universe with no systems. Yet, 
the division into systems is imperfect. As a consequence, 
the Universe is a collection of
open (interacting) quantum systems. As the interpretation problem does not
arise in quantum theory unless interacting systems exist, we shall also feel
free to assume that an environment exists when looking for a resolution.

Decoherence and einselection fit comfortably in the context of the
Many Worlds Interpretation where they define the ``branches'' of 
the universal state vector. Decoherence makes MWI complete:
It allows one to analyze the Universe as it is seen by an observer,
who is also subject to decoherence. Einselection justifies elements
of Bohr's CI by drawing the border between the quantum and the classical. 
This natural boundary can be sometimes shifted: Its effectiveness depends 
on the degree of isolation and on the manner in which the system 
is probed, but it is a very effective quantum - classical border nevertheless.

Einselection fits either MWI or CI framework: It supplies
a statute of limitations, putting an end to the quantum jurisdiction.
. It delineates how much of the
Universe will appear classical to observers who monitor
it from within, using their limited capacity to acquire, store,
and process information. It allows one to understand classicality 
as an idealization that holds in the limit of macroscopic open 
quantum systems.

Environment imposes superselection rules by preserving part of the
information that resides in the correlations between the system 
and the measuring apparatus (Zurek, 1981, 1982). The observer and 
the environment compete for the information about the system. 
Environment -- because of its size and its incessant interaction 
with the system -- wins that competition, acquiring information 
faster and more completely than the observer. Thus, a record useful 
for the purpose of prediction must be restricted to the observables 
that are already monitored by the environment. In that case, 
the observer and the environment no longer compete and
decoherence becomes unnoticeable. Indeed, typically {\it observers
use environment as a ``communication channel'', and monitor it to find
out about the system}.

Spreading of the information about the system through the environment is
ultimately responsible for the emergence of the ``objective reality''.
Objectivity of a state can be quantified by the redundancy with which it
is recorded throughout Universe. Intercepting fragments 
of the environment allows observers to find out (pointer) state of 
the system without perturbing it (Zurek, 1993a, 1998a, and 2000; 
see especially section VII of this paper for a preview of 
this new ``environment as a witness'' approach to the interpretation 
of quantum theory).


When an effect of a transformation acting on a system can be undone
by a suitable transformation acting on the environment, so that the
joint state of the two remains unchanged, the transformed property 
of the system is said to exhibit ``environment assisted invariance''
or {\it envariance} (Zurek, 2002b). Observer must be obviously ignorant 
of the envariant properties of the system. Pure entangled states
exhibit envariance. Thus, in quantum physics perfect information about
the joint state of the system-environment pair can be used to prove 
ignorance of the state of the system.  

Envariance offers a new fundamental view of what is information 
and what is ignorance in the quantum world. It leads to Born's rule for
the probabilities and justifies the use of reduced density matrices
as a description of a part of a larger combined system. Decoherence
and einselection rely on reduced density matrices. Envariance provides 
a fundamental resolution of many of the interpretational issues. It will
be discussed in section VI~D.


\subsection{Existential Interpretation and `Quantum Darwinism'}

What the observer knows is inseparable from what the observer is:
The physical state of his memory implies his information about 
the Universe. Its reliability depends
on the stability of the correlations with the external observables.
In this very immediate sense decoherence enforces the apparent
``collapse of the wavepacket'': After a decoherence timescale,
only the einselected memory states will exist and retain useful 
correlations (Zurek, 1991; 1998a,b; Tegmark, 2000). 
The observer described by some specific einselected
state (including a configuration of memory bits) will be able
to access (``recall'') only that state. The collapse is a consequence
of einselection and of the one-to-one correspondence between the state
of his memory and of the information encoded in it. Memory is simultaneously 
a description of the recorded information and a part of the ``identity tag'', 
defining observer as a physical system.
It is as inconsistent to imagine observer perceiving something else than
what is implied by the stable (einselected) records
in his possession as it is impossible to imagine the same person with
a different DNA: Both cases involve information encoded in a state of
a system inextricably linked with the physical identity of an individual.

Distinct memory/identity states of the observer (that are also his
``states of knowledge'') cannot be superposed: This censorship is
strictly enforced by decoherence and the resulting einselection.
Distinct memory states label and ``inhabit'' different branches of
the Everett's ``Many Worlds'' Universe.  Persistence of {\it correlations}
is all that is needed to recover ``familiar reality''. In this manner,
the distinction between epistemology and ontology is washed away: To put
it succinctly (Zurek, 1994) there can be {\it no information without
representation} in physical states.

There is usually no need to trace the collapse
all the way to observer's memory. It suffices that the states of a
decohering system quickly evolve into mixtures of the preferred (pointer)
states. All that can be known in principle about a system (or about an
observer, also introspectively, e.g., by the observer himself) is its
decoherence-resistant `identity tag' -- a description of its einselected state.

Apart from this essentially negative function of a censor the environment plays
also a very different role of a ``broadcasting agent", relentlessly cloning
the information about the einselected pointer states. This role of
the environment as a witness in determining what exists was not appreciated
until now: Throughout the past two decades, study of decoherence focused
on the effect of the environment on the system. This has led to a multitude
of technical advances we shall review, but it has also missed one crucial point
of paramount conceptual importance: Observers monitor systems indirectly, by
intercepting small fractions of their environments (e.g., a fraction of 
the photons that have been reflected or emitted by the object of interest). 
Thus, if the understanding of why we perceive quantum Universe as classical 
is the principal aim, study of the nature of accessibility of information
spread throughout the environment should be the focus of attention. 
This leads one away from the models of measurement inspired by the 
``von Neumann chain'' (1932) to studies of information  transfer involving
branching out conditional dynamics and the resulting ``fan-out'' of the
information throughout environment (Zurek, 1983, 1998a, 2000). This new 
`quantum Darwinism' view of environment selectively amplifying einselected 
pointer observables of the systems of interest is complementary to the usual 
image of the environment as the source of perturbations that destroy quantum 
coherence of the system. It suggests the redundancy of the imprint of the
system in the environment may be a quantitative measure of relative 
{\it objectivity} and hence of classicality of quantum states. It is 
introduced in Sections VI and VII of this review.

Benefits of recognition of the role of environment include not just
operational definition of the objective existence of the einselected states,
but -- as is also detailed in Section VI -- a clarification of the
connection between the quantum amplitudes and probabilities.
Einselection converts arbitrary states into mixtures of well defined
possibilities. Phases are envariant: Appreciation of envariance as 
a symmetry tied to the ignorance about the state of the system was 
the missing ingredient in the attempts of `no collapse' derivations
of Born's rule and in the probability 
interpretation. While both envariance and the ``environment as a witness''
point of view are only beginning to be investigated, the extension of the
program of einselection they offer allowes one to understand emergence
of ``classical reality'' form the quantum substrate as a consequence of 
quantum laws.

\section{QUANTUM MEASUREMENTS}

The need for a transition from quantum determinism of the global
state vector to classical definiteness of states of individual systems
is traditionally illustrated by the example of quantum measurements.
An outcome of a ``generic'' measurement of the state of
a quantum system is {\it not} deterministic. In the textbook
discussions this random element is blamed on the
``collapse of the wavepacket'', invoked
whenever a quantum system comes into contact with a classical apparatus.
In a fully quantum discussion this issue still
arises, in spite (or rather because) of the overall deterministic
quantum evolution of the state vector of the Universe:
As pointed out by von Neumann (1932),
there is no room for a `real collapse' in the
purely unitary models of measurements.

\subsection{Quantum conditional dynamics}

To illustrate the ensuing difficulties, consider
a quantum system ${\cal S}$ initially in a state $|\psi\rangle $
interacting with a quantum apparatus ${\cal A}$ initially in
a state $|A_0\rangle $:
\begin{eqnarray} |\Psi_0\rangle  = |\psi\rangle  |A_0\rangle  = 
\bigl( \sum_i a_i |s_i\rangle  \bigr)|A_0\rangle  \nonumber
\\ \longrightarrow
\sum_i a_i |s_i\rangle  |A_i\rangle  = |\Psi_t\rangle \ .
\eqnum{2.1}
\end{eqnarray}
Above, $\{|A_i\rangle \}$ and $\{|s_i\rangle \}$ are states in the Hilbert 
spaces of the apparatus and of the system, respectively, and $a_i$ are
complex coefficients. Conditional dynamics of such premeasurement
(as the step achieved by Eq. (2.1) is often called)
can be accomplished by means of a unitary Schr\"odinger evolution.
Yet it is not enough to claim that a measurement has been
achieved: Equation (2.1) leads to an uncomfortable conclusion:
$|\Psi_t\rangle $ is an EPR-like {\it entangled state}. Operationally, 
this EPR nature of the state emerging from the premeasurement
can be made more explicit by re-writing the sum in a different basis:
\begin{eqnarray}
& |\Psi_t\rangle  &=\sum_i a_i |s_i\rangle  |A_i\rangle  = \sum_i b_i |r_i\rangle  |B_i\rangle  \ .
\eqnum{2.2}
\end{eqnarray}
This freedom of basis choice -- basis ambiguity -- is guaranteed by 
the principle of superposition. Therefore, if one were to associate 
states of the apparatus (or the observer) with decompositions of
$|\Psi_t\rangle $, then even before enquiring about the specific outcome
of the measurement one would have to decide on the decomposition
of $|\Psi_t\rangle $; the change of the basis redefines the measured quantity.

\subsubsection{Controlled not and a bit-by-bit measurement}

The interaction required to entangle the measured
system and the apparatus, Eq. (2.1), is a generalization of
the basic logical operation known as a ``controlled not'' or a {\tt c-not}.
Classical, {\tt c-not} changes the state $a_t$ of the target when the
control
is 1, and does nothing otherwise:
$$ 0_{\tt c} \ a_{\tt t} \ \longrightarrow \ 0_{\tt c} \ a_{\tt t} \  ; \ \ \
1_{\tt c} \ a_{\tt t} \ \longrightarrow \ 1_{\tt c}  \ \neg a_{\tt t}
\eqno(2.3) $$
Quantum {\tt c-not} is a straightforward quantum version of Eq. (2.3).
It was known as a ``bit by bit measurement'' (Zurek, 1981; 1983) and used to
elucidate the connection between entanglement and premeasurement already
before it acquired its present name and significance in the context of quantum
computation (see e.g. Nielsen and Chuang, 2000). Arbitrary superpositions of
the control bit and of the target bit states are allowed:
\begin{eqnarray} (\alpha | 0_{\tt c} \rangle  ~ &+& ~ \beta | 1_{\tt c} \rangle ) |
a_{\tt t} \rangle  \nonumber \\
& &\ \longrightarrow  \
\alpha | 0_{\tt c} \rangle  | a_{\tt t} \rangle  ~ + ~ \beta | 1_{\tt c} \rangle  | \neg a_{\tt
t} \rangle  \eqnum{2.4}
\end{eqnarray}
Above ``negation'' $|\neg a_{\tt t}\rangle $ of a state is basis dependent;
$$\neg (\gamma|0_{\tt t}\rangle +\delta |1_{\tt t}\rangle ) =\gamma| 1_{\tt t} \rangle 
+ \delta | 0_{\tt t} \rangle   \eqno(2.5)$$
With $ |A_0\rangle  = |0_{\tt t}\rangle , \ |A_1\rangle  = |1_{\tt t}\rangle $ we have
an obvious analogy between the {\tt c-not} and a premeasurement.

In the classical controlled not the direction of information transfer is
consistent with the designations of the two bits: The state of
the control remains unchanged while it influences the target, Eq. (2.3).
Classical measurement need not influence the system. Written in
the logical basis $\{|0\rangle ,|1\rangle \}$, the truth table of the quantum
{\tt c-not} is essentially -- that is, save for the possibility
of superpositions -- the same as Eq. (2.3). One might have anticipated
that the direction of information transfer and the designations
(``control/system'' and ``target/apparatus'') of the two qubits
will be also unambiguous, as in the classical case.
This expectation is incorrect. In the conjugate basis
$\{|+\rangle , |-\rangle \}$ defined by:
$$ |\pm\rangle  = (|0\rangle  \pm |1\rangle )/\sqrt{2} \ , \eqno(2.6) $$
the truth table of Eq. (2.3) along with Eq. (2.6)
lead to a new complementary truth table:
\begin{eqnarray}|\pm\rangle |+\rangle  \longrightarrow |\pm\rangle |+\rangle ; 
\ \ |\pm\rangle |-\rangle \longrightarrow |\mp\rangle|-\rangle\ .\eqnum{2.7,~2.8}
\end{eqnarray}
In the complementary basis $\{ |+\rangle ,|-\rangle  \}$ roles of
the control and of the target are reversed: The former
(basis $\{|0\rangle ,|1\rangle \}$) target -- represented by
the second ket above -- remains unaffected,
while the state of the former control
(the first ket) is conditionally ``flipped''.

In the bit-by-bit case the measurement interaction is:
\begin{eqnarray}H_{int} \ = \ g|1\rangle 
\langle 1|_{\cal S} |-\rangle \langle -|_{\cal A}\ =\ \nonumber \\
{g \over 2}|1\rangle \langle 1|_{\cal S} \otimes ({\bf 1} - (|0\rangle 
\langle 1| + |1\rangle \langle 0|))_{\cal A} \eqnum{2.9}
\end{eqnarray}
Above, $g$ is a coupling constant, and the two operators refer to the system
(i.e., to the former control), and to the apparatus pointer (the former
target),
respectively. It is easy to see that the states $\{|0\rangle ,|1\rangle \}_{\cal S}$ of
the system are unaffected by $H_{int}$, since;
$$ [H_{int}, e_0|0\rangle \langle 0|_{\cal S} + e_1|1\rangle \langle 1|_{\cal S} ] = 0 \ \eqno(2.10)$$
The measured (control) observable $\hat \epsilon = e_0|0\rangle \langle 0| +
e_1|1\rangle \langle 1|$ is a constant of motion under $H_{int}$. {\tt c-not} requires
interaction time $t$ such that $gt= \pi/2$.

The states $\{|+\rangle ,|-\rangle \}_{\cal A}$ of the apparatus encode the information
about phase between the logical states. They have exactly the same
``immunity'':
$$ [H_{int}, f_+|+\rangle \langle +|_{\cal A} + f_-|-\rangle \langle -|_{\cal A} ] = 0 \ \eqno(2.11)$$
Hence, when the apparatus is prepared in a definite phase state (rather than
in a definite pointer/logical state), it will pass on its phase onto the
system, as Eqs. (2.7) - (2.8), show. Indeed, $H_{int}$ can be written as:
\begin{eqnarray}&  H_{int} \ = \ g|1\rangle \langle 1|_{\cal S} |-\rangle \langle -|_{\cal A} &
\nonumber \\
& = {g \over 2}
({\bf 1} - (|-\rangle \langle +| + |+\rangle \langle -|))_{\cal S}\otimes |-\rangle \langle -|_{\cal A} & \eqnum{2.12}
\end{eqnarray}
making this ``immunity'' obvious.

This basis-dependent direction of information flow in a quantum
{\tt c-not} (or in a premeasurement) is a consequence of complementarity.
While the information about the observable with the eigenstates
$\{|0\rangle ,|1\rangle \}$ travels from the system to the apparatus, in
the complementary $\{|+\rangle ,|-\rangle \}$ basis it seems that the apparatus is
measured by the system. This remark (Zurek 1998a,~b; see also 
Beckman, Gottesman, and Nielsen, 2001) clarifies the sense in which 
phases are inevitably ``disturbed'' in measurements. They are not really 
destroyed, but, rather, as the apparatus measures a certain observable 
of the system, the system simultaneously ``measures'' phases between 
the possible outcome states of the apparatus. This leads to loss of 
phase coherence: Phases become ``shared property'' as we shall see 
in more detail in the discussion of envariance.

The question ``what measures what?"
(decided by the direction of the information flow)
depends on the initial states.
In ``the classical practice'' this ambiguity does not arise.
Einselection limits the set of possible states of
the apparatus to a small subset.

\subsubsection{Measurements and controlled shifts.}

The truth table of a whole class of {\tt c-not} like transformations that
includes general premeasurement, Eq. (2.1), can be written as:
$$ |s_j\rangle  |A_k\rangle  \longrightarrow |s_j\rangle  |A_{k+j}\rangle  \ \eqno(2.13)$$
Equation (2.1) follows when $k=0$. One can therefore model measurements
as controlled shifts -- {\tt c-shift}s -- generalizations of the
{\tt c-not}. In the bases $\{ |s_j\rangle \}$ and $\{|A_k\rangle \}$, the
direction of the information flow appears to be unambiguous -- from the
system ${\cal S}$ to the apparatus ${\cal A}$. However, a complementary
basis can be readily defined (Ivanovic, 1981; Wootters and Fields, 1989);
$$|B_k\rangle \ =\ N^{ - {1 \over 2 }} \sum_{l=0}^{N-1}\exp ({{2 \pi i} \over N}
kl) ~ |A_l\rangle 
\ . \eqno(2.14a)$$
Above $N$ is the dimensionality of the Hilbert space.
Analogous transformation can be carried out on the basis $\{|s_i\rangle \}$ of the
system, yielding states $\{ | r_j \rangle  \}$.

Orthogonality of $\{ | A_k\rangle \}$ implies:
$$ \langle  B_l | B_m \rangle  = \delta _{lm} \ . \eqno (2.15)$$
$$|A_k\rangle \ =\ N^{ - {1 \over 2 }} \sum_{l=0}^{N-1}\exp (-{{2 \pi i} \over
N} kl) ~ |B_l\rangle  \eqno(2.14b)$$
inverts of the transformtion of Eq. (2.14a). Hence:
$$ |\psi \rangle  = \sum_l \alpha_l |A_l\rangle  = \sum_k \beta_k |B_k\rangle  \ , \eqno(2.16) $$
where the coefficients $\beta_k$ are:
$$ \beta_k = N^{ - {1 \over 2 }} \sum_{l=0}^{N-1} \exp(-{{2 \pi i } \over
N} kl) \alpha_l \ . \eqno(2.17)$$
Hadamard transform of Eq. (2.6) is a special case 
of the more general transformation considered here.

To implement the truth tables
involved in premeasurements we define observable $\hat A$
and its conjugate:
$$ \hat A = \sum_{k=0}^{N-1} k |A_k \rangle  \langle  A_k|; \
    \hat B = \sum_{l=0}^{N-1} l |B_l \rangle  \langle  B_l| \ .\eqno(2.18a,~b)$$
The interaction Hamiltonian:
$$ H_{int} = g \hat s \hat B \ \eqno(2.19)$$
is an obvious generalization of Eqs. (2.9) and (2.12), with
$g$ the coupling strength and $\hat s$:
$$ \hat s =  \sum_{l=0}^{N-1} l |s_l\rangle \langle s_l| \ \eqno(2.20)$$
In the $\{ |A_k\rangle  \}$ basis $\hat B$ is a shift operator,
$$\hat B = {{i N} \over {2 \pi}} {\partial \over {\partial\hat
A}} \ . \eqno(2.21)$$
To show how $H_{int}$ works, we compute:
\begin{eqnarray}
&\exp(-i H_{int} t / \hbar) |s_j\rangle  |A_k\rangle   =   & \nonumber \\
&|s_j\rangle  N^{-{1 \over 2}}
\sum_{l=0}^{N-1}\exp[-i(jgt/\hbar + 2\pi k /N)l] |B_l\rangle 
\eqnum{2.22}
\end{eqnarray}
We now adjust the coupling $g$ and the duration of the interaction
$t$ so that the action $\iota$ expressed in Planck units $2 \pi \hbar$
is a multiple of $1/N$;
$$ \iota = gt/\hbar =  G * 2\pi / N \ . \eqno(2.23a)$$
For an integer $G$, Eq. (2.22) can be readily evaluated:
\begin{eqnarray}
\exp(-i H_{int} t / \hbar) |s_j\rangle  |A_k\rangle 
= |s_j\rangle  |A_{{\{k+G * j\}}_N}\rangle   .
\eqnum{2.24}\end{eqnarray}
This is a shift of the apparatus state by an amount
$G * j$ proportional to the eigenvalue $j$ of the state of
the system. $G$ plays the role of gain. The index $\{k+G * j\}_N$
is evaluated modulo $N$, where $N$ is the number of the
possible outcomes, that is, the dimensionality of
the Hilbert space of the apparatus pointer ${\cal A}$:
When $G * j > N$, the pointer will just rotate through
the initial ``zero''. The truth table for $G=1$ defines
a {\tt c-shift}, Eq. (2.13), and with $k=0$ leads to
a premeasurement, Eq. (2.1).

The form of the interaction, Eq. (2.19), in conjunction with
the initial state decide the direction of information transfer.
Note that -- as was the case with
the {\tt c-not}s -- the observable that
commutes with the interaction Hamiltonian will not be
perturbed:
$$ [H_{int}, \hat s] = 0 \eqno(2.25)$$
$\hat s$ commutes with $H_{int}$, and is therefore a non-demolition
observable (Braginsky, Vorontsov and Thorne, 1980; Caves et al, 1980;
Bocko and Onofrio, 1996).

\subsubsection{Amplification}

Amplification was often regarded as the process forcing
quantum potentialities to become classical reality.
Its example is an extension of the model of measurement described above.

Assume the Hilbert space of the apparatus pointer
is large compared with the space spanned by the eigenstates of the
measured observable $\hat s$:
$$ N= Dim ({\cal H}_{\cal A}) \gg Dim ({\cal H}_{\cal S}) = n
\eqno(2.26)$$
Then one can increase $\iota$ to an integer multiple $G$ of
$2\pi /N$. This was implicit in Eqs. (2.23a) and (2.24).
However, larger $\iota$ will lead to redundancy only when the
apparatus the Hilbert space has many more dimensions in then 
there are possible outcomes. Otherwise,
only ``wrapping'' of the same record will ensue. The simplest example 
of such wrapping: ({\tt c-not})$^2$ is the identity.
For $N \gg n$, however, one can attain gain:
$$ G = {{N g t } / {2 \pi \hbar}} \ . \eqno(2.23b)$$
The outcomes are now separated by $G-1$ ``empty'' eigenstates
of the record observable. In this sense $G\gg 1$ achieves redundancy,
providing that wrapping of the record is avoided, which is guaranteed
when:
$$ n G < N \ . \eqno(2.27)$$

Amplification is useful in presence of noise. For example,
it may be difficult to initiate the apparatus in $|A_0\rangle $,
so the initial state may be a superposition;
$$ |a_l\rangle  = \sum_k \alpha_l(k) |A_k\rangle  \ . \eqno(2.28a)$$
Indeed, typically a mixture of such superpositions;
$$ \rho_{\cal A}^0 = \sum_i w_i |a_i\rangle \langle a_i| \eqno(2.28b)$$
may be the starting point for a premeasurement.
Then:
\begin{eqnarray}|s_k\rangle \langle s_{k'}|\rho_{\cal A} = |s_k\rangle \langle s_{k'}|\sum_l w_l
|a_l\rangle \langle a_l| \nonumber \\
\longrightarrow |s_k\rangle \langle s_{k'}|\sum_l w_l|a_{l + Gk}\rangle \langle a_{l + Gk'}|
\eqnum{2.29}\end{eqnarray}
where $|a_{l+Gk}\rangle $ obtains from $|a_l\rangle $, Eq. (2.28a), through:
$$ |a_{l+Gk}\rangle  \ = \ \sum_j \alpha_l(j) |A_{j+Gk}\rangle  \ , \eqno(2.30)$$
and the simplifying assumption about the coefficients;
$$\alpha_l(j) = \alpha(j-l) \eqno(2.31)$$
has been made. Its aim to focus on
the case when the apparatus states are
peaked around a certain value $l$ (e.g., $\alpha_l(j) \sim
\exp(-(j-l)^2/2 \Delta^2)$, and where the form of their distribution
over $\{|A_k\rangle \}$ does not depend on $l$.

A good measurement allows one to distinguish states of the system. Hence,
it must satisfy:
\begin{eqnarray} & |\langle a_{l+Gk}|a_{l+Gk'}\rangle |^2 & \nonumber \\
& = |\sum_j \alpha(j+G(k-k'))
\alpha^*(j)|^2
\approx \delta_{k',k} \ . & \eqnum{2.32}\end{eqnarray}
States of the system that need to be distinguished should
rotate the pointer of the apparatus to the correlated outcome states
that are approximately orthogonal. When the coefficients
$\alpha(k)$ are peaked around $k= 0$ with dispersion $\Delta$, this implies:
$$ \Delta \ll G \ . \eqno(2.33)$$

In the general case of an initial mixture, Eq. (2.29),
one can evaluate the dispersion of the expectation value of
the record observable $\hat A$ as:
$$ \langle \hat A^2\rangle  - \langle \hat A\rangle ^2 \ = \ Tr \rho^0_{\cal A} \hat A^2 \ - \ (Tr
\rho^0_{\cal A} \hat A)^2 \eqno(2.34) $$
The outcomes are distinguishable when:
$$ \langle \hat A^2\rangle  - \langle \hat A\rangle ^2  \ll G \eqno(2.35)$$
Interaction with the environment yields a mixture
of the form of Eq. (2.29). Amplification
can protect measurement outcomes from noise through
redundancy.\footnote{The above model of amplification 
is unitary. Yet, it contains seeds of irreversibility. Reversibility
of {\tt c-shift} is evident: As the interaction continues, the two
systems will eventually disentangle.
For instance, it takes $t_e=2\pi\hbar/(g N)$ (see Eq. (2.23b) with $G=1$)
to entangle ${\cal S}$ $(Dim({\cal H_S}) = n)$ with an ${\cal A}$
with $Dim({\cal H_A}) = N \geq n$ pointer states. However, as the interaction
continues, ${\cal A}$ and ${\cal S}$ disentangle.
For a {\tt c-shift} this recurrence timescale is:
$ t_{Rec}=Nt_e=2\pi\hbar/g$. It corresponds to gain $G=N$. Thus, for
an instant of less than $t_e$ at
$t=t_{Rec}$ the apparatus disentangles
from the system, as $\{k+N*j\}_{N} = k$.
Reversibility results in recurrences of the initial state, 
but for $N \gg 1$, they are rare.
\\
For less regular interactions (e.g, involving environment)
recurrence time is much longer.
In that case, $t_{Rec}$ is, in effect, a Poincar\'e time:
$t_{Rec} \sim t_{Poincar\acute e} \approx N! t_e$.
In any case $t_{Rec} \gg t_e$ for large $N$. Undoing entanglement in this
manner would be exceedingly difficult because one would need to know
precisely when to look, and because one would need to isolate
the apparatus or the immediate environment from other degrees of
freedom -- their environments.
\\
The price of letting the entanglement undo itself by waiting for
an appropriate time interval is at the very least given  by the cost
of storing the information on how long is it necessary to wait.
In the special {\tt c-shift} case this is $\sim \log N$ memory bits.
In situations when eigenvalues of the interaction Hamiltonian
are not commensurate, it will be more like $\sim \log N! \approx N \log N$,
as the entanglement will get undone only after a Poincar\'e time.
Both classical and quantum case can be analyzed using algorithmic 
information . For related discussions see Zurek, (1989), Caves, (1994);
Schack and Caves (1996) and Zurek (1998b).
\\
Amplified correlations are hard to contain.
The return to purity after $t_{Rec}$ in the manner described above can be
hoped for only when either the apparatus or the immediate environment
${\cal E}$ (i.e., the environment directly interacting with the system)
cannot ``pass on" the information to their more remote environments 
${\cal E}'$. The degree of isolation required puts a stringent limit
on the coupling $g_{{\cal EE}'}$ between the two environments:
Return to purity can be accomplished in this manner only if
$t_{Rec} < t_{e^\prime} = 2 \pi \hbar /(N' g_{{\cal EE}'})$, where $N'$
is the dimension of the Hilbert space of the environment ${\cal E}'$.
Hence, the two estimates of $t_{Rec}$ translate into:
$ g_{{\cal EE}'} < {g / {N'}} $
for the regular spectrum, and the much tighter;
$ g_{{\cal EE}'} < {g / {N! N'}} $
for the random case more relevant for decoherence.
\\
In short, once information ``leaks'' into the correlations
between the system and the apparatus or the environment, keeping it from
spreading further ranges between very hard and next to impossible.
With the exception of very special
cases (small $N$, regular spectrum), the strategy of
``enlarging the system, so that it includes
the environment'' -- occasionally mentioned as an argument against
decoherence -- is doomed to fail, unless
the Universe as a whole is included. This is a questionable setting
(as the observers are {\it inside} this ``isolated'' system) and
in any case makes the relevant Poincar\'e time absurdly long!}

\subsection{Information transfer in measurements}

Information transfer is the objective of the measurement process. Yet,
quantum measurements were only rarely analyzed from that point of view.
As a result of the interaction of the system ${\cal S}$ with the apparatus
${\cal A}$, their joint state is still pure $|\Psi_t\rangle $, Eq. (2.1),
but each of the subsystems is in a mixture:
\begin{eqnarray}
    \rho_{\cal S} \ = \ Tr_{\cal A} |\Psi_t\rangle \langle \Psi_t| \ = \ \sum_{i=0}^{N-1}
|a_i|^2 |s_i\rangle \langle s_i| \ ; \ \  \eqnum{2.36a}  \\
    \rho_{\cal A} \ = \ Tr_{\cal S} |\Psi_t\rangle \langle \Psi_t| \ = \
\sum_{i=0}^{N-1}  |a_i|^2 |A_i\rangle \langle A_i| \ . \eqnum{2.36b}
\end{eqnarray}
Partial trace leads to {\it reduced density matrices}, here $\rho_{\cal S}$
and $\rho_{\cal A}$, important for what follows. They describe
subsystems to the observer who, before the premeasurement, knew pure
states of the system and of the apparatus, but who has access to only one
of them afterwards.

Reduced density matrix is a technical tool of paramount importance. It was
introduced by Landau (1927) as the only density matrix that gives rise to the
correct measurement statistics given the usual formalism that includes Born's
rule for calculating probabilities (see e.g. p. 107 of Nielsen and Chuang,
2000, for an insightful discussion). This remark will come to haunt us later
when in Section VI we shall consider the relation between decoherence and
probabilities: In order to derive Born's rule it will be important not to
assume it in some guise!

Following premeasurement, the information about the subsystems available to
the observer locally decreases. This is quantified by the increase of the entropies:
\begin{eqnarray} H_{\cal S} & = &
-Tr\rho_{\cal S} \log \rho_{\cal S} = -\sum_{i=0}^{N-1}|a_i|^2 \log
|a_i|^2\nonumber \\ &   = &
-Tr\rho_{\cal A} \log \rho_{\cal A} =  H_{\cal A}
\eqnum{2.37}\end{eqnarray}
As the evolution of the whole ${\cal SA}$
is unitary, the increase of entropies in the
subsystems is compensated by the increase of the mutual information:
\begin{eqnarray} {\cal I (S:A)} = H_{\cal S} + H_{\cal A} -
H_{\cal SA}
    = -2\sum_{i=0}^{N-1}|a_i|^2 \log |a_i|^2
\eqnum{2.38}\end{eqnarray}
It was used in quantum theory as a measure of
entanglement (Zurek, 1983; Barnett and Phoenix, 1989).

\subsubsection{Action per bit}

An often raised question concerns the price of information in units of some
other ``physical currency'', (Brillouin, 1962; 1964; Landauer, 1991).
Here we shall establish that the least action
necessary to transfer one bit is of the order
of a fraction of $\hbar$ for quantum systems with
two-dimensional Hilbert spaces. Information transfer can be
made cheaper on the ``wholesale'' level, when
the systems involved have large Hilbert spaces.

Consider Eq. (2.1). It evolves initial product state of the two
subsystems into a superposition of product states,
$ ( \sum_j \alpha_j |s_j\rangle  ) |A_0\rangle  \longrightarrow \sum_j\alpha_j|s_j\rangle |A_j\rangle $.
The expectation value of the action involved is no less than:
$$ I = \sum_{j=0}^{N-1} |\alpha_j|^2 \arccos |\langle A_0|A_j\rangle | \eqno(2.39)$$
When $\{ |A_j\rangle \}$ are mutually orthogonal, the action is:
$$ I = \pi / 2 \eqno(2.40)$$
in Planck ($\hbar$) units. This estimate can be lowered by using
the initial $|A_0\rangle $, a superposition
of the outcomes $| A_j\rangle $.
In general, interaction of the form:
\begin{eqnarray} H_{\cal SA} = i g \sum_{k=0}^{N-1} |s_k\rangle \langle s_k|
\sum_{l=0}^{N-1} (|A_k\rangle \langle A_l| - h.c.) \ \eqnum{2.41} \end{eqnarray}
saturates the lower bound given by:
$$ I = \arcsin\sqrt {1-1/N} \ , \eqno(2.42)$$
For a two-dimensional Hilbert
space the average action can be thus brought down to
$\pi \hbar/4$ (Zurek, 1981; 1983).

As the size of the Hilbert space increases, action involved
approaches the asymptotic estimate of Eq. (2.40).
The entropy of entanglement can be as large as $\log N$
where $N$ is the dimension of the Hilbert space of the smaller
of the two systems. Thus, the least action per bit
of information decreases with the increase of $N$:
$$ \iota =  {I \over {\log_2 N}} \approx {\pi \over {2 \log_2 N}}
\eqno(2.43)$$
This may be one reason why information appears ``free''
in the macroscopic domain, but expensive (close to $ \hbar$/bit)
in the quantum case of small Hilbert spaces.

\subsection{``Collapse'' analogue in a classical measurement}

Definite outcomes we perceive appear to be at odds with the principle
of superposition. They can nevertheless occur also in quantum physics
when the initial state of the measured system is -- already before the
measurement -- in one of the eigenstates of the measured observable.
Then Eq. (2.1) will deterministically rotate the pointer of the apparatus to
the appropriate record state. The result of such a measurement can be
predicted by an {\it insider} -- an observer aware of the initial state
of the system. This {\it a priori} knowledge can be represented by the
preexisting record $|A_i\rangle $, which is only corroborated by an additional
measurement:
$$ |A_i\rangle  |A_0\rangle  |\sigma_i\rangle  \longrightarrow |A_i\rangle  |A_i\rangle |\sigma_i\rangle \
.\eqno(2.44a)$$
In classical physics complete information about the initial state of
an isolated system always allows for an exact prediction of its future state:
A well-informed observer will be even able to predict future of the
classical Universe as a whole (``Laplace's demon''). 

Any element of surprise (any use
of probabilities) must be therefore blamed on partial ignorance. Thus, when
the information available initially does not include the exact initial state
of the system, observer can use an ensemble described by $\rho_{\cal S}$ 
-- by a list of possible initial states $\{|\sigma_i\rangle \}$ and 
their probabilities $p_i$. This is the {\it ignorance interpretation} of
probabilities. We shall see in section VI that -- using quantum envariance --
one can justify ignorance about a part of the system by relying on perfect
knowledge of the whole. 

Through measurement observer finds out which of the potential outcomes
consistent with his prior (lack of) information actually happens. This
act of acquisition of information changes physical state of the observer 
-- the state of his memory: The initial memory state
containing description $A_{\rho_S}$ of an ensemble and  a ``blank'' $A_0$,
$|A_{\rho_{\cal S}}\rangle  \langle  A_{\rho_{\cal S}}|~|A_0\rangle  \langle  A_0|$,
is transformed into record of a specific outcome:
$|A_{\rho_{\cal S}}\rangle  \langle  A_{\rho_{\cal S}}|~|A_i\rangle  \langle  A_i|$.
In quantum notation this process will be described by such a {\it discoverer} 
as a random ``collapse'':
\begin{eqnarray} |A_{\rho_{\cal S}}\rangle \langle A_{\rho_{\cal S}}|~|A_0\rangle \langle A_0|~
\sum_i p_i |\sigma_i\rangle \langle \sigma_i| \nonumber \\
\longrightarrow
|A_{\rho_{\cal S}}\rangle \langle A_{\rho_{\cal S}}|~|A_i\rangle \langle A_i|~
|\sigma_i\rangle \langle \sigma_i| \ .
\eqnum{2.44b}
\end{eqnarray}
This is only the description of what happens ``as reported by the
discoverer''.
Deterministic representation of this very same process by Eq. (2.44a) is still
possible. In other words, in classical physics discoverer can be always
convinced that the system was in a state $|\sigma_i\rangle $ already before the record
is made in accord with Eq. (2.44b). 

This sequence of events as seen by the discoverer looks like a ``collapse'' 
(see also Zurek, 1998a,b): For instance, insider who 
knew the state of the system before discoverer carried out his measurement 
need not notice any change of that state when he makes further ``confirmatory'' 
measurements. This property is the cornerstone of the ``reality''  of classical 
states -- they need not ever change as a consequence of measurements.
We emphasize, however, that while the state of the system may remain unchanged,
the state of the observer must change to reflect the acquired information.

Last not least, an {\it outsider} -- someone who knows about the measurement,
but (in contrast to the insider) not about the initial state of the system nor
(in contrast to the discoverer) about the measurement outcome, will describe
the same process still differently:
\begin{eqnarray}|A_{\rho_{\cal S}}\rangle \langle  A_{\rho_{\cal S}}|~|A_0\rangle \langle A_0|~
\sum_i p_i |\sigma_i\rangle \langle \sigma_i|   \longrightarrow \nonumber \\
    \longrightarrow |A_{\rho_{\cal S}}\rangle \langle  A_{\rho_{\cal S}}|
( \sum_i p_i |A_i\rangle \langle A_i|~ |\sigma_i\rangle \langle \sigma_i| ) \ . \eqnum{2.44c}
\end{eqnarray}
This view of the outsider, Eq. (2.44c), combines one-to-one classical
correlation of the states of the system and the records with the indefiniteness
of the outcome.

We have just seen three distinct quantum-looking descriptions of the very same
{\it classical} process (see Zurek, 1989, and Caves, 1994 for previous
studies of the insider - outsider theme). They differ only in the information
available {\it ab initio} to the observer. 
The information in possession of the observer prior to the
measurement determines in turn whether -- to the observer -- the evolution
appears to be (a) a confirmation of the preexisting data, Eq. (2.44a),
(b) a ``collapse'' associated with the information gain, Eq. (2.44b)
-- and with the entropy decrease translated into algorithmic randomness of
the acquired data (Zurek, 1989; 1998b) -- or (c) an entropy-preserving
establishment of a correlation, Eq. (2.44c). All three descriptions are
classically compatible, and can be implemented by the same (deterministic
and reversible) dynamics.

{\it In classical physics the insider view always exists
in principle. In quantum physics it does not.} Every observer in
a classical Universe could in principle aspire to be an ultimate insider.
The fundamental contradiction between every observer knowing precisely the
state of the rest of the Universe (including the other observers) can
be swept under the rug (if not really resolved) in the Universe where the
states are infinitely precisely determined and the observer's records
(as a consequence of the $\hbar \rightarrow 0$ limit) may have an infinite
capacity for information storage. However -- given a set value of
$\hbar$ -- information storage resources of any finite physical system are
finite. Hence, in quantum physics observers remain largely ignorant of the 
detailed state of the Universe as there can be {\it no information without
representation} (Zurek, 1994).

Classical ``collapse'' is described by Eq. (2.44b): The observer discovers
the state of the system. From then on, the state of the system will remain
correlated with his record, so that all the future outcomes can be predicted,
in effect by iterating Eq. (2.44a). This disappearance of all the potential
alternatives save for one that becomes a ``reality'' is the essence of
the collapse. There need not be anything quantum about it.

Einselection in observers memory provides many of the ingredients
of the ``classical collapse'' in the quantum
context. In presence of einselection, one-to-one
correspondence between the state of the observer and his knowledge
about the rest of the Universe can be firmly established, and (at least in
principle) operationally verified: One could measure bits in observers
memory and determine what he knows without altering his records -- without 
altering his state. After all, one can do so with a classical computer.
Existential interpretation recognizes that the information possessed by 
the observer is reflected in his einselected state, explaining his
perception of a single ``branch'' -- ``his'' classical Universe.

\section{CHAOS AND LOSS OF CORRESPONDENCE}

The study of the relationship between the quantum and the classical 
has been -- for a long time -- focused almost entirely on measurements.
However, the problem of measurement is difficult to discuss
without observers. And once observer enters,
it is often hard to avoid its ill-understood anthropic attributes such
as consciousness, awareness, and the ability to perceive.

We shall sidestep these presently ``metaphysical'' problems 
and focus on the information-processing underpinnings of the 
``observership''. It is nevertheless fortunate that there is another
problem with the quantum - classical correspondence that leads to 
interesting questions not motivated by measurements. As was anticipated 
by Einstein (1917) before the advent of modern quantum theory, 
chaotic motion presents such a challenge. Full implications of classical
dynamical chaos were understood much later. The concern about the
quantum-classical correspondence in this modern context dates to
Berman and Zaslavsky (1978) and Berry and Balazs (1979) (see Haake, 1991
and Casati and Chirikov, 1995a, for references).  It has even led some 
to question validity of quantum theory (Ford and Mantica, 1992).

\subsection{Loss of the quantum-classical correspondence}

The interplay between quantum interference and chaotic exponential 
instability leads to the rapid loss of the quantum-classical correspondence.
Chaos in dynamics is characterized by the exponential divergence
of the classical trajectories. As a consequence, a small
patch representing the probability density in phase space is 
exponentially stretching in unstable directions and to exponentially
compressing in the stable directions. The rates of stretching and compression
are given by positive and negative Lyapunov exponents $\Lambda_i$.
Hamiltonian evolution demands that the sum of all
the Lyapunov exponents be zero. In fact, they appear in $\pm \Lambda_i$
pairs. 

Loss of the correspondence in chaotic
systems is a consequence of the exponential stretching of the effective
support of the probability distribution in the unstable direction (say, $x$)
and its exponential narrowing in the complementary direction
(Zurek and Paz, 1994; Zurek, 1998b). As a consequence, classical
probability distribution will develop structures on the scale:
$$ \Delta p \sim \Delta p_0 ~ \exp(- \Lambda t) \ . \eqno(3.1)$$
Above, $\Delta p_0 $ is the measure of the initial momentum spread.
$\Lambda $ is the net rate of contraction in the direction of momentum
given by the Lyapunov exponents (but see Boccaletti, Farini,
and Arecchi, 1997). In the real chaotic system stretching and narrowing
of the probability distribution in both $x$ and $p$ occur simultaneously,
as the initial patch is rotated and folded. Eventually, the envelope
of its effective support will swell to fill in the available phase
space, resulting in the wavepacket coherently spread over the spatial region
of no less than;
$$ \Delta x \sim ( \hbar/\Delta p_0 ) ~ \exp(\Lambda t) \  \eqno(3.2)$$
until it becomes confined by the potential, while the small-scale structure 
will continue to descend to ever smaller scales (Fig. 1). Breakdown of
the quantum-classical correspondence can be understood in two complementary
ways, either as a consequence of small $\Delta p$ (see the discussion of 
Moyal bracket below), or as a result of large $\Delta x$.

Coherent exponential spreading of the wavepacket -- large $\Delta x$ --
must cause problems with correspondence. This is inevitable, as classical
evolution appeals to the idealization of a point in phase space acted upon 
by a force given by the gradient $\partial_x V$ of the potential
$V(x)$ evaluated at that point. But quantum wavefunction
can be coherent over a region larger than the nonlinearity
scale $\chi$ over which the gradient of the potential changes
significantly. $\chi$ can be usually estimated by:
$$\chi \simeq \sqrt{\partial_x V/\partial_{xxx} V} \ , \eqno(3.3)$$
and is typically of the order of the size $L$ of the system:
$$ L \sim \chi \ . \eqno(3.4)$$

An initially localized state evolving in accord with Eqs. (3.1) and (3.2)
will spread over such scales after:
$$t_{\hbar} \simeq \Lambda^{-1} \ln {{\Delta p_0 \chi} \over \hbar} \ .
\eqno(3.5)$$
It is then impossible to tell what is the force acting on the system,
as it is not located in any specific $x$. This estimate of what
can be thought of as Ehrenfest time -- the time over which a quantum system
that has started in a localized state will continue to be sufficiently
localized for the quantum corrections to the equations of motion obeyed by
its expectation values to be negligible (Gottfried, 1966) -- is valid for
chaotic systems. Logarithmic dependence is the result of inverting of
the exponential sensitivity. In the absence of the exponential instability
($\Lambda = 0$) divergence of trajectories is typically polynomial,
and leads to a power law dependence, $t_{\hbar}\sim (I/\hbar)^{\alpha}$,
where $I$ is the classical action. Thus, macroscopic (large $I$) integrable
systems can follow classical dynamics for a very long time, providing they
were initiated in a localized state. For chaotic systems $t_{\hbar}$ also
becomes infinite in the limit $\hbar \rightarrow 0$, but that happens only
logarithmically slowly. As we shall see below, in the context of
quantum-classical correspondence this is too slow for comfort.

Another way of describing the ``root cause'' of the correspondence breakdown
is to note that after the timescale of the order of $t_{\hbar}$ quantum wave
function of the system would have spread over all of the available space, and
is being forced to ``fold'' onto itself. Fragments of the wavepacket arrive at
the same location (although with different momenta, and having followed
different paths). The ensuing evolution critically depends on whether
they have retained phase coherence. When coherence persists, a complicated
interference event decides the subsequent evolution. And -- as can be
anticipated from the double slit experiment -- there is a big difference
between coherent and incoherent folding in the configuration space. This
translates into the loss of correspondence, which sets in surprisingly
quickly, at $t_{\hbar}$.

To find out how quickly, we
estimate $t_{\hbar}$ for an obviously macroscopic Hyperion, chaotically
tumbling moon of Saturn (Wisdom, 1985). Hyperion has a prolate shape
of a potato and moves on an eccentric orbit
with a period $t_O=$21 days. Interaction
between its gravitational quadrupole and the tidal field of Saturn leads to
chaotic tumbling with Lyapunov time $\Lambda^{-1}\simeq $42 days.

To estimate the time over which orientation of Hyperion becomes
delocalized, we use a formula (Berman and Zaslavsky, 1978,
Berry and Balazs, 1979):
$$t_r=\Lambda^{-1}\ln{{LP} \over \hbar}=\Lambda^{-1} \ln {{I} \over
\hbar} \
\eqno(3.6)$$
Above $L$ and $P$ give the range of values of the coordinate and
momentum in
phase space of the system. Since $L \simeq \chi$ and $P > \Delta p_0$,
it follows that
$t_r \geq t_{\hbar}$. On the other hand, $LP \simeq I$, the action of the
system.

The advantage of Eq. (3.6) is its insensitivity to initial conditions,
and the ease with which the estimate can be obtained. For Hyperion,
a generous overestimate of the classical action $I$ can be obtained from
its binding energy $ E_B $ and its orbital time $t_{O}$:
$$ I/\hbar \simeq  E_B t_O/\hbar \simeq 10^{77} \eqno(3.7)$$
The above estimate (Zurek, 1998b) is ``astronomically" large.
However, in the calculation of the loss of correspondence,
Eq. (3.6), only its logarithm enters. Thus,
$$ t_{r}^{Hyperion} \simeq 42 [{\rm days}] \ln 10^{77} \simeq 20 [{\rm
yrs}] \eqno(3.8)$$
After approximately 20 years Hyperion would be in a coherent superposition
of orientations that differ by $2 \pi$!

We conclude that after a relatively short time an obviously macroscopic
chaotic system becomes forced into a flagrantly non-local
``Schr\"odinger cat" state. In the original discussion (Schr\"odinger,
1935a,b) an intermediate step
in which decay products of the nucleus were {\it measured} to determine
the fate of the cat was essential. Thus, it was possible to maintain that
the preposterous superposition of the dead and alive cat could be avoided
providing that quantum measurement (with the ``collapse" it presumably
induces) was properly understood.

This cannot be the resolution for chaotic quantum systems.
They can evolve -- as the example of Hyperion demonstrates -- into states
that are non-local, and, therefore, extravagantly quantum,
simply as a result of the
exponentially unstable dynamics. Moreover, this happens surprisingly
quickly,
even for very macroscopic examples. Hyperion is not the only chaotic system.
There are asteroids that have
chaotically unstable orbits (e.g., Chiron), and even indications that
the solar system as a whole is chaotic (Laskar, 1989; Sussman
and Wisdom, 1992). In all of these cases straightforward estimates of
$t_{\hbar}$ yield answers much smaller than the age of the solar system.
Thus, if unitary evolution of closed subsystems was a complete
description of planetary dynamics, planets would be delocalized
along their orbits.

\subsection{Moyal bracket and Liouville flow}

Heuristic argument about breakdown of the quantum-classical
correspondence can be made more rigorous with the help of the Wigner
function. We start with the von Neumann equation:
$$ i \hbar \dot \rho = [ H, \rho] \ . \eqno(3.9)$$
It can be transformed into the equation for Wigner
function $W$, which is defined in phase space:
$$ W(x,p) = {1 \over {2 \pi \hbar}} \int \exp({{ipy}\over \hbar})
\rho(x-{y \over 2},
x +{y \over 2}) dy  \ . \eqno(3.10)$$
The result is:
$$\dot W = \{ H, W\}_{MB} \ . \eqno(3.11)$$
Here $\{...,...\}_{MB}$ stands for Moyal bracket,
the Wigner transform of the von Neumann bracket (Moyal, 1949).

Moyal bracket can be expressed in terms of the Poisson bracket $\{...,...\}$,
which generates Liouville flow in the classical phase space, by the formula:
$$ i \hbar \{...,...\}_{MB} = \sin ( i \hbar \{..., ...\} ) \ . \eqno(3.12)$$
When the potential $V(x)$ is analytic, Moyal bracket can be expanded
(Hillery et al, 1984) in powers of $\hbar$:
\begin{eqnarray} \dot W = \{H, W\}
+ \sum_{n \geq 1}{ {\hbar^{2n} (-)^n} \over
{2^{2n} (2n + 1)!}} \partial_x^{2n+1}V \partial_p^{2n+1}W \ .
\eqnum{3.13} \end{eqnarray}
The first term is just the Poisson bracket. Alone, it would
generate classical motion in phase space. However, when the evolution is
chaotic, quantum corrections (proportional to the odd order momentum
derivatives of the Wigner function) will eventually dominate the
right hand side of Eq. (3.10). This is because the exponential squeezing
of the initially regular patch in phase space (which begins its evolution
in the classical regime, where the Poisson bracket dominates) leads to an
exponential explosion of the momentum derivatives. Consequently, after
a time logarithmic in $\hbar$, (Eqs. (3.5), (3.6))
the Poisson bracket will cease to be a good estimate of the right hand side of
Eq. (3.13).

The physical reason for the ensuing breakdown of the quantum-classical
correspondence was already explained before: Exponential instability of the
chaotic evolution delocalizes the wavepacket. As a result, the force acting on
the system is no longer given by the gradient of the potential evaluated at
the location of the system: It is not even possible to say where the system
is, since it is in a superposition of many distinct locations. Consequently,
phase space distribution and even the typical observables of the system
noticeably differ when evaluated classically and quantum mechanically
(Haake, Ku\'s and Sharf, 1987;  Habib, Shizume and Zurek, 1998; Karkuszewski,
Zakrzewski, and Zurek, 2002). Moreover, this will happen
after an uncomfortably short time $t_{\hbar}$.

\subsection{Symptoms of correspondence loss}

Wavepacket becomes rapidly delocalized in a chaotic system, and the
correspondence between classical and quantum is quickly lost. Schr\"odinger
cat states appear after $t_{\hbar}$ and this is the overarching
interpretational as well as physical problem: In the familiar real
world we never seem to encounter such smearing of the wavefunction even 
in the examples of chaotic dynamics it is predicted by quantum theory.

\subsubsection{Expectation values}

Measurements often average out fine phase space structures, which may 
be striking, but experimentally inaccessible symptom of breakdown of 
the correspondence.
Thus, one might hope that when interference patterns in the Wigner
function are ignored by looking at the coarse-grained distribution,
the quantum results should be in accord with the classical.
This would not exorcise the `chaotic cat' problem. Moreover,
breakdown of correspondence can be also seen in the expectation
values of quantities that are smooth in phase space.

Trajectories diverge exponentially in a chaotic system. A comparison 
between expectation values for a single trajectory 
and for a delocalized quantum state (which is how the Ehrenfest
theorem mentioned above is usually stated) would clearly lead to
a rapid loss of correspondence. One may object to the use
of a single trajectory and argue that both quantum and classical state
be prepared and accessed only through measurements that are subject 
to the Heisenberg indeterminacy. Thus, it should be fair to compare 
averages over an evolving Wigner function with an initially 
identical classical probability distribution (Ballentine, Yang, 
and Zibin, 1994; Haake, Ku\'s \& Sharf, 1987; Fox and Elston, 1994; 
Miller, Sarkar and Zarum, 1998). These are shown in Fig. 2 for 
an example of a driven chaotic system.
Clearly, there is reason for concern: Fig. 2 
(corroborated by other studies, see e.g.
Karkuszewski, Zakrzewski, and Zurek, 2002, for references)
demonstrates that not just phase space portrait but also averages 
diverge at a time $\sim t_{\hbar}$.

In integrable systems rapid breakdown of correspondence may still
occur, but only for very special initial conditions. It is due to the local
instability in phase space.  Indeed, a double - slit experiment is
an example of a regular system in which a local instability
(splitting of the paths) leads to correspondence loss, but 
only for judiciously selected initial conditions. Thus, one may 
dismiss it as a consequence of a rare pathological starting point, 
and argue that the conditions that lead to
discrepancies between classical and quantum behavior
exist, but are of measure zero in the classical limit.

In the chaotic case the loss of correspondence is typical. As shown
in Fig. 2, it happens after a disturbingly short $t_{\hbar}$ for
generic initial conditions. The time at which the quantum and classical
expectation values diverge in the example studied here is consistent
with the estimates of $t_{\hbar}$, Eq. (3.5), but exhibits a significant
scatter. This is not too surprising -- exponents characterizing
local instability vary noticeably with the location in phase space.
Hence, stretching and contraction in phase space  will occur
at a rate that depends on the specific trajectory. Dependence of its
typical magnitude as a function of $\hbar$ is still not clear. Emerson
and Ballentine (2001a\&b) study coupled spins and argue that it is of order 
of $\hbar$, but Fig. 2 suggests it decreases more slowly than that, and 
that it may be only logarithmic in $\hbar$ (Karkuszewski et al. 2002).

\subsubsection{Structure saturation}

Evolution of the Wigner function exhibits rapid buildup of interference
fringes. These fringes become progressively smaller, until saturation
when the wavepacket is spread over the available phase space. At that
time their scales in momentum and position are typically:
$$ dp = \hbar / L \ \ ; \ \ dx = \hbar / P \eqno(3.14,~3.15)$$
where $L$ ($P$) defines the range of positions (momenta) of the effective 
support of $W$ in phase space.

Hence, smallest structures in the Wigner function occur (Zurek, 2001)
on scales corresponding to action:
$$  a \ = \ dx~ dp \ = \ \hbar \times { \hbar /  {LP}} \ = \
    \hbar^2  / I \ , \eqno(3.16)$$
where $I \simeq L P$ is the classical action of the system.
Action $a \ll \hbar$ for macroscopic $I$. 

Sub-Planck structure is a kinematic property of quantum states. 
It limits their sensitivity to perturbations, and has applications
outside quantum chaos or decoherene. For instance, Schr\"odinger cat state 
can be used as a weak force detector (Zurek, 2001), and its sensitivity
is determined by Eqs. (3.14)-(3.16).

Structure saturation on scale  $a$ is an important
distinction between quantum and classical. In chaotic systems
smallest structures in the classical probability density exponentially 
shrink with time, in accord with Eq. (3.1) (see Fig. 1). 
Equation (3.16) has implications for decoherence, as $a$ controls 
sensitivity of the systems as well as environments (Zurek, 2001; Karkuszewski, 
Jarzynski, and Zurek, 2002): As a result of smallness of $a$, Eq. (3.16),
and as anticipated by Peres (1993), quantum systems are more sensitive to
perturbations when their classical counterparts are chaotic 
(see also Jalabert and Pastawski, 2001). But -- in contrast 
to classical chaotic systems -- they are not exponentially sensitive 
to infinitesimally small perturbations: Rather, the smallest perturbations 
that can effective are set by Eq. (3.16).

Emergence of Schr\"odinger cats through dynamics is a challenge
to quantum - classical correspondence. It is not yet clear to what extent
one should be concerned about the discrepancies between quantum
and classical averages: The size of this discrepancy may or may not be
negligible. But in the original Schr\"odinger cat problem quantum
and classical expectation values (for the survival of the cat) were
also in accord. In both cases it is ultimately the state of the cat which
is most worrisome. 

Note that we have not dealt with dynamical localization (Casati 
and Chirikov, 1995a). This is because it appears after too long 
a time ($\sim \hbar^{-1}$) to be a primary concern in the macroscopic 
limit, and is quite sensitive to small perturbations of the potential 
(Karkuszewski, Zakrzewski, and Zurek, 2002).

\section{ENVIRONMENT -- INDUCED SUPERSELECTION}

The principle of superposition applies only when the quantum system is closed.
When the system is open, interaction with the environment results in 
an incessant monitoring of some of its observables. As a result, pure states 
turn into mixtures that rapidly diagonalize in the einselected states.
These pointer states are chosen with the help of the interaction Hamiltonian
and are independent of the initial state of the system. Their predictability
is key to the effective classicality (Zurek, 1993a; Zurek, Habib, and Paz,
1993).

Environments can be external (such as particles of the air or
photons that scatter off, say, the apparatus pointer) or internal
(collections of phonons or other internal excitations). Often,
environmental degrees of freedom emerge from the split of the original
set of degrees of freedom into the ``system of interest'' which may be
a collective observable (e.g., an order parameter in a phase transition),
and the ``microscopic remainder''.

The set of einselected states is called {\it the pointer basis} (Zurek, 1981)
in recognition of its role in measurements. The criterion for
the einselection of states goes well beyond the often repeated
characterizations based on the instantaneous eigenstates of the density
matrix. What is of the essence is the ability of the einselected states
to survive monitoring by the environment. This heuristic
criterion can be made rigorous by quantifying predictability of the evolution
of the candidate states, or of the associated observables. Einselected
states provide optimal initial conditions: They can be employed for
the purpose of prediction better than other Hilbert space alternatives
-- they retain correlations in spite of their immersion in the environment.

Three quantum systems -- the measured system ${\cal S}$, the apparatus
${\cal A}$, and the environment ${\cal E}$ -- and correlations
between them are the subject of our study. In pre-measurements ${\cal S}$
and ${\cal A}$ interact. Their resulting entanglement transforms into
an effectively classical correlation as a result of the interaction
between ${\cal A}$ and ${\cal E}$. 

This ${\cal SAE}$ triangle helps analyze decoherence and study 
its consequences. By keeping all
three corners of this triangle in mind, one can avoid confusion,
and maintain focus on the {\it correlations} between, e.g., the memory
of the observer and the state of the measured system.
The evolution from a quantum entanglement to the classical correlation
may be the easiest relevant theme to define operationally. In the language
of the last part of Section II, we are about to justify the ``outsider''
point of view, Eq. (2.44c), before considering the ``discoverer'', Eq. (2.44b)
and the issue of the ``collapse''. In spite of this focus
on correlations, we shall often suppress one of the corners of the ${\cal SAE}$
triangle to simplify notation. All three parts will however play
a role in formulating questions and in motivating criteria
for classicality.

\subsection{Models of einselection}


The simplest case of a single act of decoherence involves just three
one-bit systems (Zurek, 1981; 1983). They are denoted by ${\cal S, \ A}$,
and ${\cal E}$ in an obvious reference to their roles.
The measurement starts with the interaction of the measured system with
the apparatus:
$$|\uparrow \rangle  | A_0 \rangle  \longrightarrow | \uparrow \rangle  | A_1 \rangle  ,\
|\downarrow \rangle  | A_0 \rangle  \longrightarrow | \downarrow \rangle  | A_0 \rangle ; \eqno(4.1a,~b)$$
where $\langle  A_0 | A_1 \rangle  = 0 $. For a general state:
$$(\alpha | \uparrow \rangle  + \beta | \downarrow \rangle ) |A_0 \rangle 
\longrightarrow
\alpha | \uparrow \rangle  | A_1 \rangle + \beta | \downarrow \rangle  | A_0 \rangle   = | \Phi \rangle 
. \eqno(4.2) $$
These formulae are an example of a {\tt c-not} like the premeasurement
discussed in section 2.

The basis ambiguity -- the ability to re-write $|\Phi\rangle $, Eq. (4.2),
in any basis of, say, the system, with the superposition principle 
guaranteeing existence of the corresponding pure states of the apparatus --
disappears when an additional system, ${\cal E}$, performs a premeasurement 
on ${\cal A}$:
\begin{eqnarray} & & ( \alpha | \uparrow \rangle  | A_1 \rangle  + \beta | \downarrow \rangle  |
A_0\rangle 
)|\varepsilon_0\rangle  \nonumber \\
& & \longrightarrow \alpha | \uparrow \rangle  | A_1 \rangle   |\varepsilon_1\rangle 
    +
\beta | \downarrow \rangle  | A_0 \rangle  |\varepsilon_0\rangle  = | \Psi \rangle  . \eqnum{4.3}
\end{eqnarray}
A collection of three correlated quantum systems is no longer
subject to the basis ambiguity we have pointed out in connection with the
EPR-like state $|\Phi\rangle $, Eq. (4.2). This is especially true when states
of the environment are correlated with the simple products of the states
of the apparatus - system combination (Zurek, 1981; Elby and Bub, 1994).
In Eq. (4.3) this can be
guaranteed (irrespective of the value of $\alpha$ and $\beta$) providing that:
$$ \langle  \varepsilon_0 | \varepsilon_1 \rangle  = 0 \ . \eqno(4.4) $$
When this orthogonality condition is satisfied, the state
of the ${\cal A-S }$ pair is given by a reduced density matrix:
\begin{eqnarray} &  \rho_{\cal AS} \ = \ Tr_{\cal E}| \Psi \rangle \langle  \Psi|= &
\nonumber \\
&  |\alpha|^2 | \uparrow \rangle \langle  \uparrow | |A_1 \rangle \langle  A_1| +
|\beta|^2 |\downarrow \rangle \langle  \downarrow | |A_0 \rangle \langle  A_0 |& \eqnum {4.5{\it a}}
\end{eqnarray}
containing only classical correlations.

If the condition of Eq. (4.4) did not hold -- that is, if the orthogonal
states
of the environment were not correlated with the apparatus in the basis
in which the original premeasurement was carried out -- then the eigenstates
of the reduced density matrix $\rho_{\cal AS}$ would be sums of
products rather than simply products of states of ${\cal S}$ and ${\cal A}$.
Extreme example of this situation is the pre-decoherence density matrix
of the pure state:
\begin{eqnarray} | \Phi \rangle  \langle  \Phi | & &  = 
|\alpha|^2 | \uparrow \rangle  \langle  \uparrow | |A_1 \rangle  \langle  A_1| \ + 
\alpha \beta^* | \uparrow \rangle  \langle  \downarrow | | A_1 \rangle  \langle  A_0 | 
\nonumber \\ & & 
+  \alpha^* \beta | \downarrow \rangle  \langle  \uparrow | | A_0 \rangle  \langle  A_1 |
+  |\beta|^2 |\downarrow \rangle  \langle  \downarrow | |A_0 \rangle  \langle  A_0 |
    \eqnum {4.5{\it b}}
\end{eqnarray}
Its eigenstate is $|\Phi\rangle $. When expanded, $|\Phi \rangle \langle  \Phi |$ contains
terms that are off-diagonal when expressed in any of the natural bases
consisting of the tensor products of states in the two systems. Their
disappearance as a result of tracing over the environment removes the
basis ambiguity. Thus, for example, the reduced density matrix
$\rho_{\cal AS}$, Eq. (4.5a), has the same form as the outsider
description of the classical measurement, Eq. (2.44c).

In our simple model pointer states are easy to characterize: To leave
pointer states untouched, the Hamiltonian of interaction $H_{\cal AE}$ 
should have the same structure as for the {\tt c-not}, Eqs. (2.9) 
- (2.10): It should be a function of the pointer observable;
$ \hat A = a_0 |A_0\rangle \langle A_0| + a_1 |A_1\rangle \langle A_1| $
of the apparatus. Then the states of the environment will bear an
imprint of the pointer states $\{ |A_0\rangle ,|A_1\rangle \}$. As  noted in
section II:
$$ [ H_{\cal AE}, \hat A] = 0 \eqno(4.6)$$
immediately implies that $\hat A$ is a control, and its eigenstates will
be preserved.

\subsubsection{Decoherence of a single qubit}

An example of continuous decoherence is afforded by a two-state apparatus
${\cal A}$ interacting with an environment of $N$ other spins (Zurek, 1982).  
The two apparatus states are $\{|\Uparrow\rangle ,|\Downarrow\rangle \}$.
For the simplest, yet already interesting example the self-Hamiltonian 
of the apparatus disappears, $ H_{\cal A} =0$, and the interaction has the form:
\begin{eqnarray} H_{\cal AE} &=& (|\Uparrow\rangle \langle \Uparrow| -
|\Downarrow\rangle \langle \Downarrow|)
\nonumber \\
& &\otimes
\sum_k g_k (|\uparrow\rangle \langle \uparrow| - |\downarrow\rangle \langle \downarrow|)_k \ ,
\eqnum{4.7}
\end{eqnarray}
Under the influence of this Hamiltonian the initial state:
\begin{eqnarray} |\Phi(0)\rangle  = (a|\Uparrow\rangle  + b |\Downarrow\rangle ) ~
\prod_{k=1}^N
(\alpha_k |\uparrow\rangle _k + \beta_k |\downarrow\rangle _k) \ \eqnum{4.8}
\end{eqnarray}
evolves into:
$$ |\Phi(t)\rangle  = a |\Uparrow\rangle  |{\cal E}_{\Uparrow}(t)\rangle  +
b |\Downarrow\rangle  |{\cal E}_{\Downarrow}(t)\rangle  \ ; \eqno(4.9) $$
\begin{eqnarray} |{\cal E}_{\Uparrow}(t)\rangle  &=& \prod_{k=1}^N
(\alpha_k e^{i g_k t} |\uparrow\rangle _k + \beta_k e^{- i g_k t}
|\downarrow\rangle _k)  = |{\cal E}_{\Downarrow}(-t)\rangle .
\eqnum{4.10} \end{eqnarray}
The reduced density matrix is:
\begin{eqnarray}\rho_{\cal A} &=& |a|^2 |\Uparrow\rangle \langle \Uparrow| +
ab^*r(t)|\Uparrow\rangle \langle \Downarrow|
\nonumber \\
&+& a^*br^*(t) |\Downarrow\rangle \langle \Uparrow| + |b|^2
|\Downarrow\rangle \langle \Downarrow|
\ . \eqnum{4.11} \end{eqnarray}
The coefficient $r(t) =\langle {\cal E}_{\Uparrow}|{\cal E}_{\Downarrow}\rangle  $
determines the relative size of the off-diagonal terms.
It is given by:
\begin{eqnarray}r(t) =
    \prod_{k=1}^N  [\cos2g_kt + i (|\alpha_k|^2-|\beta_k|^2)\sin 2 g_k t ] \
.
\eqnum{4.12}\end{eqnarray}
For large environments consisting of many ($N$) spins
    at large times the off-diagonal terms are typically small:
$$ |r(t)|^2 \simeq 2^{-N} \prod_{k=1}^N[1 + (|\alpha_k|^2 - |\beta_k|^2)^2]
\eqno(4.13)$$

The density matrix of any two-state system can be represented
by a point in  the 3-D space. In terms of the coefficients
$a,\ b,$ and $r(t)$ coordinates of the point representing it are;
$ z = (|a|^2 - |b|^2), $ $ x = \Re (ab^*r)  , $ $ y = \Im (a b^* r)  .$
When the state is pure, $x^2+y^2+z^2=1$: Pure states lie
on the surface of the Bloch sphere (Fig. 3).

Any conceivable (unitary or non-unitary) quantum evolution can be
thought of as a transformation of the surface of the pure states into
the ellipsoid contained inside the Bloch sphere. Deformation of
the Bloch sphere surface caused by decoherence is a special case of such
general evolutions (Zurek, 1982, 1983; Berry, 1995). Decoherence does 
not affect $|a|$ or $|b|$. Hence, evolution due to decoherence alone occurs 
in plane $z$=const. Such a ``slice'' through the Bloch sphere would show the
point representing the state at a fraction $|r(t)|$ of its maximum distance.
The complex $r(t)$ can be expressed as the sum of complex phase
factors rotating with the frequencies given by differences $\Delta \omega_j$
between the energy eigenvalues of the interaction Hamiltonian, weighted
with the
probabilities  of finding them in the initial state:
$$ r(t) = \sum_{j=1}^{2^N} p_j \exp(-i \Delta \omega_j t) \ . \eqno(4.14)$$
The index $j$ denotes the environment part of the energy eigenstates of
the interaction Hamiltonian, Eq. (4.7):
$|j\rangle  =|\uparrow\rangle _1 \otimes |\downarrow\rangle _2 \otimes \dots \otimes
|\uparrow\rangle_N  $.
The corresponding differences between the energies of the
eigenstates $|\Uparrow\rangle  |j\rangle $ and $|\Downarrow\rangle |j\rangle $ are; 
$ \Delta \omega_j = \langle \Uparrow|\langle j|H_{\cal AE}|j\rangle |\Downarrow\rangle . $ 
There are $2^N$ distinct  $|j\rangle $'s, and, barring degeneracies, the same
number of different $\Delta \omega_j$'s. Probabilities $p_j$ are:
$$ p_j = | \langle j | {\cal E}(t=0)\rangle  |^2 \ , \eqno(4.15) $$
which is in turn easily expressed in terms of the appropriate squares of
$\alpha_k$ and  $\beta_k$.

The evolution of $r(t)$, Eq. (4.14), is a consequence of the
rotations of the complex vectors $p_k\exp (-i \Delta \omega_j t)$
with different frequencies. The resultant $r(t)$ will then start with
the amplitude 1, and quickly ``crumble''to
$$\langle |r(t)|^2\rangle  \sim \sum_{j=1}^{2^N} p_j^2 \sim 2^{-N} \ . \eqno(4.16)$$
In this sense, decoherence is exponentially effective -- the
magnitude of the off-diagonal terms decreases exponentially fast
with the physical size $N$ of the environment effectively coupled to
the state of the system.

We note that the effectiveness of einselection depends on the initial state of
the environment: When ${\cal E}$ is in the $k$'th eigenstate
of $H_{\cal  AE}$, $p_j = \delta_{jk}$, the coherence in the system
will be retained.  This special environment state is, however,
unlikely in realistic circumstances.

\subsubsection{The classical domain and a quantum halo}

Geometry of flows induced by decoherence in the Bloch sphere
exhibits characteristics encountered in general:

(i) The classical set of the einselected pointer states
($\{|\Uparrow\rangle , |\Downarrow\rangle \}$ in our case). Pointer states are
the pure states least affected  by decoherence.

(ii) Classical domain consisting of all the pointer states and their
mixtures. In Fig. 3 it corresponds to the section [-1,+1] of $z$-axis.

(iii) The quantum domain -- the rest of the volume of the Bloch sphere --
consisting of more general density matrices.

Visualizing decoherence-induced decomposition of the Hilbert space may be
possible only in the simple case studied here, but whenever decoherence leads
to classicality, emergence of generalized and often approximate version of
the elements (i) -- (iii) is an expected feature.

As a result of decoherence the part of the Hilbert space outside of the
classical domain is ``ruled out'' by einselection. The severity of
the prohibition on its states varies. One may measure 
``non-classicality'' of (pure or mixed) states by quantifying their distance 
from this classical state with the rate of entropy production and
comparing it with the much lower rate in the classical domain. Classical
pointer states would be then enveloped by a ``quantum halo'' (Anglin and
Zurek, 1996) of nearby, relatively decoherence - resistant but still 
somewhat quantum states, with a more flagrantly quantum (and more fragile)
Schr\"odinger cat states further away.

By the same token, one can define einselection - induced
metric in the classical domain, with the distance between two pointer
states given by the rate of entropy production of their
superposition. This is not the only way to define a distance: 
As we shall see in Section VII, redundancy of the record of
a state imprinted on the environment is a very natural measure of its
classicality. In course of decoherence, pointer states tend to be recorded
redundantly and can be deduced by intercepting a very small fraction of
the environment (Zurek, 2000; Dalvit, Dziarmaga and Zurek, 2001). One can
then define distance using the fraction of the environment 
that needs to be intercepted to distinguish between two pointer states 
of the system (Ollivier, Poulin, and Zurek, 2002). 

\subsubsection{Einselection and controlled shifts}

Discussion of decoherence can be generalized
to the situation where the system, the apparatus,
and the environment have many states, and their interactions
are more complicated. Here we assume that the system is isolated,
and that it interacts with the apparatus in a {\tt c-shift} manner
discussed in Section II. As a result of that interaction the state
of the apparatus becomes entangled with the state of the system:
$(\sum_i\alpha_i |s_i\rangle )|A_0\rangle \ \longrightarrow\ \sum_i \alpha_i |s_i\rangle  |A_i\rangle 
$.
This state suffers from the basis ambiguity:
The ${\cal S-A}$ entanglement implies that for any state of either there
exists a corresponding pure state of its partner. Indeed, when the
initial state of ${\cal S}$ is chosen to be one of the eigenstates of the
conjugate basis:
$$ |r_l\rangle  = N^{-{1 \over 2}} \sum_{k=0}^{N-1} \exp(2 \pi i k l / N) |s_k\rangle  \ ,
\eqno(4.17)$$
{\tt c-shift} could equally well represent a measurement of the
apparatus (in the basis conjugate to $\{ |A_k \rangle  \}$) by the system.
Thus, it is not just the basis which is ambiguous: Also the roles of
the control (system) and of the target (apparatus) can be reversed
when the conjugate basis is selected. These ambiguities can be
removed by recognizing the role of the environment.

Figure 4 captures the essence of the idealized decoherence process,
that allows the apparatus to be -- in spite of the interaction
with the environment -- a noiseless classical communication channel
(Schumacher, 1996; Lloyd, 1997). This is possible because the
${\cal A-E}$ {\tt c-shift}s do not disturb pointer states.

The advantage of this caricature of the decoherence process as
a sequence of {\tt c-shift}s lies in its simplicity. However,
the actual process of decoherence is usually continuous
(so that it can be only approximately broken up into discrete {\tt c-shift}s).
Moreover, in contrast to the {\tt c-not}s used in quantum logic circuits,
the record inscribed in the environment is usually distributed over
many degrees of freedom. Last not least, the observable of the apparatus
(or any other open system) may be subject to noise (and not just
decoherence) or its self-Hamiltonian may rotate instantaneous
pointer states into their superpositions. These very likely complications
will be investigated in specific models below.

Decoherence is caused by a premeasurement - like process carried out
by the environment ${\cal E}$:
\begin{eqnarray} & & |\Psi_{\cal SA}\rangle  |\varepsilon_0\rangle  = (\sum_j \alpha_j
|s_j\rangle  |A_j\rangle )
|\varepsilon_0\rangle  \ \nonumber \\
& & \longrightarrow \sum_j \alpha_j |s_j\rangle  |A_j\rangle  |\varepsilon_j\rangle  \
= \ |\Phi_{\cal SAE}\rangle  \eqnum{4.18}\end{eqnarray}
Decoherence leads to einselection when the states of the environment
$|\varepsilon_j\rangle $ corresponding to different pointer states become
orthogonal:
$$ \langle \varepsilon_i | \varepsilon_j \rangle  = \delta_{ij} \eqno(4.19)$$
Then the Schmidt decomposition of the state vector $|\Phi_{\cal SAE}\rangle  
$ into a composite subsystem ${\cal SA}$
and ${\cal E}$ yields product states $|s_j\rangle  |A_j\rangle $ as partners 
of the orthogonal environment states. The decohered density matrix describing  
${\cal SA}$ pair is then diagonal in  product states:
For simplicity we shall often discard reference to the object that
\begin{eqnarray} \rho_{\cal SA}^D &=& \sum_j |\alpha_j|^2 |s_j \rangle \langle 
s_j||A_j\rangle \langle A_j| \nonumber \\
&=& Tr_{\cal E} |\Phi_{\cal SAE}\rangle \langle \Phi_{\cal SAE}| \ .
\eqnum{4.20}\end{eqnarray}
does not interact with the environment (here -- the system ${\cal S}$).
Nevertheless,  {\it preservation of the ${\cal SA}$ correlations
is the criterion defining the pointer basis}. Invoking it would get rid
of many a confusion (see, e.g, discussions in Halliwell, Perez-Mercader,
and Zurek, 1994; Venugopalan, 1994). The density matrix of a
single object in contact with the environment will be always
diagonal in an (instantaneous) Schmidt basis. This instantaneous diagonality
should not be used the {\it sole} criterion for classicality (although see Zeh,
1973,~1990; Albrecht, 1992\&1993). Rather, ability of certain states to
retain correlations in spite of the coupling to the environment
is decisive.

When the interaction with the apparatus has the form:
$$ H_{\cal AE} = \sum_{k,l,m} g^{\cal AE}_{klm}
|A_k\rangle \langle A_k||\varepsilon_l\rangle \langle \varepsilon_m| + h.c. \ , \eqno(4.21)$$
the basis $\{|A_k\rangle \}$ is left unperturbed and any correlation with the
states $\{|A_k\rangle \}$  is preserved. But, by definition, pointer states 
preserve correlations in spite of decoherence, so that any observable $A$ 
co-diagonal with the interaction Hamiltonian will be pointer observable.
For, when the interaction is a function of $A$, it can be expanded in $A$
as a power series, so it commutes with $A$:
$$ [ H_{\cal AE}(A), A] = 0 \eqno(4.22)$$
The dependence of the interaction Hamiltonian on the observable
is an obvious precondition for the monitoring of that observable
by the environment. This admits existence of degenerate pointer
eigenspaces of $A$.

\subsection{Einselection as the selective loss of information}

Establishment of the measurement-like correlation between the apparatus
and the
environment changes the density matrix from the premeasurement
$\rho^P_{\cal SA}$ to the decohered $\rho^D_{\cal SA}$, Eq. (4.20). 
For the initially pure $|\Psi_{\cal SA}\rangle $, Eq. (4.18), this transition 
is represented by:
\begin{eqnarray} & & \rho^P_{\cal SA} = \sum_{i,j} \alpha_i \alpha_j^*
|s_i\rangle \langle s_j||A_i\rangle \langle A_j| \longrightarrow
\nonumber \\
& &\longrightarrow
\sum_i |\alpha_i|^2 |s_i\rangle \langle s_i||A_i\rangle \langle A_i| = \rho_{\cal SA}^D
\eqnum{4.23}\end{eqnarray}
Einselection is accompanied by the increase of entropy:
$$\Delta H(\rho_{\cal SA})=H(\rho_{\cal SA}^D) - H({\rho^P_{\cal
SA}}) \geq 0 \eqno(4.24)$$
and by the disappearance of the ambiguity in what was
measured
(Zurek, 1981, 1993a). Thus, before decoherence the conditional density
matrices
of the system $\rho_{{\cal S}|C_j\rangle }$ are pure for any state $|C_j\rangle $ of
the apparatus pointer. They are defined using the
unnormalized:
$$ \tilde \rho_{{\cal S}|\Pi_j} = Tr_{\cal A} \Pi_j \rho_{\cal SA}
\eqno(4.25)$$
where in the simplest case $\Pi_j = |C_j\rangle \langle C_j|$ projects onto 
a pure state of the apparatus.\footnote{This can be generalized to
projections onto multidimensional subspaces of the apparatus. In that case,
purity of the conditional density matrix will be usually lost during the trace
over the states of the pointer. This is not surprising: When the observer
reads off the pointer of the apparatus only in a coarse-grained manner,
he will forgo part of the information about the system. Amplification we
have considered before can prevent some of such loss of resolution due to
coarse graining in the apparatus. Generalizations to density matrices 
conditioned upon projection operator valued measures (POVM's)
(Kraus, 1983) are also possible.}

Normalized $\rho_{{\cal S}|\Pi_j}$ can be obtained by using the probability
of the outcome:
\begin{eqnarray}
\rho_{{\cal S}|\Pi_j} =p_j^{-1}\tilde \rho_{{\cal S}|\Pi_j} \ ;
\ \ p_j = Tr \tilde \rho_{{\cal S}|\Pi_j} \ .  \eqnum{4.26}
\end{eqnarray}
Conditional density matrix represents the description of the system ${\cal
S}$
available to the observer who knows that the apparatus ${\cal A}$ is in a
subspace defined by $\Pi_j$.

Before decoherence, $\rho^P_{{\cal S}|C_j\rangle }$ is pure for any state $|C_j\rangle $:
$$(\rho^P_{{\cal S}|\Pi_j})^2\ =\ \rho^P_{{\cal S}|\Pi_j} \ \ \  \forall
|C_j\rangle  \eqno(4.27a)$$
providing the initial premeasurement state, Eq. (4.23), was pure as well.
It follows that:
$$ H(\rho^P_{{\cal SA}|C_j\rangle }) = 0 \ \ \ \forall |C_j\rangle  \ . \eqno(4.28a)$$
For this same case given by the initially pure $\rho^P_{\cal SA}$ of Eq.
(4.23), conditional density matrices obtained from the decohered
$\rho^D_{\cal SA}$ will be pure if and only if they are conditioned
upon the pointer states $\{|A_k\rangle \}$;
\begin{eqnarray}(\rho^D_{{\cal S}|C_j\rangle })^2=\rho^P_{{\cal S}|C_j\rangle } =
|s_k\rangle  \langle  s_k| \iff |C_j\rangle  = |A_j\rangle  ; \eqnum{4.27b}\end{eqnarray}
$$ H(\rho^D_{{\cal S}|A_j\rangle } ) = H (\rho^P_{{\cal S}|A_j\rangle }) \ .
\eqno(4.28b)$$
This last equation is valid even when the initial states
of the system and of the apparatus are not pure. Thus, only in
the pointer basis the pre-decoherence strength of the
${\cal SA}$ correlation will be maintained. In all other bases:
$$ Tr(\rho^D_{{\cal S}|C_j\rangle })^2 \ < \   Tr\rho^D_{{\cal S}|C_j\rangle } ;
\ \ \ |C_j\rangle  \notin \{|A_j\rangle \} \eqno(4.27c)$$
$$ H (\rho^P_{{\cal S}|C_j\rangle }) \ < \  H( \rho^D_{{\cal S}|C_j\rangle } );
\ \ \ |C_j\rangle  \notin \{|A_j\rangle \} \eqno(4.28c)$$
In particular, in the basis $\{|B_j\rangle \}$ conjugate to the pointer
states $\{|A_j\rangle \}$, Eq. (2.14), there is no correlation left with the state
of the system, that is:
$$ \rho^D_{{\cal S}|B_j\rangle } = N^{-1}\sum_k |s_k\rangle \langle s_k| = {\bf 1}/N
\eqno(4.29)$$
where ${\bf 1}$ is a unit density matrix. Consequently;
$$ (\rho^D_{{\cal S}|B_j\rangle })^2 =  \rho_{{\cal S}|B_j\rangle }/N , \eqno(4.27d)$$
$$ H (\rho^D_{{\cal S}|B_j\rangle }) = H( \rho^P_{{\cal S}|B_j\rangle } ) - \lg N = - \lg
N \ .
\eqno(4.28d)  $$
Note that, initially, conditional density matrices were pure also in the
conjugate (and any other) basis, provided that the initial state was
the pure entangled projection operator
$\rho_{\cal SA}^P  = |\Psi_{\cal SA}\rangle \langle \Psi_{\cal SA}|$, Eq. (4.23).

\subsubsection{Mutual information and discord}

Selective loss of information everywhere except in the pointer
states is the essence of einselection. It is reflected in the change 
of the mutual information which starts from:
\begin{eqnarray} {\cal I}^P({\cal S:A}) &=&  H (\rho^P_{{\cal S}}) + H
(\rho^P_{{\cal A}}) -
H(\rho^P_{{\cal S,A}}) \nonumber \\
&=& - 2 \sum_i |\alpha_i|^2 \lg |\alpha_i|^2 \eqnum{4.30a}\end{eqnarray}
As a result of einselection, for initially pure cases, it 
decreases to at most half its initial value:
\begin{eqnarray} {\cal I}^D({\cal S:A}) &=&  H (\rho^D_{{\cal S}}) + H
(\rho^D_{{\cal A}}) -
H(\rho^D_{{\cal S,A}}) \nonumber \\
    &=& - \sum_i |\alpha_i|^2 \lg |\alpha_i|^2 \eqnum{4.30b}\end{eqnarray}
This level is reached when the pointer basis coincides with the Schmidt basis
of $|\Psi_{\cal SA}\rangle $. The decrease in the mutual information is due to
the increase of the joint entropy $H(\rho_{\cal S,A})$:
\begin{eqnarray} \Delta {\cal I}({\cal S:A}) &=& {\cal I}^P({\cal S:A})-
{\cal I}^D({\cal
S:A}) \nonumber \\
    &=& H(\rho^D_{{\cal S,A}}) - H(\rho^P_{\cal S,A})
= \Delta H(\rho_{\cal S,A}) \eqnum{4.31}\end{eqnarray}
Classically, equivalent definition of the mutual information obtains
from the asymmetric formula:
$$ {\cal J}_{\cal A}({\cal S:A}) = H(\rho_{\cal S}) - H (\rho_{\cal S|A})
\eqno(4.32)$$
with the help of the conditional entropy $H(\rho_{\cal S|A})$. Above,
subscript ${\cal A}$ indicates the member of the correlated pair that will
be the source of the information about its partner. A symmetric counterpart
of the above equation, 
${\cal J}_{\cal S}({\cal S:A}) = H(\rho_{\cal A}) - H (\rho_{\cal A|S})$,
can be also written. 

In the quantum case, definition of Eq. (4.32) is so
far incomplete, as a quantum analogue of the classical conditional
information has not been yet specified. Indeed, Eqs. (4.30a) and
(4.32) jointly imply that in the case of entanglement a quantum conditional
entropy  $H(\rho_{\cal S|A})$  would have to be negative! For, in this case;
$$ H(\rho_{\cal S|A}) = \sum_i |\alpha_i|^2 \lg |\alpha_i|^2 < 0
\eqno(4.33)$$
would be required to allow for ${\cal I}({\cal S:A}) = {\cal J_A}({\cal S:A})$.
Various quantum redefinitions of ${\cal I} ({\cal S:A}) $ or
$H(\rho_{\cal S|A})$ have been proposed to address this
(Lieb, 1975; Schumacher and Nielsen, 1996; Lloyd, 1997; Cerf and Adami, 1997).
We shall simply regard this fact as an illustration of the strength of
quantum correlations (i. e., entanglement), which allow
${\cal I} ({\cal S:A}) $ to violate the inequality:
$$ {\cal I}({\cal S:A}) \leq min (H_{\cal S}, H_{\cal A}) \eqno(4.34)$$
This inequality follows directly from Eq. (4.32) and the non-negativity of
classical conditional entropy (see e.g. Cover and Thomas, 1991).

Decoherence decreases ${\cal I}({\cal S:A})$ to this allowed level
(Zurek, 1983). Moreover, now the conditional entropy can be defined in
the classical pointer basis as the average of partial entropies computed
from the conditional $\rho^D_{{\cal S}|A_i\rangle }$ over the probabilities
of different outcomes:
$$ H(\rho_{\cal S|A})
= \sum_i p_{|A_i\rangle } H(\rho^D_{{\cal S}|A_i\rangle }) \eqno(4.35)$$
Prior to decoherence, the use of probabilities would not have been legal.

For the case considered here, Eq. (4.18), the conditional entropy
$ H(\rho_{\cal S|A})=0$: In the pointer basis there is a perfect
correlation between the system and the
apparatus, providing that the premeasurement Schmidt basis and the pointer
basis coincide. Indeed, it is tempting to define a good apparatus
or a {\it classical} correlation by insisting on such a coincidence.

The difficulties with conditional entropy and mutual
information are highly symptomatic.  The trouble with $H(\rho_{\cal S|A})$ 
arises for states that exhibit quantum
correlations -- entanglement of $|\Psi_{\cal SA}\rangle $ being an extreme
example -- and, thus,
do not admit an interpretation based on probabilities. A useful sufficient
condition for the classicality of correlations is then the existence of
an apparatus basis that allows quantum versions of the two classically 
identical expressions for the mutual information to coincide,
${\cal I}({\cal S:A})={\cal J}_{\cal A}({\cal S:A})$,
(Zurek, 2000b; 2002a; Ollivier and Zurek, 2002): 
Equivalently, the {\it discord}
$$ \delta{\cal I_A} ({\cal S}|{\cal A}) \ = \ {\cal I}({\cal S:A})
- {\cal J}_{\cal A}({\cal S:A}) \eqno(4.36)$$
must vanish: Unless $\delta {\cal I_A}({\cal S} | {\cal A}) = 0$, 
probabilities of the distinct apparatus pointer states cannot exist.

We end this subsection with a part summary, part anticipatory remarks:
Pointer states retain undiminished correlations with
the measured system ${\cal S}$, or with any other system,
including observers. The loss of information caused by decoherence
is given by Eq. (4.31). It was precisely such as to lift
conditional information from the paradoxical (negative) values to
the classically allowed level, Eq. (4.33).
This is equal to the information gained by the observer when
he consults the  apparatus pointer. This is no coincidence -- the environment
has `measured' (became correlated with) the apparatus
in the very same pointer basis in which observers have to access ${\cal A}$
to take advantage of the remaining (classical) correlation between
the pointer and the system. Only when observers and the environment
monitor co-diagonal observables they do not get in each others way.

In the idealized case, preferred basis was distinguished by its ability to
retain perfect correlations with the system in spite of decoherence. This
remark shall serve as a guide in other situations: It will lead to
a criterion -- predictability sieve -- used to identify preferred
states in less idealized circumstances. For example, when the self-Hamiltonian
of the system is non-trivial, or when the commutation relation, Eq. (4.22),
does not hold exactly for any observable, we shall seek states that are
best in retaining correlations with the other systems.

\subsection{Decoherence, entanglement, dephasing, and noise}

In the symbolic representation of Fig. 4, noise is
the process in which environment acts as a control,
inscribing information about its  state in the state of the system,
that assumes the role of the target. However, the
direction of the  information flow in  {\tt c-nots} and {\tt c-shifts}
depends on the choice of initial states. Control and target switch
roles when, for a given Hamiltonian of interaction, one prepares input
of the {\tt c-not} in the basis conjugate to the logical ``pointer'' states.
Einselected states correspond to the set which
-- when used in {\tt c-not}s  or {\tt c-shift}s -- minimizes the effect
of interactions directed from the environment to the system.

Einselection is caused by the premeasurement carried out
by the environment on the pointer states. Decoherence
follows from the Heisenberg's indeterminacy: Pointer observable is
measured by the environment. Therefore, the complementary observable
must become at least as indeterminate as is demanded by
the Heisenberg's principle. As the environment and
the systems  entangle through an interaction that favors a set of
pointer states, their phases become indeterminate (see Eq.
(4.29) and discussion of envariance in section VI). Decoherence 
can be thought of as the resulting loss of phase relations.

Observers can be ignorant of phases for reasons that do not lead to
an imprint of the state of the system on the environment. Classical
noise can cause such {\it dephasing} when the observer does not know
the time-dependent classical perturbation Hamiltonian  responsible
for this unitary, but unknown evolution.
For example, in the pre-decoherence state vector, Eq. (4.18),
random phase noise will  cause a transition:
\begin{eqnarray} & & |\Psi_{\cal SA}\rangle  = (\sum_j \alpha_j
|s_j\rangle  |A_j\rangle )
\ \nonumber \\
& & \longrightarrow \sum_j \alpha_j \exp( i \phi_j^{(n)}) |s_j\rangle  |A_j\rangle  \
= \ | \Psi_{\cal SA}^{(n)}\rangle  \ . \eqnum{4.37}\end{eqnarray}
A dephasing Hamiltonian acting either on the system or on the apparatus
can lead to such an effect. In this second case its form could be:
$$H_D^{(n)}=\sum_j \dot \phi_j^{(n)} (t) |A_j\rangle \langle A_j| \ . \eqno(4.38)$$
In contrast to interactions
causing premeasurements, entanglement, and decoherence, $H_D$ cannot
influence the nature or the degree of the ${\cal SA}$ correlations:
$H_D$ does not imprint the state of ${\cal S}$ or ${\cal A}$ 
anywhere else in the Universe:
For each individual realization $n$ of the phase noise (each selection of
$\{ \phi_j^{(n)}(t) \}$ in Eq. (4.37)) the state $|\Phi_{\cal SA}^{(n)}\rangle $
remains pure. Given only $\{ \phi_j^{(n)} \}$ one could restore
pre-dephasing state on a case - by - case basis. However, in absence
of such detailed information, one is often forced to represent
${\cal SA}$ by the density matrix averaged over the
ensemble of  noise realizations:
\begin{eqnarray}
& & \overline\rho_{\cal SA} = \langle  |\Psi_{\cal SA}\rangle  \langle  \Psi_{\cal SA}| \rangle 
= \sum_j |\alpha_j|^2 |s_j \rangle  \langle  s_j||A_j \rangle  \langle  A_j| \nonumber \\
& & + \sum_{j,k} \sum_n
e^{i(\phi_j^{(n)} - \phi_k^{(n)})}\alpha_j \alpha_k
|s_j\rangle  \langle  s_k||A_j \rangle  \langle  A_k| \eqnum{4.39}
\end{eqnarray}
In this phase - averaged density matrix off-diagonal terms
may disappear. Nevertheless, each member of the ensemble may exist in
a state as pure as it was before dephasing. NMR offers examples of
dephasing (which can be reversed using spin echo). Dephasing is a loss
of phase coherence due to noise in phases. It does not result in an
information transfer to the environment.

Dephasing cannot be used to justify existence of preferred basis
in individual quantum systems. Nevertheless, the ensemble as a whole
may obey the same master equation as individual systems entangling with
the environment. Indeed, many of the symptoms of decoherence arise in
this setting. Thus, in spite of the light shed on this issue by
the discussion of simple cases (Wootters and Zurek, 1979;
Stern, Aharonov, and Imry, 1989) more remains to be understood, 
perhaps by considering implications of envariance (see Section VI).

Noise is an even more familiar and less subtle effect represented
by transitions that break one-to-one correspondence in Eq. (4.39).
Noise in the apparatus would cause a random rotation  of states $|A_j\rangle $.
It could be modelled by a collection of Hamiltonians similar to $H_D^{(n)}$
but not co-diagonal with the observable of interest. Then, after an ensemble
average similar to Eq. (4.39), one-to-one correspondence
between ${\cal S}$ and ${\cal A}$ would be lost.  However -- as before --
the evolution is unitary for each $n$, and the unperturbed state could be
reconstructed from the information observer could have in advance.

Hence, in case of dephasing or noise information about their cause obtained
either in advance, or afterwards, suffices to undo their effect.
Decoherence relies on entangling interactions (although, strictly 
speaking, it need not invlove entanglement (Eisert and Plenio, 2002). 
Thus, neither prior nor posterior
knowledge of the state of environment is enough. Transfer of
information about the decohering system to the environment is essential,
and plays key role in the interpretation.

We note that, while nomenclature used here seems most sensible
to this author and is widely used, it is unfortunately not
universal. For example, in the context of quantum computation
``decoherence'' is sometimes used to describe any process that
can cause errors (but see related discussion in Nielsen and Chuang, 2000).

\subsection{Predictability sieve and einselection}

Evolution of a quantum system prepared in a classical state should emulate
classical evolution that can be idealized as a ``trajectory'' --
a predictable  sequence of objectively existing states. For a purely
unitary evolution, all of the states in the Hilbert space retain their
purity and are therefore equally predictable. However, in the presence 
of an interaction with
the environment, a generic superposition representing correlated
states of the system and of the apparatus will decay into a mixture
diagonal in pointer states, Eq. (4.23). Only when the pre-decoherence
state of ${\cal SA}$ is a product of a single apparatus pointer state
$|A_i\rangle $ with the corresponding outcome state of the system (or a mixture
of such product states), decoherence has no effect:
$$ \rho^P_{\cal SA} = |s_i\rangle \langle s_i||A_i\rangle \langle A_i| = \rho^D_{\cal SA}
\eqno(4.40)$$
A correlation of a pointer state with any state of an isolated system
is untouched by the environment. By the same token, when the observer
prepares ${\cal A}$ in the pointer state $|A_i\rangle $, he can count on it
remaining pure. One can even think of $|s_i\rangle $ as the record
of the pointer state of ${\cal A}$. Einselected states are predictable:
they preserve correlations, and hence are effectively classical.

In the above idealized cases predictability of some states follows
directly from the structure of the relevant Hamiltonians (Zurek, 1981).
Correlation with a subspace associated with a projection operator
$P_A$ will be immune to decoherence providing that:
$$ [H_{\cal A} + H_{\cal AE}, P_A] = 0 \ . \eqno(4.41)$$
In more realistic cases it is difficult to demand exact conservation
guaranteed by such a commutation condition. Looking for approximate
conservation may still be a good strategy. Various densities used
in hydrodynamics are one obvious choice (Gell-Mann and Hartle, 1990, 1994).

In general, it is useful to invoke a more fundamental predictability
criterion (Zurek, 1993a). One can measure the loss of predictability
caused by the evolution for every pure state $|\Psi\rangle $ by von Neumann
entropy or some other measure of predictability such as the purity:
$$ \varsigma_{\Psi} (t) = Tr \rho_{\Psi}^2(t)  \ .\eqno(4.42)$$
In either case, predictability is a function of time and a functional
of the initial state as $\rho_{\Psi}(0)=|\Psi\rangle  \langle  \Psi  |$.
Pointer states are obtained by maximizing predictability functional
over $|\Psi\rangle $. When decoherence leads to classicality, good
pointer states exist, and the answer is robust.

Predictability sieve sifts all of the Hilbert space, ordering states
according to their predictability. The top of the list will be the most
classical. This point of view allows for unification of the simple
definition of the pointer states in terms of the commutation
relation Eq. (4.41), with the more general criteria required to discuss
classicality in other situations. The eigenstates of the exact pointer
observable are selected by the sieve:  Eq. (4.41) guarantees that they
will retain their  purity in spite of the environment, and are (somewhat
trivially) predictable.

Predictability sieve can be generalized to situations where the initial
states are mixed (Paraonanu, 2002). Often whole subspaces emerge from the
predictability sieve, naturally leading to ``decoherence-free subspaces''
(see e. g. Lidar et al., 1999) and can be adapted to yield
``noiseless subsystems'' (which are a non-Abelian generalization of pointer
states; see e.g. Zanardi, 2000; Knill, Laflamme, and Viola, 2000). However,
calculations are in general quite difficult even for the initial pure state
cases.

The idea of the ``sieve'' selecting preferred to-be-classical states is novel
and only partly explored. We shall see it ``in action'' below. 
We have outlined two criteria for sifting through the Hilbert space
in search for classicality: von Neumann entropy and purity define, after
all, two distinct functionals. Entropy is arguably an obvious
information-theoretic measure of predictability loss. Purity is much
easier to compute and is often used as a ``cheap substitute'', and has a
physical significance of its own. It seems unlikely that pointer states
selected by the predictability and purity sieves could substantially
differ. After all,
$ - Tr \rho \ln \rho = Tr \rho \{ ({\bf 1} - \rho) - ({\bf 1} - 
\rho)^2/2 + ... $
so that one can expect the most predictable states to also remain purest
(Zurek, 1993a). However, the expansion, Eq. (4.43), is very slowly convergent.
Therefore, a more mathematically satisfying treatment of the differences
between the states selected by these two criteria would be desirable,
especially in cases where (as we shall see in the next section for
the harmonic oscillator) preferred states are coherent, and
hence the classical domain forms relatively broad ``mesa''
in the Hilbert space.

The possible discrepancy between the states selected by the sieves based
on the predictability and on the purity raises a more general question:
Will all the sensible criteria yield identical answer? After all,
one can imagine other reasonable criteria for classicality, such as the
yet-to-be-explored ``distinguishability sieve'' of Schumacher (1999) which
picks out states whose descendants are most distinguishable in spite of
decoherence. Moreover, as we shall see in Section VII (also, Zurek 2000)
one can ascribe classicality to states that are most redundantly recorded
by the environment. The menu of various classicality criteria already
contains several positions, and more may be added in the future. There is no
{\it a priori} reason to expect that {\it all} of these criteria will lead
to {\it identical} sets of preferred states. It is nevertheless reasonable
to hope that, in the macroscopic limit in which classicality is indeed
expected, differences between various sieves should be negligible. The same
stability in the selection of the classical domain is expected with respect
to the changes of, say, the time of the evolution from the initial
pure state: Reasonable changes of such details within the time interval
where einselection is expected to be effective should lead to more or less
similar preferred states, and certainly to preferred states contained
within each others ``quantum halo'' (Anglin and Zurek, 1996). As noted
above, this seems to be the case in the examples explored to date. It remains
to be seen whether all criteria will agree in other situations of interest.

\section{EINSELECTION IN PHASE SPACE}

Einselection in phase space is a special, yet very important topic.
It should lead to phase space points, trajectories, and to classical
(Newtonian) dynamics. Special role of position in classical physics 
can be traced to the nature of interactions (Zurek, 1981; 1982; 1991) 
that depend on distance, and, therefore, commute with position 
(see Eq. (4.22)). Evolution of open systems includes, however,
the flow in phase space induced by the self-Hamiltonian.
Consequently, the set of preferred states turns out to be a compromise,
localized in both position and momentum -- localized in phase space.

Einselection is responsible for the classical structure of phase space.
States selected by the predictability sieve
become phase space ``points'', and their time-ordered sequences
are ``trajectories''. In the underdamped,
classically regular systems one can recover this phase space
structure along with the (almost) reversible evolution.
In chaotic systems there is a price to be paid for classicality:
Combination of decoherence with the exponential divergence
of classical trajectories (which is the defining feature of chaos) 
leads to entropy production at a rate given -- in the classical
limit -- by the sum of positive Lyapunov exponents.
Thus, the dynamical second law can emerge from the interplay
of classical dynamics and quantum decoherence, with the entropy production
caused by the information  ``leaking'' into the environment (Zurek and Paz
1994 \& 1995a; Zurek, 1998b; Paz and Zurek 2001).

\subsection{Quantum Brownian motion}

Quantum Brownian motion model consists of an environment ${\cal E}$
-- a collection of harmonic oscillators (coordinates $q_n$,
masses $m_n$, frequencies $\omega_n$,
and coupling constants $c_n$) interacting with the system ${\cal S}$
(coordinate $x$), with a mass $M$ and a potential $V(x)$.
We shall often consider harmonic $V(x) = M \Omega_0^2 x^2/2$,
so that the whole ${\cal SE}$ is linear and one can obtain
an exact solution. This assumption will be relaxed later. 

The Lagrangian of the system-environment entity is:
$$ L(x, q_n) = L_{\cal S}(x) + L_{\cal SE}(x, \{q_n\})  \ ; \eqno(5.1)$$
The system alone has the Lagrangian:
$$ L_{\cal S}(x) = {M \over 2} \dot x^2 - V(x) \
= \ {M \over 2}(\dot x^2 - \Omega_0^2 x^2)\ . \eqno(5.2)$$
The effect of the environment is modelled by the sum of the Lagrangians
of individual oscillators and of the system-environment interaction
terms:
$$ L_{\cal SE} = \sum_n { m_n \over 2} \biggl( \dot q_n^2 - \omega_n^2
\bigl(q_n - {c_n x \over m_n \omega_n^2 } \bigr)^2 \biggr) \  . \eqno(5.3)$$
This Lagrangian takes into account renormalization of potential
energy of the Brownian particle. The interaction depends (linearly)
on the position $x$ of the
harmonic oscillator. Hence, we expect $x$ to be an instantaneous pointer
observable. In combination with the harmonic
evolution this leads Gaussian pointer states -- well-localized
in both $x$ and $p$. An important characteristic
of the model is the spectral density of the environment:
$$ C (\omega) \ = \ \sum_n {c_n^2 \over { 2 m_n  \omega_n} } \delta
(\omega - \omega_n) \ . \eqno(5.4) $$

The effect of the environment can be expressed through the
propagator acting on the reduced $\rho_{\cal S}$:
$$ \rho_{\cal S}(x,x',t)  =  \int dx_0 dx_0' J(x,x',t|x_0,x_0',t_0)
\rho_{\cal S}
(x_0,x_0',t_0) . \eqno(5.5)$$
We focus on the case when the system and the environment
are initially statistically independent, so that their density matrices 
in a product state:
$$ \rho_{\cal SE} \ = \ \rho_{\cal S} \rho_{\cal E} \ . \eqno(5.6) $$
This is a  restrictive assumption. One can try to justify it
as an idealization of a measurement that correlates ${\cal S}$
with the observer and destroys correlations of ${\cal S}$ with ${\cal E}$,
but that is only an approximation,  as realistic measurements
leave partial correlations with the environment intact. Fortunately,
pre-existing post-measurement correlations lead only to minor differences
in the salient features of the subsequent evolution of the system
(Romero and Paz, 1997; Anglin, Paz and Zurek, 1997).

Evolution of the whole $\rho_{\cal S E}$ can be represented as:
\begin{eqnarray} & & \rho_{\cal SE} (x,q,x',q',t)  =
    \int dx_0 dx'_0 dq_0 dq'_0 ~
\rho_{\cal SE} (x_0,q_0,x'_0,q'_0,t_0)  \nonumber \\
& & \ \ \ \ \ \ \ \ \ \ \ \ \ \ \ \ \ \ K(x,q,t,x_0,q_0) K^*(x',q',t,x'_0,q'_0)
\eqnum{5.7}\end{eqnarray}
Above, we suppress the sum over the indices of the
individual environment oscillators. The evolution operator
$K(x,q,t,x_0,q_0)$ can be expressed as a path integral:
$$ K(x,q,t,x_0,q_0) = \int Dx Dq \exp ({i \over \hbar} I [x,q])
\eqno(5.8)$$
where $I[x,q]$ is the action functional that depends on the trajectories
$x,~q$. The integration must satisfy boundary conditions:
$$ x(0) = x_0; \ x(t) = x; \ q(0) = q_0; \ q(t) = q \ . \eqno(5.9)$$
The expression for the propagator of the density matrix can be now written
in terms of actions corresponding to the two Lagrangians, Eqs. (5.1-3):
\begin{eqnarray} & &J(x,x',t | x_0,x'_0,t_0)  =  \int ~ Dx ~ Dx'
\exp {i \over \hbar}(I_{\cal S}[x] - I_{\cal S}[x']) \nonumber \\
& &\ \ \ \ \ \ \times \int~ dq~ dq_0~ dq_0'~ \rho_{\cal E}(q_0,q'_0)
\nonumber \\
& &\ \ \ \ \ \int ~ Dq ~ Dq'
\exp {i \over \hbar}(I_{\cal SE}[x,q] - I_{\cal SE}[x',q'])  \ .  \eqnum{5.10}
\end{eqnarray}
The separability of the initial conditions, Eq. (5.6), was used
to make propagator depend only on the initial conditions of
the environment. Collecting all terms containing integrals over 
${\cal E}$ in the above expression leads to the influence functional
(Feynman and Vernon, 1963):
\begin{eqnarray} F(x,x') \ & & = \ \int~ dq~ dq_0~ dq_0'~ \rho_{\cal
E}(q_0,q'_0) \nonumber \\
& & \int ~ Dq ~ Dq'
\exp {i \over \hbar}(I_{\cal SE}[x,q] - I_{\cal SE}[x',q'])  \ .
\eqnum{5.11}\end{eqnarray}
It can be evaluated explicitly for specific models of the initial density
matrix of the environment.

Environment in thermal equilibrium provides a useful and tractable model
for
the initial state. The density matrix of the $n$'th mode of
the thermal environment is:
\begin{eqnarray}& & \rho_{{\cal E}_n}(q,q')\
=\ {m_n \omega_n \over {2 \pi \hbar \sinh({{\hbar \omega_n } \over { k_B
T}} )}} \times \exp - \Bigl\{ {m_n \omega_n \over {2 \pi \hbar \sinh( { {
\hbar \omega_n} \over
{k_B T}})}} \nonumber \\
& &
\ \ \ \ \ \times \Bigl[ (q_n^2 + q_n'^2) \cosh ({{\hbar \omega_n} \over { k_B
T}}) - 2q_nq_n'
\Bigr]
\Bigr\} \ . \eqnum{5.12}\end{eqnarray}
The influence functional can be written
as (Grabert, Schramm, and Ingold, 1988):
\begin{eqnarray} & & i \ln F(x,x') = \int_0^tds~(x-x')(s)~
\int_0^sdu \nonumber \\
& & \Bigl(\eta(s-s')(x+x')(s') - i \nu (s-s')(x-x')(s')\Bigr) \ ,
\eqnum{5.13}\end{eqnarray} where $\nu(s)$ and $\eta(s)$ are dissipation and
noise kernels,
respectively, defined in terms of the spectral density:
$$ \nu(s) = \int_0^{\infty} d\omega ~ C(\omega) \coth ( {{\hbar \omega}
\beta /2} ) \cos (\omega s) ; \eqno(5.14)$$
$$ \eta(s) = \int_0^{\infty} d\omega ~ C(\omega) \sin (\omega s) \ .
\eqno(5.15)$$

With the assumption of thermal equilibrium at $k_BT= 1/\beta$, and 
in the harmonic oscillator case $V(x) = M \Omega_0^2 x^2 /2$, the
integrand of Eq. (5.10) for the propagator is Gaussian:  The integral can
be computed exactly, and should also have a Gaussian form.  The result
can be conveniently written in terms of  the diagonal and off-diagonal
coordinates of the density matrix in the  position
representation, $X=x+x'$, $Y=x-x'$:
\begin{eqnarray} & & \ \ \ \ \ \ \ \ J(X, Y, t | X_0, Y_0, t_0) \ = \
\nonumber \\
& &  {b_3 \over {2 \pi}}
{{ \exp i (b_1 XY + b_2 X_0 Y - b_3 XY_0 -b_4 X_0 Y_0) }
\over { \exp ( a_{11} Y^2 + 2a_{12} YY_0 + a_{22}Y_0^2)}} \ .
\eqnum{5.16} \end{eqnarray}
The time-dependent coefficients $b_k$ and $a_{ij}$ are computed from
the noise and dissipation kernels, that reflect the properties of the
environment. They obtain from the solutions of the equation:
$$ \ddot u(s) + \Omega_0^2 u(s) + 2 \int_0^s ds \eta (s - s' ) u (s' ) = 0 \ .
\eqno(5.17)$$
Two such solutions that satisfy boundary conditions  $u_1(0) = u_2(t) =
1$
and $u_1(t) = u_2(0) = 0$ can be used for this purpose.
They yield the coefficients of the Gaussian propagator through;
$$ b_{1(2)} (t) = \dot u_{2(1)}(t)/2 \ ,\ \ b_{3(4)}(t) = \dot
u_{2(1)}(0)/2 \ ,
\eqno(5.18a)$$
\begin{eqnarray} a_{ij}(t) = { 1 \over{1+ \delta_{ij}}} \int_0^t ds
\int_0^t ds' u_i(s)
u_j(s')
\nu (s-s'). \eqnum{5.18b}\end{eqnarray}

The master equation can be now obtained by taking the time derivative of
the
Eq. (5.5), which in effect reduces to the computation of the derivative
of the propagator, Eq. (5.16) above:
\begin{eqnarray} \dot J \ = & &\ \{ \dot b_3/b_3 + i \dot b_1 XY + i \dot
b_2 X_0 Y
- i\dot b_3XY_0 \nonumber \\
& &- i \dot b_4X_0Y_0 - \dot a_{11}Y^2 - \dot a_{12} YY_0
-\dot a_{22} Y_0^2 \} J \ . \eqnum{5.19}\end{eqnarray}
The time derivative of $\rho_{\cal S}$
can be obtained by multiplying the operator on the right hand side by an
initial
density matrix and integrating over the initial coordinates $X_0, ~Y_0$.
Given the form of Eq. (5.19), one may expect that this procedure will yield
an integro-differential (non-local in time) evolution operator for
$\rho_{\cal S}$. However, time dependence of the evolution operator
disappears
as a result of two identities satisfied by the propagator:
$$ Y_0 J= \Bigr({b_1 \over b_3}Y+{i\over b_3 } \partial_X \Bigl)J\
,\eqno(5.20a)$$
\begin{eqnarray} X_0 J= & &\Bigr(-{b_1 \over b_2}X - {i \over b_2}
\partial_Y \nonumber \\
& &- i \bigr( {2 a_{11} \over b_2}
+ {{a_{12} b_1} \over {b_2b_3}} \bigl) Y + {a_{12} \over {b_2b_3}}
\partial_X
\Bigl) J \ . \eqnum{5.20{\it b}}\end{eqnarray}
After the appropriate substitutions, the resulting equation
with renormalized Hamiltonian has a form:
\begin{eqnarray} & & \dot \rho_{\cal S}(x,x',t)\ = \ -{i \over \hbar} \langle x|[
H_{ren}(t),\rho_{\cal
S}]|x'\rangle  \nonumber \\
& & -\bigl(\gamma(t) (x-x')
(\partial_x + \partial_{x'}) - D(t)(x-x')^2 \bigr)
\rho_{\cal S}(x,x',t)  \nonumber \\
& & -i f(t)(x-x')(\partial_x+\partial_{x'})
\rho_{\cal
S}(x,x',t) .
\eqnum{5.21}
\end{eqnarray}
The calculations leading to this master equation are non-trivial. They
involve use of relations between the coefficients $b_k$ and $a_{ij}$. The
final result leads to explicit formulae for its coefficients:
$$ \Omega_{ren}(t) / 2 = b_1 \dot b_2/b_2 - \dot b_1 \ , \eqno(5.22a)$$
$$ \gamma(t) = - b_1 - \dot b_2 /2 b_2 \ , \eqno(5.23a)$$
\begin{eqnarray} D(t) =& & \dot a_{11} - 4 a_{11} b_1  \nonumber \\
& & + \dot a_{12} b_1/b_3
- \dot b_2(2a_{11}
+a_{12} b_1 / b_3)/b_2 \ , \eqnum{5.24{\it a}}\end{eqnarray}
$$2 f(t) = \dot a_{12}/b_3 - \dot b_2 a_{12}/(b_2b_3) - 4a_{11} \ .
\eqno(5.25a)$$
The fact that the exact master equation (5.21) is local
in time for an arbitrary spectrum of the environment is remarkable.
It was demonstrated by Hu, Paz, and Zhang (1992) following
discussions carried out under more restrictive assumptions by
Caldeira and Leggett (1983), Haake and Reibold (1985),
Grabert, Schramm, and Ingold (1988), and Unruh and Zurek (1989). It
depends on the linearity of the problem, that allows
one to anticipate (Gaussian) form of the propagator.

The above derivation of the exact master equation used the method of Paz (1994)
(see also Paz and Zurek, 2001).  Explicit formulae for the time-dependent
coefficients can be obtained when one focuses on the perturbative master
equation. It can be derived {\it ab initio} (see Paz and Zurek, 2001) 
but can be also obtained from the above results
by finding a perturbative solution to Eq. (5.17), and then substituting
it in Eqs. (5.22{\it a} - 5.25{\it a}). The resulting master equation
in the operator form is:
\begin{eqnarray} \dot \rho_{\cal S}\ = & &\ -{i \over \hbar}
[  H_{\cal S} + M \tilde\Omega (t) ^2 x^2/2 ,\rho_{\cal S}] -
{{i \gamma(t)} \over \hbar}  [x, \{ p, \rho_{\cal S}\}] \nonumber \\
& & - D(t) [x,[x,\rho_{\cal S}]] -{{f(t)} \over \hbar} [x,[p,\rho_{\cal
S}]] \ .
\eqnum{5.26}
\end{eqnarray}
Coefficients such as the frequency renormalization
$\tilde \Omega$, the relaxation coefficient $\gamma(t)$, and the normal
and anomalous diffusion coefficients $D(t)$ and $f(t)$ are given by:
$$ \tilde \Omega^2(t) = - {{2 } \over M}
\int_0^tds \cos(\Omega s) \eta(s) \eqno(5.22b)$$
$$ \gamma(t) = {{2} \over{ M \Omega}}
\int_0^tds \sin(\Omega s) \eta(s) \eqno(5.23b)$$
$$ D(t) =
{1 \over \hbar} \int_0^tds \cos(\Omega s) \nu(s) \eqno(5.24b)$$
$$f(t) = -{1 \over {M \Omega}} \int_0^tds \sin(\Omega s) \eta(s) \eqno(5.25b)$$

These coefficient can be made even more explicit when a convenient specific
model for the spectral density:
$$ C(\omega) = 2 M \gamma_0 {\omega \over \pi} {{\Gamma^2} \over {\Gamma^2
+ \omega^2}} \eqno(5.27)$$
is adopted. Above, $\gamma_0$ characterizes the strength of the interaction,
and $\Gamma$ is the high-frequency cutoff. Then:
$$ \tilde \Omega^2 = -{{2 \gamma_0 \Gamma^3} \over {\Gamma^2 + \Omega^2}}
\biggl(1-\bigl(\cos\Omega t - {\Omega \over \Gamma} \sin \Omega t \bigr)
e^{-\Gamma t} \biggr) \ ; \eqno(5.22c)$$
$$ \gamma(t) =  {{\gamma_0 \Gamma^2} \over {\Gamma^2 + \Omega^2}}
\biggl(1-\bigl(\cos\Omega t - {\Gamma \over \Omega} \sin \Omega t \bigr)
e^{-\Gamma t}  \biggr) \ . \eqno(5.23c)$$
Note that both of these coefficients
are initially zero. They grow to their asymptotic values on a timescale
set by the inverse of the cutoff frequency $\Gamma$.

The two diffusion coefficients can be also studied,
but it is more convenient to evaluate them numerically. In Fig. 5
we show their behavior. The normal
diffusion coefficient quickly settles to its long-time asymptotic value:
$$ D_{\infty} = M \gamma_0 \Omega \hbar^{-1} \coth (\hbar \Omega \beta/2)
\Gamma^2/(\Gamma^2 + \Omega^2) \ .  \eqno(5.28)$$
The anomalous diffusion coefficient $f(t)$ also approaches asymptotic value.
For high temperature it is suppressed by a cutoff $\Gamma$ with respect
to $D_{\infty}$, but the approach to $f_{\infty}$ is more gradual, 
algebraic rather than exponential.  Environments with different spectral 
content exhibit different behavior (Hu, Paz, and Zhang, 1992; 
Paz, Habib, and Zurek, 1993; Paz, 1994; Anglin, Paz and Zurek, 1997).

\subsection{Decoherence in quantum Brownian motion}

The coefficients of the master equation we have just derived can
be computed under a variety of different assumptions. The two obvious
characteristics of the environment one can change are its temperature $T$
and its spectral density $C(\omega)$. One general conclusion: In case of
high temperatures, $D(t)$ tends to a temperature-dependent constant,
and dominates over $f(t)$. Indeed, in this case all of the coefficients
settle to asymptotic values after an initial transient. Thus:
\begin{eqnarray} \dot \rho_{\cal S} =& & - {i \over \hbar}
[ H_{ren}, \rho_{\cal S}]
- \gamma (x-x')(\partial_x-\partial_{x'})\rho_{\cal S} \nonumber \\
& & - {{2 M \gamma k_B T} \over \hbar^2}(x-x')^2 \rho_{\cal S} \ .
\eqnum{5.29}\end{eqnarray}
This master equation for $\rho(x,x')$ obtains in the unrealistic but
convenient limit known as the {\it high temperature approximation} valid
when $k_B T$ is much higher than all the other relevant energy scales,
including the energy content of the initial state and the frequency
cutoff in $C(\omega)$ (see, e.g, Caldeira and Leggett, 1983). However,
when these restrictive conditions hold, Eq. (5.29) can be written
for an arbitrary $V(x)$. To see why, we give a derivation patterned on
Hu, Paz, and Zhang (1993).

We start with the propagator, Eq. (5.5),
$\rho_{\cal S}(x,x',t) = J(x,x',t|x_0,x'_0,t_0) \rho_{\cal S} (x_0,x_0',t_0)$,
which we shall treat as
if it were an equation for {\it state vector} of the {\it two dimensional}
system with coordinates $x,x'$. The propagator is then given by the
high-temperature version of Eq. (5.10);
\begin{eqnarray} & & J(x,x',t|x_0,x_0',t_0) = \int ~ Dx Dx' \exp{ i \over
\hbar}
\Bigr\{ I_R(x) - I_R(x') \Bigl\} \nonumber \\
& & \ \ \ \ \ \ \ e^{- M \gamma \bigl(\int_0^t ~ ds [x \dot x - x' \dot x'
+ x \dot x' - x' \dot x] + {{2  k_B T} \over \hbar^2}
    [x-x']^2 \bigr) } .  \eqnum{5.30} \end{eqnarray}
The term in the exponent can be interpreted as an
effective Lagrangian of a two-dimensional system:
\begin{eqnarray} L_{eff}(x,x')  = M \dot x^2/2 - V_R(x) -M \dot x'^2/2 +
V_R(x')
\nonumber \\
  + \gamma (x-x') (\dot x + \dot x') + i {{2 M \gamma k_B T} \over 
\hbar^2}(x-x')^2
\ . \eqnum{5.31} \end{eqnarray}
One can readily obtain the corresponding Hamiltonian;
$$ H_{eff} \ = \ \dot x \partial L_{eff}/\partial \dot x
+ \ \dot x' \partial L_{eff}/\partial \dot x' - L_{eff} \ . \eqno(5.32)$$
Conjugate momenta $p=p_x=M \dot x + \gamma (x-x')$ and $p'=p_{x'} = -
M\dot x'
+ \gamma (x-x')$ are used to express the kinetic term of $H_{eff}$.
After evaluating $\dot x$ and $\dot x'$ in terms of $p$ and $p'$ in
the expression for $H_{eff}$ one obtains:
\begin{eqnarray} H_{eff} = & & \bigr(p-\gamma (x-x') \bigl)^2/2M
- \bigr(p'-\gamma (x-x') \bigl)^2 /2M  \nonumber \\
& & + V(x) - V(x') - i 2 M \gamma k_B T
(x'-x)^2 / \hbar^2 \ . \eqnum{5.33}\end{eqnarray}
This expression yields the operator that
generates the evolution of the density matrix, Eq. (5.29).

The coefficients of Eq. (5.21) approach their high-temperature 
values quickly (see Fig. 5). Already for $T$ well
below what the rigorous derivation would demand high-temperature limit
appears to be an excellent approximation. The discrepancy -- manifested
by symptoms such as some of the diagonal terms of $\rho_{\cal S}(x',x)$
assuming negative values when the evolutions starts from an initial state
that is so sharply localized in position to have kinetic energy in excess
of the values allowed by the high-temperature approximation -- is limited to
the initial instant of order $1/\Gamma$, and is known to be essentially 
unphysical for other reasons (Unruh and Zurek, 1989; Ambegoakar, 1991; 
Romero and Paz, 1997; Anglin, Paz, and Zurek, 1997). This short - time anomaly
is closely tied to the fact that Eq. (5.33) (and, indeed, many of the
exact or approximate master equations derived to date) does not have the
Lindblad form (Lindblad, 1976; see also Gorini, Kossakowski and Sudarshan,
1976, Alicki and Lendi, 1987) of a dynamical semigroup.

High temperature master equation (5.29) is a good approximation in a wider
range of circumstances than the one for which it was derived
(Feynman and Vernon, 1963; Dekker, 1977; Caldeira and Leggett, 1983).
Moreover, our key qualitative conclusion -- rapid decoherence in the macroscopic
limit -- does not crucially depend on the approximations leading to Eq. (5.29).
We shall therefore use it in our further studies.

\subsubsection{Decoherence timescale}

In the macroscopic limit (that is, when $\hbar$ is small 
compared to other quantities with dimensions of action, such as 
$\sqrt{2 M k_B T \langle  (x-x')^2 \rangle  }$ in the last term) 
the high-temperature master equation is dominated by:
$$ \partial_t  \rho_{\cal S} (x,x',t) =
- \gamma \bigr\{ {{(x-x')} \over \lambda_T} \bigl\}^2
\rho_{\cal S} (x,x',t)  \ .
\eqno(5.34)$$
Above;
$$ \lambda_T \ = \ {\hbar \over \sqrt{2 M k_B T}} \  \eqno(5.35)$$
is thermal de Broglie wavelength. Thus, the density matrix
looses off-diagonal terms in position representation:
$$\rho_{\cal S}(x,x', t) \ = \ \rho_{\cal S}(x,x', 0)
e^{ - \gamma t ( { {x-x'} \over \lambda_T})^2} \ . \eqno(5.36)$$
while the diagonal ($x=x')$ remains untouched.

Quantum coherence decays exponentially at the
rate given by the relaxation rate times the square of the distance measured
in units of thermal de Broglie wavelength (Zurek, 1984a).
Position is the ``instantaneous pointer observable'':
If Eq. (5.36) was always valid, eigenstates of position would
attain the classical status.

The importance of position can be traced to the nature of the
interaction Hamiltonian between the system and the environment. According
to Eq. (5.3):
$$ H_{\cal SE} \ = \ x \sum_n c_n q_n \ . \eqno(5.37)$$
This form of $H_{\cal SE}$ is motivated by physics (Zurek, 1982; 1991):
Interactions depend on the distance. However, had we endeavored
to find a situation where a different form of the
interaction Hamiltonian -- say, a momentum-dependent interaction --
was justified, the form and consequently predictions
of the master equation would have been analogous to Eq. (5.36), but with
a substitution of the relevant observable ``monitored'' by the environment
for $x$. Such situations may be experimentally accessible 
(Poyatos, Cirac, and Zoller, 1996) providing
a test of one of the key ideas of einselection: the
relation between the form of interaction and the preferred basis.

The effect of the evolution, Eqs. (5.34) - (5.36), on the density
matrix in the position representation is easy to envisage. Consider
a superposition of two minimum uncertainty Gaussians. Off-diagonal peaks
represent coherence. They decay on a {\it decoherence timescale}
$\tau_D$, or with a {\it decoherence rate} (Zurek, 1984a, 1991);
$${\tau_D}^{-1} \ = \ \gamma ( {{ x-x'} \over \lambda_T } )^2 \ .
\eqno(5.38)$$
Thermal de Broglie wavelength, $\lambda_T$, is microscopic for massive
bodies and for the environments at reasonable temperatures. For a mass 1g
at room temperature and for the separation $x'-x=$1cm, Eq. (5.38)
predicts decoherence approximately $10^{40}$ times faster than relaxation!
Even cosmic microwave background suffices to cause rapid loss of quantum
coherence in objects as small as dust grains (Joos and Zeh, 1985).
These estimates for the rates of decoherence and relaxation should
be taken with a grain of salt: Often the assumptions that have led
to the simple high temperature master equation (5.29) are not
valid (Gallis \& Fleming, 1990; Gallis, 1992; Anglin, Paz \& Zurek, 1997).
For example, the decoherence rate cannot be faster than inverse of the
spectral cutoff in Eq. (5.27), or than the rate with which
the superposition is created. Moreover, for large separations quadratic
dependence of decoherence rate may saturate (Gallis and Fleming, 1990;
Anglin, Paz, and Zurek, 1997) as seen in the ``simulated decoherence''
experiments of Cheng and Raymer (1999). Nevertheless, in the macroscopic
domain decoherence of widely delocalized ``Schr\"odinger cat" states
will occur very much faster than relaxation, which proceeds
at the rate given by $\gamma$.

\subsubsection{Phase space view of decoherence}

A useful alternative way of illustrating decoherence
is afforded by the Wigner function representation:
\begin{eqnarray} W(x,p) = { 1 \over {2 \pi \hbar}} \int_{ -
\infty}^{+ \infty} dy ~
e^{(ipy/\hbar) } \rho(x+{y \over 2},x-{y \over 2})  \ . \eqnum{5.39}
\end{eqnarray}
Evolution equation followed by the Wigner function obtains through the
Wigner transform of the corresponding master equation.
In the high temperature limit, Eq. (5.29) (valid for general potentials)
this yields:
$$ \partial_t W = \{ H_{ren}, W \}_{MB} + 2 \gamma \partial_p (p W) +
D \partial_{pp} W \ . \eqno(5.40)$$
The first term -- Moyal bracket -- is the Wigner transform of
the von Neumann equation (see Section III). In the linear case it
reduces to the Poisson bracket. The second term is responsible for
relaxation. The last diffusive term is responsible for decoherence.

Diffusion in momentum occurs at the rate set by $D=2M\gamma k_B T$.
Its origin can be traced to the continuous ``measurement'' of the position of
the system by the environment: In accord with the Heisenberg
indeterminacy, measurement of position results in the increase of
the uncertainty in momentum (see Section IV).

Decoherence in phase space can be explained on the example of a
superposition of two Gaussian wavepackets. Wigner
function in this case is given by:
\begin{eqnarray} W(x,p) =  G(x+x_0,p) \ + \ G(x-x_0,p) \  \nonumber \\ +
(\pi \hbar)^{-1}
\exp(-p^2\xi^2/\hbar^2 - x^2/\xi^2) \cos (\Delta x p / \hbar) \ ,
    \eqnum{5.41}
\end{eqnarray}
where;
\begin{eqnarray}
    G(x \pm x_0, p-p_0) = {{ e^{ - (x \mp x_0)^2/\xi^2
- (p-p_0)^2 \xi^2/\hbar^2 }} \over {\pi \hbar}} \ . \eqnum{5.42}
\end{eqnarray}
We have assumed the Gaussians are not moving ($p_0=0$).

The oscillatory term in Eq. (5.41) is the signature
of superposition. The frequency of the oscillations is proportional to the
distance between the peaks. When the separation is only in position $x$,
this frequency is:
$$ f = \Delta x / \hbar = 2 x_0 / \hbar \ . \eqno(5.43)$$
Ridges and valleys of the interference pattern are parallel to the
separation between the two peaks. This, and the fact that $\hbar$
appears in the interference term in $W$ is important for
phase space derivation of the decoherence time. We focus on the dominant
effect and direct our attention on the last term
of Eq. (5.40). Its effect on the rapidly
oscillating interference term will be very different from its effect on the
two Gaussians: The interference term is dominated by the cosine:
$$ W_{int} \sim \cos ({ \Delta x \over \hbar} p) \ . \eqno(5.44)$$
This is an eigenfunction of the diffusion operator. Decoherence timescale
emerges (Zurek, 1991) from the corresponding eigenvalue;
$$ \dot W_{int} \approx - \{ D \Delta x^2 / \hbar^2 \}\times W_{int}\
.\eqno(5.45)$$
We have recovered the formula for $\tau_D$, Eq. (5.38), from a different
looking argument:  Eq. (5.40) has no explicit dependence on $\hbar$ for linear
potentials (in the nonlinear case $\hbar$ enters through the Moyal bracket).
Yet, the decoherence timescale contains $\hbar$ explicitly -- $\hbar$
enters through Eq. (5.43), that is through its role in determining
the frequency of the interference  pattern  $W_{int}$.

The evolution of the pure initial state of the type considered here is shown
in Fig. 6. There we illustrate evolution of the Wigner function for two
initial pure states: Superposition of two positions and superposition of
two momenta. There is a noticeable difference in the rate at which
the interference term disappears between these two cases. This was
anticipated. The interaction in Eq. (5.3) is a function of $x$.
Therefore, $x$ is monitored by the environment directly, and the superposition
of positions decoheres almost instantly.
By contrast, the superposition of momenta is initially insensitive
to the monitoring by the environment -- the corresponding initial state
is already well localized in the observable singled out by the interaction.
However, a superposition of momenta leads to a superposition of positions,
and hence to decoherence, albeit on a dynamical (rather than $\tau_D$)
timescale.

An intriguing example of a long-lived superposition of two seemingly distant
Gaussians was pointed out by Braun, Braun and Haake (2000) in the context of
superradiance. As they note, the relevant decohering interaction cannot
distinguish between some such superpositions, leading to a `Schr\"odinger 
cat' pointer subspace.

\subsection{Predictability sieve in phase space}

Decoherence rapidly destroys non-local superpositions. Obviously, states
that survive must be localized. However, they cannot be localized to a point
in $x$, as this would imply -- by Heisenberg's indeterminacy -- an infinite
range in momenta, and, hence, of velocities. As a result, a wave function
localized too well at one instant would become very non-local a moment later.

Einselected pointer states minimize the damage done by decoherence over 
the ``timescale of interest'' (usually associated
with predictability or with dynamics). They can be found through
the application of the predictability sieve outlined at the end
of Section IV. To implement it, we compute entropy
increase or purity loss for all initially pure states in the Hilbert space
of the system under the cumulative evolution caused by
the self-Hamiltonian and by the interaction with the environment.  It
would be a tall order to carry out the requisite calculations for an arbitrary
quantum system interacting with a general environment. We  focus
on the exactly solvable case.

In the high temperature limit the master equations (5.26) and (5.29)
can be expressed in the operator form:
\begin{eqnarray} \dot \rho = & & {1 \over {i \hbar}}[H_{ren},\rho] +
{\gamma \over {i \hbar}}
[\{p,x\},\rho] - {{\eta k_B T} \over \hbar^2}[x,[x,\rho]] \nonumber \\
& & \ \ \ \ \ \ \ \ \ \ - {{i \gamma} \over \hbar} \bigl( [x,\rho p] 
- [p, \rho x] \bigr)
\eqnum{5.46}
\end{eqnarray}
Above, $\eta = 2 M \gamma$ is the viscosity.
Only the last two terms can change entropy.
Terms of the form:
$$ \dot \rho = [ \hat O, \rho] \eqno(5.47)$$
where $\hat O$ is Hermitean leave the purity $\varsigma = Tr \rho^2$ and
the von Neumann entropy $H = - Tr \rho \lg \rho$ unaffected. This follows from
the cyclic property of the trace:
$$ {d \over {dt}} Tr \rho^N = \sum_{k=1}^{N} (Tr \rho^{k-1} [\hat O \rho]
\rho^{N-k}) = 0 \eqno(5.48)$$
Constancy of $Tr \rho^2$ is obvious, while for $Tr \rho \lg \rho$
it follows when the logarithm is expanded in powers of $\rho$.

Equation (5.46) leads to the loss of purity at the rate (Zurek, 1993a):
$${d \over {dt}} Tr \rho^2 = - {{4 \eta k_B T} \over \hbar^2}
Tr \bigl(\rho^2 x^2 - (\rho x)^2 \bigr) + 2 \gamma Tr \rho^2 \eqno(5.49)$$
The second term increases purity -- decreases entropy -- as the system is
damped from an initial highly mixed state. For predictability sieve this
term is usually unimportant, as for a vast majority of the initially pure
states its effect will be negligible when compared to the first,
decoherence - related term. Thus, in the case of pure initial states:
$${d \over {dt}} Tr \rho^2 = - {{4 \eta k_B T} \over \hbar^2} (\langle x^2\rangle  -
\langle x\rangle ^2),
\eqno(5.50)$$
Therefore, the instantaneous loss of purity is minimized for perfectly
localized states (Zurek, 1993a). The second term of Eq. (5.49) allows
for the equilibrium. (Nevertheless, early on, and for
very localized states, its presence causes an (unphysical)
{\it increase} of purity to {\it above} unity. This is a well-known
artifact of
the high-temperature approximation (see discussion following Eq. (5.33)).

To find most predictable states relevant for dynamics consider entropy
increase over a period of the oscillator. For a harmonic oscillator with
mass $M$ and frequency $\Omega$, one can compute purity loss averaged
over $\tau = 2 \pi / \Omega$:
$$ \Delta \varsigma |_0^{2\pi/\Omega} = - 2D \bigl( \Delta x^2 + \Delta p^2
/ (M \Omega)^2 \bigr)  \ . \eqno(5.51)$$
Above, $\Delta x$ and $\Delta p$ are dispersions of the state at the initial
time. By the Heisenberg indeterminacy, $\Delta x \Delta p \ge \hbar/2$.
The loss of purity will be smallest when:
$$ \Delta x^2 = \hbar / 2 M \Omega, \ \Delta p^2 = \hbar M \Omega /2  \ .
\eqno(5.52)$$
Coherent quantum states are selected by the predictability sieve
in an underdamped harmonic oscillator (Zurek, 1993a; Zurek, Habib and
Paz, 1993; Gallis, 1996; Tegmark and Shapiro, 1994; 
Wiseman and Vaccaro, 1998; Paraoanu, 2002).
Rotation induced by the self-Hamiltonian turns preference for states
localized in position into preference for localization in phase space.
This is illustrated in Fig. 7.

We conclude that for an underdamped harmonic oscillator
coherent Gaussians are the best quantum theory
has to offer as an approximation to a classical point. Similar
localization in phase space should obtain in the reversible classical limit
in which the familiar symptoms of the ``openness'' of the system --
such as the finite relaxation rate $\gamma = \eta / 2 M$ -- become
vanishingly small. This limit can be attained for large mass
$M \longrightarrow \infty$, while the viscosity $\eta$ remains
fixed and sufficiently large to assure localization (Zurek, 1991; 1993a).
This is, of course, not the only possible situation. Haake and Walls (1987)
discussed the overdamped case, where pointer states are still localized,
but relatively more narrow in position. On the other hand, "adiabatic"
environment enforces einselection in energy eigenstates (Paz and Zurek, 1999).

\subsection{Classical limit in phase space}

There are three strategies that allow one to simultaneously recover
both the classical phase space structure and the classical equations
of motion.

\subsubsection{Mathematical approach ($\hbar \rightarrow 0$)}

This ``mathematical'' classical limit could not be implemented without
decoherence, since the oscillatory terms associated with interference
do not have an analytic $\hbar \rightarrow 0$ limit (see, e.g, Peres, 1993).
However, in the presence of the environment, the relevant terms in
the master equations increase as ${\cal O}(\hbar^{-2})$, and make the
non-analytic manifestations of interference disappear. Thus, phase space
distributions can be always represented by localized coherent state
``points'', or by the distributions over the basis consisting of such points.

This strategy is easiest to implement starting from the phase-space
formulation. It follows from Eq. (5.45) that the interference term
in Eq. (5.41) will decay (Paz, Habib, and Zurek, 1993)
over the time interval $\Delta t$ as:
$$ W_{int} \sim \exp\bigl( - \Delta t {{D \Delta x^2} \over \hbar^2} \bigr)
\cos \bigl( {{\Delta x} \over \hbar} p \bigr) \ . \eqno(5.53)$$
As long as $\Delta t$ is large compared to decoherence
timescale $\tau_D \simeq \hbar^2 / D \Delta x^2$, the oscillatory
contributions
to the Wigner function $W(x,p) $ shall disappear with $\hbar \rightarrow 0$.
Simultaneously, Gaussians representing likely locations of the system
become
narrower, approaching $\delta$-functions in phase space. For
instance,
in Eq. (5.42);
$$ \lim_{\hbar \rightarrow 0} G(x-x_0, p-p_0) \ = \ \delta(x - x_0, p-p_0) \ ,
\eqno(5.54)$$
providing half-widths of the coherent states in $x$ and $p$
decrease to zero as $\hbar \rightarrow 0$. This would be
assured when, for instance, in Eqs. (5.41)-(5.42):
$$ \xi^2 \sim \hbar \ . \eqno(5.55)$$
Thus, individual coherent-state Gaussians approach phase space points.
This behavior indicates that in a macroscopic open system nothing but
probability distributions over localized phase space points can survive
in the $\hbar \rightarrow 0$ limit for any time of dynamical
or predictive significance. (Coherence between immediately adjacent
points separated only by $\sim \xi$, Eq. (5.55),
can last longer. This is no threat to the classical
limit. Small scale coherence is a part of a ``quantum halo'' of
the classical pointer states (Anglin and Zurek, 1996).)

The {\it mathematical classical limit} implemented by letting
$\hbar \rightarrow 0$ becomes possible in presence of decoherence.
It is tempting to take this strategy to its logical conclusion, and
represent every probability density in phase space in the point(er) basis
of narrowing coherent states. Such a program is beyond the scope
of this review, but the reader should be by now convinced that it
is possible. Indeed, Perelomov (1986) shows that a general quantum state can
be represented in a sparse basis of coherent states that occupy sites
of a regular lattice, providing that the volume per coherent state
``point'' is no more than $(2 \pi \hbar)^d$ in the $d-$dimensional
configuration space. In presence of decoherence arising from the
coordinate-dependent interaction, evolution of a general quantum
superposition should  be -- after a few decoherence times -- well
approximated by the probability distribution over such  Gaussian ``points''.

\subsubsection{Physical approach: The macroscopic limit}

The possibility of the $\hbar \rightarrow 0$ classical limit in presence
of decoherence is of interest. But $\hbar=1.05459 \times 10^{-27}$erg~s.
Therefore, a physically more reasonable approach increases the size of
the object, and, hence, its susceptibility to decoherence. This strategy
can be implemented starting with Eq. (5.40). Reversible dynamics obtains
as $\gamma \rightarrow 0$ while $D= 2M \gamma k_B T =\eta k_B T$ increases.

The decrease of $\gamma$ and the simultaneous increase of $\eta k_B T$
can be anticipated with the increase of the size and mass.
Assume that density of the object is independent of
its size $R$, and that the environment quanta
scatter from its  surface (as would photons or
air molecules). Then $M\sim R^3$ and $\eta \sim R^2$.  Hence:
$$ \eta \sim {\cal O}(R^2) \longrightarrow \infty \ , \eqno(5.56)$$
$$ \gamma= \eta/2M \sim {\cal O}(1/R) \longrightarrow 0 \ , \eqno(5.57)$$
as $R \rightarrow \infty $:
Localization in phase space and reversibility can be
simultaneously achieved in a macroscopic limit.

Existence of macroscopic classical limit in simple cases has been
pointed out some time ago (Zurek 1984a, 1991; Gell-Mann and Hartle, 1993).
We shall analyze it in the next section in a more complicated chaotic setting,
where reversibility can no longer be taken for granted. In the harmonic
oscillator case approximate reversibility is effectively guaranteed,
as the action associated with the 1-$\sigma$ contour of the
Gaussian state increases with time at the rate (Zurek, Habib, and Paz, 1993):
$$ \dot I = \gamma {{ k_B T} \over {\hbar \Omega}} \eqno(5.58)$$
Action $I$ is a measure of the lack of information about phase space
location. Hence, its rate of increase is a measure of the rate of
predictability loss. Trajectory is a limit of the ``tube'' swept in phase
space by the moving volume representing instantaneous uncertainty
of the observer about the state of the system. Evolution is
approximately deterministic when the area of this contour is nearly
constant. In accord with Eqs. (5.56)-(5.57) $\dot I$ tends to zero
in the reversible macroscopic limit:
$$ \dot I \sim {\cal O}(1/R) \eqno(5.59)$$
The existence of an approximately reversible trajectory-like thin tubes
provides an assurance that, having localized the system within a regular
phase space volume at $t=0$, we can expect that it can be found later inside
the Liouville - transported contour of nearly the same measure. Similar
conclusions follow for integrable systems.

\subsubsection{Ignorance inspires confidence in classicality}

Dynamical reversibility can be achieved with einselection
in the macroscopic limit. Moreover, $\dot I / I$ or other measures of
predictability loss decrease with the increase of $I$. This is
especially dramatic when quantified in terms of the von Neumann entropy,
that, for Gaussian states, increases at the rate (Zurek, Habib and Paz, 1993):
$$ \dot H = \dot I \lg {{I+1} \over {I-1}} \eqno(5.60a)$$
The resulting $\dot H$ is infinite for pure coherent states ($I=1$), but
quickly decreases with increasing $I$. Similarly, the rate of purity
loss for Gaussians is:
$$ \dot \varsigma = \dot I / I^2 \ . \eqno(5.60b) $$
Again, it tapers off for more mixed states.

This behavior is reassuring. It leads us to conclude that irreversibility
quantified through, say, von Neumann entropy production, Eq. (5.60a), will
approach $ \dot H \approx 2 \dot I / I  ~ ,$ vanishing in the limit
of large $I$. When in the spirit of the  macroscopic limit we do not
insist on the maximal resolution allowed by the quantum indeterminacy,
the subsequent predictability losses measured by the increase of
entropy or through the loss of purity will diminish.
Illusions of reversibility, determinism, and exact classical predictability
become easier to maintain in presence of ignorance about the initial state!

To think about phase space points one may not even need to invoke
a specific quantum state. Rather, a point can be regarded as a limit of an
abstract recursive procedure, in which phase space coordinates of the
system are determined better and better in a succession of increasingly
accurate measurements. One may be tempted to extrapolate this limiting
process {\it ad infintessimum} which would lead beyond Heisenberg's 
indeterminacy principle and to a false conclusion
that idealized points and trajectories ``exist objectively'', and that
the insider view of Section II can be always justified. While in our
quantum Universe this conclusion is wrong, and the extrapolation described
above illegal, the presence, within the Hilbert space, of localized 
wavepackets near the minimum uncertainty end of such imagined sequences 
of measurements is reassuring.  Ultimately, the ability to represent
motion in terms of points and their time - ordered 
sequences (trajectories) is the essence of classical mechanics.

\subsection{Decoherence, chaos, and the Second Law}

Breakdown of correspondence in this chaotic setting was described in
section III. It is anticipated to occur in all non-linear systems, as
the stretching of the wavepacket by the dynamics is a generic feature,
absent only in a harmonic oscillator. However, exponential instability
of chaotic dynamics implies rapid loss of quantum - classical correspondence
after the Ehrenfest time, $t_{\hbar}=\Lambda^{-1}\ln \chi \Delta p/ \hbar $.
Here $\Lambda$ is the Lyapunov exponent, $\chi = \sqrt{V_x/V_{xxx}}$
typically characterizes the dominant scale of nonlinearities in the potential
$V(x)$, and $\Delta p$ gives the coherence scale in the initial wavepacket.
The above estimate, Eq. (3.5) depends on the initial conditions.
It is smaller than, but typically close to, $t_r=\Lambda^{-1}\ln I/\hbar$,
Eq. (3.6), where $I$ is the characteristic action of the system.
By contrast, phase space patches of regular systems undergo
stretching with a power of time. Consequently, loss of correspondence
occurs only over a much longer $t_r \sim (I/\hbar)^{\alpha}$,
that depends polynomially on $\hbar$.

\subsubsection{Restoration of correspondence}

Exponential instability spreads the wavepacket to a ``paradoxical'' extent
at the rate given by the positive Lyapunov exponents
$\Lambda_+^{(i)}$. Einselection attempts to enforce localization in phase
space by tapering off interference terms at the rate given by the inverse of
the
decoherence timescale $\tau_D = \gamma^{-1} (\lambda_T / \Delta x)^2$.
The two processes reach {\it status quo} when the coherence length
$\ell_c$ of the wavepacket makes their rates comparable, that is:
$$ \tau_D \Lambda_+ \simeq 1 \eqno(5.61)$$
This yields an equation for the steady-state coherence length and for the
corresponding momentum dispersion:
$$ \ell_c \simeq \lambda_T \sqrt{{\Lambda_+} / {2 \gamma}} \ ;
\eqno(5.62)$$
$$ \sigma_c =  \hbar/ \ell_c = \sqrt{2 D/ \Lambda_+ } \ . \eqno(5.63)$$
Above, we have quoted results (Zurek and Paz, 1994) that follow
from a more rigorous derivation of the coherence length $\ell_c$ than
the ``rough and ready'' approach that led to Eq. (5.61).
They embody the same physical argument, but seek asymptotic behavior of
the Wigner function that evolves according to the equation:
\begin{eqnarray} \dot W = & & \{H, W\} + \sum_{n \geq 1}{ {\hbar^{2n}
(-)^n} \over
{2^{2n} (2n + 1)!}} \partial_x^{2n+1}V \partial_p^{2n+1}W \nonumber \\
& &  + \ D \partial_p^2 W
\ . \eqnum{5.64}\end{eqnarray}
The classical Liouville evolution generated by Poisson bracket ceases to be 
a good approximation of the decohering quantum evolution when the leading 
quantum correction becomes comparable to the classical force:
$$ {{\hbar^2} \over {24}} V_{xxx} W_{ppp} \approx {{\hbar^2} \over
{24}}
{{V_x} \over \chi^2}{{W_p} \over \sigma_c^2} \eqno(5.65)$$
The term $\partial_x V \partial_p W$ represents the classical force in
Poisson bracket. Quantum corrections are small when;
$$ \sigma_c \chi \gg \hbar \eqno(5.66)$$
Equivalently, Moyal bracket generates approximately Liouville flow when
the coherence length satisfies:
$$ \ell_c \ll \chi \eqno(5.67)$$
This last inequality has an obvious interpretation: It is a demand for the
localization to within a region $\ell_c$ small compared to the scale $\chi$
of the nonlinearities of the potential. When this condition holds, classical
force will dominate over quantum corrections.

Restoration of correspondence is illustrated in Fig. 8 where Wigner 
functions are compared with classical probability distributions in 
a chaotic system. The difference between the classical and quantum 
expectation values in same chaotic system is shown in Fig. 9. Even
relatively weak decoherence suppresses the discrepancy, helping 
reestablish the correspondence: $D=0.025$ translates through Eq. (5.62) 
into coherence over $\ell_c \simeq 0.3$, not much smaller than 
the nonlinearity scale $\chi \simeq 1$ for the investigated Hamiltonian 
of Fig. 8. 

\subsubsection{Entropy production}

Irreversibility is the price for the restoration of quantum-classical
correspondence in chaotic dynamics. It can be quantified through the
entropy production rate. The simplest argument recognizes
that decoherence restricts spatial coherence to $\ell_c$.
Consequently, as the exponential instability stretches the size $L^{(i)}$
of the distribution in directions corresponding to
the positive Lyapunov  exponents $\Lambda_+^{(i)}$,
$ L^{(i)} \sim \exp (\Lambda_+^{(i)} t) $ the squeezing mandated
by the Liouville theorem
in the complementary  directions corresponding to $\Lambda_-^{(i)}$
will halt at $\sigma_c^{(i)}$, Eq. (5.63). In this limit, the number of
pure states needed to represent resulting mixture increases exponentially
$$ N^{(i)} \simeq L^{(i)} / \ell_c^{(i)} \eqno(5.68)$$
in each dimension. The least number of pure states overlapped by $W$ will be
then $ {\cal N} = \Pi_iN^{(i)} $. This implies:
$$ \dot H \simeq \partial_t \ln {\cal N} \simeq \sum_i \Lambda_+^{(i)} \ .
\eqno(5.69)$$

This estimate for entropy production rate becomes accurate as the width 
of the Wigner function reaches saturation at $\sigma_c^{(i)}$. When a patch in
phase space corresponding to the initial $W$ is regular and smooth
on scales large compared to $\sigma_c^{(i)}$, evolution will start
nearly reversibly (Zurek and Paz, 1994). However, as squeezing brings 
the extent of the effective support of $W$ close to $\sigma_c^{(i)}$, 
diffusion bounds from below the size of the smallest features of $W$.
Stretching in the unstable directions continues unabated.
As a consequence, the volume of the support of $W$ will grow exponentially,
resulting in an entropy production rate set by Eq. (5.69), the sum of
the classical Lyapunov exponents. Yet, it has an obviously quantum
origin in decoherence. This quantum origin may be apparent initially,
as the rate of Eq. (5.69) will be approached from above when the initial
state is a non-local.  On the other hand, in a multidimensional
system different Lyapunov exponents may begin to contribute to entropy
production at different instants (as the saturation condition, Eq. (5.61),
may not be met simultaneously for all $\Lambda_+^{(i)}$). Then the entropy
production rate can accelerate (before subsiding as a consequence of
approaching equilibrium).

The timescales on which this estimates of entropy production apply are still
subject to investigation (Zurek and Paz, 1995a; Zurek, 1998b; Monteoliva
and Paz 2000 \& 2001) and even controversy (Casati and Chirikov, 1995b;
Zurek and Paz, 1995b). The instant when Eq. (5.69) becomes a good approximation
corresponds to the moment when the exponentially unstable evolution forces
the Wigner function to develop small phase space structures
on the scale of the effective decoherence - imposed ``coarse graining'',
Eq. (5.63). This will be a good approximation until the time $t_{EQ}$
at which equilibrium sets in. Both $t_{\hbar}$ have a logarithmic
dependence on the corresponding (initial and equilibrium) phase space volumes
$I_0$ and $I_{EQ}$, so the validity of Eq. (5.69) will be limited
to $t_{EQ}-t_{\hbar} \simeq \Lambda^{-1} \ln I_{EQ}/I_0$.

There is a simple and conceptually appealing way to extend the interval
over which entropy is produced at the rate given by Eq. (5.69): Imagine
an observer monitoring a decohering chaotic system, finding out its state
at time intervals small compared to $\Lambda^{-1}$, but large compared to
the decoherence timescale. One can show (Zurek, 1998b) that the average
increase of the size of the algorithmically
compressed records of measurement of a decohering
chaotic system (that is, the algorithmic randomness of the acquired data, 
see e.g. Cover and Thomas, 1991) is given -- after conversion
into bits from ``nats'' -- by Eq. (5.69). This conclusion holds
providing that the effect of the ``collapses of the wavepacket'' caused
by the repeated measurements is negligible -- i.e., the observer
is ``skillful''. A possible strategy a skillful observer may adopt
is that of indirect measurements, of monitoring a fraction of
the environment responsible for decoherence to find out the state
of the system. As we shall see in more detail in the following sections of
the paper, this is a very natural strategy, often employed by the observers.

A classical analogue of Eq. (5.69) has been obtained 
by Kolmogorov (1960) and Sinai (1960) starting from very different,
mathematical arguments that in effect relied on an arbitrary but fixed
coarse graining imposed on phase space (see Wehrl, 1978).
Decoherence leads to a similar looking {\it quantum} result in a very
different fashion: Coarse graining is imposed by the coupling to
the environment, but only in the sense implied by the einselection.
Its ``graininess'' (resolution) is set by the accuracy of the monitoring
environment. This is especially obvious when the indirect monitoring
strategy mentioned immediately above is adopted by the observers. Preferred
states will be partly localized in $x$ and $p$, but (in contrast to
the harmonic oscillator case with its coherent states) details of this
environment - imposed coarse graining will likely depend on phase space
location, precise nature of the coupling to the environment, etc. Yet,
in the appropriate limit, Eqs. (5.66)-(5.67), the asymptotic entropy production
rate defined with the help of the algorithmic contribution discussed
above (i.e., in a manner of the {\it physical entropy}, Zurek (1989))
does not depend on the strength or nature of the coupling, but is instead
given by the sum of the positive Lyapunov exponents.

Von Neumann entropy production consistent with the above discussion has been
now seen in numerical studies of decohering chaotic systems (Shiokawa and Hu,
1995; Furuya, Nemes, and Pellegrino, 1998; Miller and Sarkar, 1999;
Schack, 1998; Monteoliva and Paz, 2000 \& 2001). Extensions to situations
where relaxation matters, as well as in the opposite direction -- to where
decoherence is relatively gentle -- have been also discussed (Brun, Percival,
and Schack, 1995; Miller, Sarkar, and Zarum, 1998; Pattanyak, 2000). An
exciting development is the experimental study of the Loschmidt echo using NMR
techiques (Levstein, Usaj, and Pastawski, 1998; Levstein et al, 2000,
Jalabert and Pastawski, 2001) which sheds a new light on the irreversibility 
in decohering complex dynamical systems. We shall briefly return to this 
subject in Section VIII, while discussing experimental investigations relevant 
for decoherence and quantum chaos.

\subsubsection{Quantum predictability horizon}

The cross section $I$ of the trajectory-like tube containing the state
of the harmonic oscillator in phase space increases only slowly, Eq. (5.58), 
at a rate which -- once the limiting Gaussian is reached -- does not depend 
on $I$. By contrast, in chaotic quantum systems this rate is:
$$ \dot I \ \simeq \ I \sum_i \Lambda_+^{(i)} \ . \eqno(5.70)$$
A fixed rate of entropy production implies an exponential increase of
the cross-section of the tube of, say, the 1-$\sigma$ contour containing
points consistent with the initial conditions: Phase space support 
expands exponentially.

This quantum view of chaotic evolution can be compared with the classical
``deterministic chaos''. In both cases, in the appropriate classical limit
-- which may involve either mathematical $\hbar \rightarrow 0$, or
a macroscopic limit -- the future state of the system can be in principle
predicted to a set accuracy for an arbitrarily long time. However, such
predictability can be accomplished only when the initial conditions are given
with the resolution that increases exponentially with the time interval
over which the predictions are to be valid. Given the fixed value of 
$\hbar$, there is therefore a
{\it quantum predictability horizon} after which the Wigner function of the
system starting from an initial minimum uncertainty Gaussian becomes
stretched to the size of the order of the characteristic dimension of
the system (Zurek, 1998b). The ability to predict the location of the system
in the  phase space is then lost after $t\sim t_{\hbar}$, Eq. (3.5),
regardless of whether evolution is generated by the Poisson or Moyal bracket,
or, indeed, whether the system is closed or open.

The case of regular systems
is closer to the harmonic oscillator. The rate of increase of
the cross-section of phase space ``trajectory tube'' consistent with
the initial patch in the phase space will asymptote to $ \dot I \simeq const$:
$$ \dot H = \dot I / I \sim 1/t \ . \eqno(5.71)$$
Thus, initial conditions allow one to predict future of a regular system for
time intervals that are exponentially longer than in the chaotic case.
The rate of entropy production of an open quantum system is therefore
a very good indicator of the nature of its dynamics, as was conjectured
some time ago (Zurek and Paz, 1995a), and as seems born out in the
numerical simulations (Shiokawa and Hu, 1995; Miller, Sarkar and Zarum,
1998; Miller and Sarkar, 1999; Monteoliva and Paz, 2000 \& 2001).

\section{EINSELECTION AND MEASUREMENTS}

It is often said that quantum states play only an {\it epistemological}
role in describing observers knowledge about the past measurement outcomes
that have prepared the system (Jammer, 1974; d'Espagnat, 1976, 1995;
Fuchs and Peres, 2000). In particular -- and this is a key argument
against their objective existence (against their {\it ontological} status)
-- it is impossible to find out what the state of an isolated quantum
system {\it is} without prior information about the observables used
to prepare it: Measurement of observables that do not commute with
this original set will inevitably create a {\it different} state.

The incessant monitoring of the einselected observables by the environment
allows pointer states to {\it exist} in much the same way classical
states do. This ontological role of the einselected quantum states
can be justified operationally, by showing that {\it in presence of
einselection one can find out what the quantum state is, without
inevitably re-preparing it by the measurement}. Thus, einselected
quantum states are no longer  just epistemological. In a system
monitored by the environment {\it what is} -- the einselected states --
coincides with {\it what is known to be} -- what is recorded by the
environment (Zurek, 1993a,b; 1998a).

The conflict between the quantum and the classical was originally noted
and discussed almost exclusively in the context of quantum measurements
(Bohr, 1928; Mott, 1929; von Neumann, 1932; Dirac, 1947; Zeh, 1971, 1973, 1993;
d'Espagnat, 1976, 1995; Zurek 1981, 1982, 1983, 1991, 1993a\&b, 1998a;
Omn\`es 1992, 1994; Elby, 1993, 1998; Butterfield, 1996; Donald, 1995;
Giulini et al., 1996;
Bub, 1997; Saunders, 1998; Healey, 1998; Bacciagaluppi and Hemmo, 1998;
Healey and Hellman, 1998). Here I shall consider quantum measurements,
and, more to the point, acquisition of information in quantum theory
from the point of view of decoherence and einselection.

\subsection{Objective existence of einselected states}

To demonstrate {\it objective existence} of {\it einselected} states we now
develop an operational definition of existence and show how, in the open
system, one can find out what the state {\it was and is}, rather than just
``prepare'' it. This point was made before (Zurek, 1993a; 1998a), but
this is first time I discuss it in more detail.

Objective existence of states can be defined operationally
by considering two observers. The first of them is the {\it record keeper}
{\bf R}. He prepares the states with the original measurement and will use
his records to find out if they were disturbed by the measurements carried out
by other observers, e.g. the {\it spy} {\bf S}. The goal of {\bf S} is to
discover the state of the system without perturbing it. When an observer
can consistently find out the state of the system without changing it,
that state -- by our operational definition -- will be said to
{\it exist objectively}.

In absence of einselection the situation of the spy {\bf S} is hopeless:
{\bf R} prepares states by measuring sets of commuting observables.
Unless {\bf S} picks, by sheer luck, the same observables
{\it in case of each state}, his measurements will re-prepare states
of the systems. Thus, when {\bf R} re-measures using the original observables,
he will likely find answers different from his records. The spy {\bf S}
will ``get caught'', because it is impossible to find out an initially
unknown state of an isolated quantum system.

In presence of environmental monitoring the nature of the ``game'' between
{\bf R} and {\bf S} is dramatically altered. Now it is no longer possible for
{\bf R} to prepare an arbitrary pure state that will persist or predictably
evolve without losing purity. Only the einselected states that are already
monitored by the environment -- that are selected by the 
predictability sieve --
will survive. By the same token, {\bf S} is no longer clueless about
the observables he can measure without leaving incriminating evidence.
For example, he can independently prepare and test survival of various states
in the same environment to establish which states are einselected,
and then measure appropriate pointer observables. Better yet,
{\bf S} can forgo direct measurements of the
system, and gather information indirectly, by monitoring the environment.

This last strategy may seem contrived, but indirect measurements -- acquisition
of information about the system by examining fragments of the environment
that have interacted with it -- is in fact more or less the only
strategy employed by the observers. Our eyes, for example, intercept
only a small fraction of the photons that scatter from various objects.
The rest of the photons constitute the environment, which retains at least
as complete a record of the same einselected observables as we can obtain
(Zurek, 1993; 1998a).

The environment ${\cal E}$ acts as a persistent observer, dominating
the game with frequent questions, always about the same observables,
compelling both {\bf R} and {\bf S} to focus on the einselected states.
Moreover, ${\cal E}$ can be persuaded to share its records of the system. 
This accessibility of the einselected states is not a violation of 
the basic tenets of quantum physics. Rather, it is a consequence of 
the fact that the data required to turn
quantum state into an ontological entity -- an einselected pointer state --
are abundantly supplied by the environment.

We emphasize the {\it operational} nature of this criterion for existence:
There may be in principle a pure state of the Universe including the
environment, the observer, and the measured system. While this may matter
to some (Zeh, 2000), real observers are forced to perceive the Universe
the way we do: We are a part of the Universe, observing it from within.
Hence, for us, {\it environment-induced superselection specifies what exists.}

Predictability emerges as a key criterion of {\it existence}.
The only states {\bf R} can rely on to store the information
are the pointer states. They are also the obvious choice for {\bf S}
to measure. Such measurements can be accomplished without danger of
re-preparation. Einselected states are insensitive to measurement
of the pointer observables -- they have already been ``measured'' by the
environment. Therefore, additional projections $P_i$ onto the einselected
basis will not perturb the density matrix (Zurek, 1993a) -- it will be
the same before and after the measurement:
$$ \rho^D_{after} = \sum_i P_i \rho^D_{before} P_i \ . \eqno(6.1)$$
Correlations with the einselected states
will be left intact (Zurek 1981; 1982).

Superselection for the observable $\hat A = \sum_i \lambda_i P_i$ with
essentially arbitrary non-degenerate eigenvalues $\lambda_i$ and
eigenspaces $P_i$ can be expressed (Bogolubov et al., 1990) through Eq. (6.1).
Einselection attains this, guaranteeing diagonality of density
matrices in the projectors $P_i$ corresponding to pointer states.
These are sometimes called {\it decoherence - free subspaces} when they
are degenerate; compare also nonabelian case of {\it noiseless subsystems}
discussed in quantum computation; see Zanardi and Rasetti, 1997; Zanardi 1998;
Duan and Guo 1998; Lidar, Bacon, and Whaley, 1999; Knill, Laflamme, and Viola, 2000;
Blanchard and Olkiewicz, 2000; Zanardi 2000).

\subsection{Measurements and memories}

The memory of an apparatus or of an observer can be modelled as
an open quantum ${\cal A}$, interacting with
${\cal S}$ through a Hamiltonian explicitly
proportional to the measured observable $\hat s$ \footnote{The observable
$\hat s$ of the system and $\hat B$ of the apparatus memory need not be
discrete with a simple spectrum as was previously assumed.
Even when $\hat s$ has a complicated spectrum, the
outcome of the measurement can be recorded in the eigenstates of the
memory observable $\hat A$, conjugate of $\hat B$, Eq. (2.21).
For the case of discrete $\hat s$ the necessary calculations that attain
premeasurement -- the quantum correlation that is the first step in
the creation of the record -- were already carried out in section II.
For the other situations they are quite similar. In either case,
they follow the general outline of the von Neumann's (1932) discussion.}
$$ H_{int} = - g \hat s \hat B \sim \hat s {\partial \over {\partial \hat
A}}\ . \eqno(6.2)$$
Von Neumann  (1932) considered an apparatus isolated from the environment.
At the instant of the interaction between the apparatus and the measured
system this is a convenient assumption. For us it suffices
to assume that, at that instant, the interaction Hamiltonian between
the system and the apparatus dominates. This can be accomplished by
taking the coupling $g$ in Eq. (6.2) to be $g(t) \sim \delta(t-t_0).$
Premeasurement happens at $t_0$:
$$ (\sum_i \alpha_i |s_i\rangle ) |A_0\rangle  \ \longrightarrow \ \sum_i \alpha_i |s_i\rangle 
|A_i\rangle  \ . \eqno(6.3)$$
In practice the action is usually large enough to accomplish amplification.
As we have seen in section II,
all this can be done without an appeal to the environment.

For a real apparatus interaction with the environment is inevitable.
Idealized effectively classical memory will retain correlations,
but will be subject to einselection: Only the einselected
memory states (rather than their superpositions) will be useful for
(or, for that matter, accessible to) the observer:
Decoherence timescale is very short compared to the time
after which memory states are typically consulted (i. e., copied or used
in information processing), which is in turn much shorter than
the relaxation timescale, on which memory ``forgets''.

Decoherence leads to {\it classical} correlation:
\begin{eqnarray} & & \rho^P_{\cal SA} = \sum_{i,j} \alpha_i \alpha_j^*
|s_i\rangle \langle s_j||A_i\rangle \langle A_j| \longrightarrow
\nonumber \\
& &\longrightarrow
\sum_i |\alpha_i|^2 |s_i\rangle \langle s_i||A_i\rangle \langle A_i| = \rho_{\cal SA}^D
\eqnum{6.4}\end{eqnarray}
following an entangling premeasurement. Left hand side of Eq. (6.4) coincides
with Eq. (2.44c), the ``outsiders'' view of the classical
measurement. We shall see how and to what extent its other aspects
-- including the insiders Eq. (2.44a) and the discoverers
Eq. (2.44b) -- can be understood through einselection.

\subsection{Axioms of quantum measurement theory}

Our goal is to establish whether the above model can
fulfill requirements expected from measurement in textbooks
(that are, essentially without exception, written in the spirit of
the Copenhagen Interpretation). There are several equivalent `textbook' 
formulations of axioms of quantum theory. We shall (approximately)
follow Farhi, Goldstone, and Gutmann (1989) and posit:
\begin{enumerate}
\item [(i)]The states of a quantum system ${\cal S}$ are associated with
the vectors $|\psi\rangle $ which are the elements of the Hilbert space
${\cal H_{\cal S}}$ that describes ${\cal S}$.

\item [(ii)] The states evolve according to $ i \hbar |\dot \psi\rangle  = H |\psi\rangle $
where $H$ is Hermitean.

\item [(iii a)] Every observable $O$ is associated with a Hermitian operator
$\hat O$.

\item [(iii b)] The only possible outcome of a measurement of $O$ is an
eigenvalue $o_i$ of $\hat O$.

\item [(iv)] Immediately after a measurement that yields the value $o_i$
the system is in the eigenstate $|o_i\rangle $ of $\hat O$.

\item [(v)] If the system is in a normalized state $|\psi\rangle $, then
a measurement of $\hat O$ will yield the value $o_i$ with
the probability $p_i = |\langle  o_i|\psi\rangle |^2$.
\end{enumerate}

The first two axioms make no reference to measurements. They state
the formalism of the theory. Axioms (iii) - (v) are, on the other hand,
at the heart of the present discussion. In the spirit, they go back to Bohr
and Born. In the letter, they follow von Neumann (1932) and Dirac (1947).
The two key issues are the {\it projection postulate}, implied by
a combination of (iv) with (iiib), and the (Born's)
{\it probability interpretation}, axiom (v).

To establish (iiib), (iv) and (v) we shall interpret -- in operational terms
-- statements such as ``the system {\it is} in the eigenstate'' and
``measurement  will yield value ... with {\it the probability} ..."
by specifying what these statements mean for the sequences of
{\it records} made and maintained by an idealized, but physical {\it memory}.

We note that the above Copenhagen-like axioms presume existence
of quantum systems and of classical measuring devices.  This (unstated)
``axiom (o)'' complements axioms (i) - (v). Our version of axiom (o)
demands that the Universe consists of systems, and asserts that 
a composite system can be described by a tensor product of the Hilbert 
spaces of the constituent systems. Some quantum systems can be measured, 
others can be used as measuring devices and/or memories, and as quantum 
environments that interact with either or both.

Axioms (iii) - (v) contain many idealizations. For instance, in real life
or in laboratory practice measurements have errors (and hence can yield
outcomes other than the eigenvalues $o_i$). Moreover, only rarely do they
``prepare'' the system in the eigenstate of the observable they are designed
to measure. Furthermore, coherent states -- often an outcome of measurements,
e. g., in quantum optics -- form an overcomplete basis. Thus, their detection
does not correspond to a measurement of a Hermitean observable. Last not least,
the measured quantity may be inferred from some other quantity (e.g. deflection 
of a beam in the Stern-Gerlach  experiment).
Yet, we shall not go beyond the idealizations of (i) - (v) above. Our goal
is to describe measurements in a quantum theory ``without collapse'', to use
axioms (o), (i) and (ii) to understand the origin of the other axioms. 
Non-ideal measurements are a fact of life incidental to this goal.

\subsubsection{Observables are Hermitean -- axiom (iiia)}

In the model of measurement considered here the observables are Hermitean
as a consequence of the assumed premeasurement interaction, e.g. Eq. (2.24).
In particular, $H_{int}$ is a product of the to-be-measured observable
of the system and of the ``shift operator'' in the pointer of the apparatus
or in the record state of the memory. Interactions involving non-Hermitean
operators (e.g., $H_{int} \sim a^{\dagger}b + a b^{\dagger}$) may be, however,
also considered.

It is tempting to speculate that one could dispose of the observables
(and, hence, of the postulate (iiia)) altogether in the formulation of 
the axioms of quantum theory. The only ingredients necessary to 
describe measurements are then the effectively classical,
but ultimately quantum apparatus, and the measured system. Observables
emerge as a derived concept, as a useful idealization, ultimately based
on the structure of the Hamiltonians. Their utility relies on
the conservation laws that relate outcomes of several measurements.
Most basic of these laws states that the system that did not (have time to)
evolve will be found in the same state when it is re-measured: This is
the content of axiom (iv). Other conservation laws are also reflected in
the patterns of correlation in the measurement records, that must in turn
arise from the underlying symmetries of the Hamiltonians.

Einselection should be included in this program, as it decides which of
the observables are accessible and useful -- which are effectively
classical. It is conceivable that also the ``fundamental'' superselection
may emerge in this manner (see Zeh, 1970; Zurek, 1982; for early speculations;
Giulini, Kiefer, and  Zeh, 1995; and Kiefer, 1996; Giulini, 2000, for
the present status of this idea).

\subsubsection{Eigenvalues as outcomes -- axiom (iiib)}

This statement is the first part  of the ``collapse'' postulate. Given
einselection, (iiib) is easy to justify: We need to show that only
the records inscribed in the einselected states of the apparatus pointer
can be read off, and that -- in a well-designed measurement --
they correlate with the eigenstates (and, therefore, eigenvalues)
of the measured observable $\hat s$.

With Dirac (1947) and von Neumann (1932) we assume that the apparatus is built
so that it satisfies the obvious ``truth table'' when the eigenstates
of the measured observable are at the input:
$$ |s_i\rangle  |A_0\rangle  \longrightarrow |s_i\rangle  |A_i\rangle  \ \eqno(6.5)$$
To assure this one can implement the interaction in accord with
Eq. (6.2) and the relevant discussion in section II. This is not
to say that there are no other ways: Braginski and Khalili (1996),
Aharonov, Anandan and Vaidman (1993), and Unruh (1994) have all
considered ``adiabatic measurements'', that correlate the apparatus
with the discrete energy eigenstates of the measured system, nearly
independently of the structure of $H_{int}$.

Truth table of Eq. (6.5) does not require collapse -- for any initial
$|s_i\rangle $ it represents a ``classical measurement in quantum notation''
in the sense of Section II. However, Eq. (6.5) typically leads to
a superposition of outcomes. This is the ``measurement problem''.
To address it, we assume that the record states $\{|A_i\rangle \}$ are einselected.
This has two related consequences:
(i) Following the measurement, the joint density matrix of
the system and the apparatus decoheres, Eq. (6.3), so that it satisfies the
superselection condition, Eq. (6.1), for $P_i = |A_i\rangle \langle A_i|$.
(ii) Einselection restricts states that can
be read of ``as if they were classical'' to pointer states.

Indeed, following decoherence only pointer states $\{|A_i\rangle \}$ of the
memory can be measured without diminishing the correlation with the states
of the system. Without decoherence, as we have seen in section II,
one could use entanglement between
${\cal S}$ and ${\cal A}$ to end up with an almost arbitrary superposition
states of either, and, hence, to violate the letter and the spirit of (iiib).

Outcomes are restricted to the eigenvalues of measured
observables because of einselection. Axiom (iiib) is then a consequence
of the effective classicality of pointer states, the only ones that can be
``found out'' without being disturbed. They can be consulted
repeatedly and remain unaffected under the joint scrutiny of the observers
and of the environment (Zurek, 1981; 1993a; 1998a).

\subsubsection{Immediate repeatability, axiom (iv)}

This axiom supplies the second half of the ``collapse'' postulate. It asserts
that in the absence of (the time for) evolution quantum system will remain
in the same state, and its re-measurement will lead to the same outcome. Hence,
once the system is found out to be in a certain state, it really is there.
As in Eq. (2.44b) observer perceives potential options ``collapse'' to 
a single actual outcome. (The association of the axiom (iv) with the collapse
advocated here seems obvious, but it is not common: Rather, some form of
our axiom (iiib) is usually regarded as the sole ``collapse postulate''.)

Immediate repeatability for Hermitean observables with discrete spectra
is straightforward to justify on the basis of Schr\"odinger evolution
generated by $H_{int}$ of Eq. (6.2) alone, although its implications depend
on whether the premeasurement is followed by einselection. Everett (1957)
used the ``no decoherence'' version as a foundation of his relative
state interpretation. On the other hand, without decoherence
and einselection one could postpone the choice of what was actually
recorded by taking advantage of the entanglement between the system
and the apparatus, evident on the left hand side of Eq. (6.3). For instance,
a measurement carried out on the apparatus in a basis different from
$\{|A_i\rangle \}$ would also exhibit a one-to-one correlation with the system:
$ \sum_i \alpha_i |s_i\rangle  | A_i\rangle =\sum_k \beta_k |r_k\rangle  |B_k\rangle  $.
This flexibility to re-write wavefunctions in different bases comes at a price
of relaxing the demand that the outcome states $\{|r_k\rangle \}$ be orthogonal
(so that there would be no associated Hermitean observable). However,
as was already noted, coherent states associated with a non-Hermitean
annihilation operator can also be an outcome of a measurement. Therefore
(and in spite of the strict interpretation of axiom (iiia)) this is not
a very serious restriction.

In presence of einselection the basis ambiguity disappears. Immediate
repeatability would apply only to the records made in the einselected
states. Other apparatus observables lose correlation with the state of
the system on decoherence timescale. In the effectively classical
limit it is natural to demand repeatability extending beyond that very
short time interval. This demand makes the role of einselection
in establishing axiom (iv) evident. Indeed, such repeatability is
-- albeit in a more general context -- the motivation for
the predictability sieve.

\subsubsection{Probabilities, einselection and records}

Density matrix alone -- without the preferred set of states -- does not
suffice as a foundation for a probability interpretation. For, any mixed
state density matrix $\rho_{\cal S}$ can be decomposed into sums
of density matrices that add up to the same resultant $\rho_{\cal S}$,
but need not share the same eigenstates.
For example, consider $\rho_{\cal S}^a$ and $\rho_{\cal S}^b$,
representing two different preparations (i.e., involving measurement of
two distinct, non-commuting observables) of two ensembles, each with multiple
copies  of a system ${\cal S}$. When they are randomly mixed in proportions
$p^a$ and  $p^b$, the resulting density matrix:
$$\rho_{\cal S}^{a \vee b}=p^a \rho_{\cal S}^a+p^b \rho_{\cal S}^b $$
is the complete description of the unified ensemble
(see Schr\"odinger (1936); Jaynes (1957)).

Unless $[\rho_{\cal S}^a, \rho_{\cal S}^b]=0$, the eigenstates of
$\rho_{\cal S}^{a \vee b}$ do not coincide with the eigenstates of
the components. This feature makes it difficult to regard any density
matrix in terms of probabilities and ignorance. Such ambiguity would be
especially troubling if it arose in the description of an observer
(or, for that matter, of any classical system). Ignorance interpretation
-- i.e., the idea that probabilities are observer's way of dealing with
uncertainty about the outcome we have briefly explored
in the discussion of the insider-outsider dichotomy, Eqs. (2.44),
-- requires at the very least that the set of events (``the sample space'')
exists independently of the information at hand -- i.e, independently of
$p^a$ and $p^b$ in the example above. Eigenstates of the density matrix
do not supply such events, since the additional loss of information associated
with mixing of the ensembles alters the candidate events.

Basis ambiguity would be disastrous for
record states. Density matrices describing a joint state of the
memory ${\cal A}$ and of the system ${\cal S}$:
$$\rho_{\cal AS}^{a \vee b}=p_a\rho_{\cal AS}^{a}+p_b\rho_{\cal AS}^{b} $$
would have to be considered. In absence of einselection the eigenstates
of such $\rho_{\cal AS}^{a \vee b}$ need not be even associated with a fixed
set of record states of the presumably classical ${\cal A}$. Indeed,
in general $\rho_{\cal AS}^{a \vee b}$ has a non-zero discord,\footnote{
As we have seen in section IV, Eqs. (4.30) -- (4.36), {\it discord}
$ \delta{\cal I}_{\cal A}({\cal S | A}) = {\cal I(S:A) - J_{\cal A}(S:A)} $
is a measure of the``quantumness'' of correlations. 
It should disappear as a result of the classical equivalence of two 
definitions of the mutual information, but is in general positive 
for quantum  correlations, including, in particular, pre-decoherence
$\rho_{\cal SA}$. Discord is asymmetric,
$\delta{\cal I}_{\cal A}({\cal S | A})\neq\delta{\cal I}_{\cal S}({\cal A| S})$.
Vanishing of $ \delta{\cal I}_{\cal A}({\cal S | A}) $ (i.e., of the discord 
in the direction exploring the classicality of the states of ${\cal A}$,
on which $H(\rho_{\cal S|A})$ in the asymmetric
$J_{\cal A}(S:A)$, Eq. (4.32), is conditioned) is necessary for the
classicality of the measurement outcome (Zurek, 2000; Ollivier and 
Zurek, 2002, Zurek, 2002a). $ \delta{\cal I}_{\cal A}({\cal S | A})$ 
can disappear as a result of decoherence in the einselected basis 
of the apparatus. Following einselection it is then possible to ascribe 
probabilities to the pointer states.
\\
In perfect measurements of Hermitean observables discord vanishes ``both ways'':
$\delta{\cal I}_{\cal A}({\cal S | A})=\delta{\cal I}_{\cal S}({\cal A| S})=0$
for the pointer basis and for the eigenbasis of the measured observable  
correlated with it.  Nevertheless, it is possible to encounter situations when 
vanishing of the discord in one direction is {\it not}
accompanied by its vanishing ``in reverse''. Such correlations
are ``classical one way'' (Zurek, 2002a) and could arise, for instance,
in measurements of non-hermitean observables (which -- as we have already
noted -- happen, in spite of the axiom (iii)).
\\
This asymmetry between classical ${\cal A}$ and quantum ${\cal S}$ arises
from the einselection. Classical record states are not arbitrary
superpositions. Observer accesses his memory in the basis in which
it is monitored by the environment. The information stored 
is effectively classical because it is being widely disseminated.
States of the observers memory exist objectively -- they can be
found out through their imprints in the environment.}
and its
eigenstates are entangled (even when the above $\rho_{\cal AS}^{a \vee b}$ is 
separable, and can be expressed as mixture of matrices that have no
entangled eigenstates). This would imply an ambiguity of what are the record
states, precluding probability interpretation of measurement outcomes.

Observer may nevertheless have records of a system that is in an ambiguous
situation described above. Thus;
$$\rho_{\cal AS}^{a \vee b} \ = \ \sum_k w_k|A_k\rangle \langle A_k|(p^a_k\rho_{{\cal
S}_k}^a
+p^b_k\rho_{{\cal S}_k}^b) $$
are admissible for an effectively classical ${\cal A}$ correlated with
a quantum ${\cal S}$. Now the discord 
$\delta_{\cal A}{\cal I}({\cal S | A})=0$.

Mixing of ensembles of pairs of correlated systems one of which is subject
to einselection does not lead to ambiguities discussed above. Discord
$\delta_{\cal A}({\cal S | A})$ disappears in the einselected basis of 
${\cal A}$, and the eigenvalues of the density matrices can behave as classical
probabilities associated with `events' -- with the records. The ``menu''
of possible events in the sample space -- e.g., records in memory --
is fixed by einselection. Whether one can really justify this interpretation
of the eigenvalues of the reducued density matrix is a separate question we 
are about to address.

\subsection{Probabilities from Envariance}

The view of `the emergence of the classical' based on the environment
-- induced superselection has been occassionally described
as ``for practical purposes only'' (see, e.g., Bell, 1990), to be 
contrasted with more fundamental (if nonexistent) solutions of
the problem one could imagine (i.e., by modifying quantum theory; see Bell, 
1987, 1990). This attitude can be in part traced to the reliance of einselection
on reduced density matrices: For, even when explanations of all aspects 
of the effectively classical behavior are accepted in the framework
of, say, Everett's MWI, and after the operational approach to
objectivity and perception of unique outcomes based on the existential
interpretation explained earlier is adopted, one major gap remains: 
Born's rule -- axiom (v) connecting probabilities with the amplitudes, 
$ p_k = |\psi_k|^2$ -- has to be postulated in addition to axioms (o) - (ii).
True, one can show that within the framework of einselection Born's rule 
emerges naturally (Zurek, 1998a). Decoherence is, however, based on 
on reduced density matrices. Now, since they were introduced by Landau (1927), 
it is known 
that ``partial trace'' leading to reduced density matrices is predicated 
on Born's rule (see Nielsen and Chuang, 2000, for discussion). Thus, 
derivations of Born's rule that employ reduced density matrices are open 
to charge of circularity (Zeh, 1997). Moreover, repeated attempts to justify 
$ p_k = |\psi_k|^2$ within the no - collapse MWI (Everett, 1957~a\&b; 
DeWitt, 1970; DeWitt and Graham, 1973; Geroch, 1984) have failed 
(see e.g., Stein, 1984; Kent, 1990; Squires, 1990): The problem is their 
circularity: An appeal to the connection (especially in certain limiting 
procedures) between the smallnes of the amplitude and the vanishing of 
probabilities has to be made to establish that relative frequencies of 
events averaged over branches of the universal state vector are consitent 
with Born's rule. In particular, one must a claim that  ``maverick'' 
branches of the MWI state vector that have ``wrong'' relative frequencies 
are of measure zero because their Hilbert space measures are small. 
This is circular, as noted even by the proponents (DeWitt, 1970). 

My aim here is to look at the origin of ignorance, information, and, therefore,
probabilities from a very quantum and fundamental perspective: Rather than
focus on probabilities for an individual isolated system I shall -- in
the spirit of einselection, but without employing its usual tools such as
trace or reduced density matrices -- consider what the observer can (and
cannot) know about a system entangled with its environment. Within this 
context I shall demonstrate that Born's rule follows from the very quantum 
fact that one can know precisely the state of the composite system and yet 
be provably ignorant of the state of its components. This is due to
{\it environment - assited invariance} or {\it envariance}, a hitherto 
unrecognised symmetry I am about to describe. Envariance of pure states 
is conspicuously missing from classical physics. It allows one to define 
ignorance as aconsequence of invariance, and thus to understand the origin 
of Born's rule, 
probabilities, and ultimately the origin of information through arguments 
based on assumptions different from Gleason's (1957) famous theorem. Rather, 
it is based on the Machian idea of `relativity of quantum states' entertained 
by this author two decades ago (see p. 772 of Wheeler and Zurek, 1983), but 
not developed untill now. Envariance (Zurek, 2002b) addresses the question 
of meaning of these probabilities by defining ``ignorance'', and justifies 
a relative frequency argument, although in a manner different from the previous 
attempts.

\subsubsection{Envariance}

Environment - assisted invariance is a symmetry exhibited by the system
${\cal S}$ correlated with the other system (usually `the environment' 
${\cal E}$). When a state of the composite ${\cal SE}$ can be transformed 
by $u_{\cal S}$ acting solely on the Hilbert space ${\cal H_S}$, 
but the effect of this transformation can be undone with an appropriate 
$u_{\cal E}$ acting only on ${\cal H_E}$, so that the joint state 
$|\psi_{\cal SE}\rangle $ remains unaltered;
$$ u_{\cal S} u_{\cal E}  |\psi_{\cal SE}\rangle  \ 
= \  |\psi_{\cal SE}\rangle  \ \eqno(6.6)$$
such $|\psi_{\cal SE}\rangle $ is envariant under $u_{\cal S}$. Generalization
to mixed $\rho_{\cal SE}$ is obvious, but we shall find it easier to assume
that ${\cal SE}$ has been purified in the usual fashion -- i.e., by enlarging 
the environment. 

Envariance is best elucidated by considering an example -- an entangled
state of ${\cal S}$ and ${\cal E}$. It can be expressed 
in the Schmidt basis as:
$$ | \psi_{\cal SE}\rangle  \ = \ 
\sum_k \alpha_k |s_k\rangle  |\varepsilon_k\rangle  \ , \eqno(6.7)$$
where $\alpha_k$ are complex, while $\{|s_k\rangle  \}$ and 
$\{ |\varepsilon_k\rangle  \}$ are orthonormal. For $|\psi_{\cal SE}\rangle $ 
(and, hence -- given our above remark about purification -- for any system 
correlated with the environment) it is easy to demonstrate:

\smallskip

\noindent {\bf Lemma 6.1}: Unitary transformations co-diagonal with 
the Schmidt basis of $|\psi_{\cal SE}\rangle $ leave it envariant. 

\smallskip

The {\it proof} relies on the form of such transformations:
$$ u_{\cal S}^{\{|s_k\rangle  \} } = \sum_k e^{i\sigma_k} |s_k \rangle  
\langle  s_k|\ ,  \eqno(6.8)$$
where $\sigma_k$ is a phase. Hence;
$$ u_{\cal S}^{\{|s_k\rangle  \} } |\psi_{\cal SE} \rangle  
\ = \ \sum_k \alpha_k e^{i\sigma_k} |s_k \rangle   |\varepsilon_k\rangle  \ ,  \eqno(6.9)$$
can be undone by:
$$ u_{\cal E}^{\{|\varepsilon_k\rangle  \} } = \sum_k e^{i\epsilon_k} \
|\varepsilon_k \rangle   \langle  \varepsilon_k|  \eqno(6.10)$$
providing that $\epsilon_k = 2 \pi l_k - \sigma_k $ for some integer $l_k$. QED.

Thus, phases associated with the Schmidt basis are envariant. We shall 
see below that they are the only envariant property of entangled states. 
The transformations defined by Eq. (6.8) are rather specific -- they share 
(Schmidt) eigenstates. Still, their existence leads us to:

\smallskip
\noindent{\bf Theorem 6.1}: Local description of the system ${\cal S}$ 
entangled with a causally disconnected environment ${\cal E}$ must not depend
on the phases of the coefficients $\alpha_k$ in the Schmidt decomposition
of $|\psi_{\cal SE}\rangle $. 
\smallskip

It follows that {\it all the measurable properties of the ${\cal S}$ are 
completely specified by the list of pairs} $\{ |\alpha_k|; |s_k\rangle  \} $.
An equivalent way of establishing this phase envariance theorem appeals even 
more directly to causality: Phases 
of $|\psi_{\cal SE}\rangle $ can be arbitrarily changed by acting on ${\cal E}$
alone (e.g, by the local Hamiltonian with eigenstates $|\varepsilon_k\rangle $, 
generating evolution of the from of Eq. (6.9)). But causality prevents 
faster than light communication. Hence, no measurable property of ${\cal S}$
can be effected. QED.

Phase envariance theorem will turn out to be the crux of our argument. It relies
on an input -- entanglement and causality -- which has not been employed to
date in discussions of the origin of probabilities. In particular,
this input is different and more ``physical" than that 
of the succesfull derivation of Born's rules by Gleason (1957). 

We also note that information contained in the ``data base'' 
$\{|\alpha_k|; |s_k\rangle \}$ implied by the Theorem 6.1 is the same as 
in the reduced density matrix of the system $\rho_{\cal S}$: Although we 
do not yet know probabilities of various $|s_k\rangle $, preferred basis 
of ${\cal S}$ has been singled out -- Schmidt states (sometimes regarded 
as instantaneous pointer states; see e.g. Albrecht, (1992;~1993)) play
a special role as the eigenstates of envariant transformations. 
Moreover, probabilities can depend on $|\alpha_k|$ (but not on the
phases). We still do not know that $p_k = |\alpha_k|^2$.

The causality argument we could have used to establis Theorem 6.1 applies of
course to arbitrary transformations one could perform on ${\cal E}$.
However, such transformations would not be in general envariant (i.e., 
could not be undone by acting on ${\cal S}$ alone). Indeed -- by the
same token -- all envariant transformations must be diagonal in Schmidt
basis:

\smallskip
\noindent {\bf Lemma 6.2}: All of the unitary envariant transformations of
$|\psi_{\cal SE}\rangle $ have Schmidt eigenstates.
\smallskip

The proof relies on the fact that other unitary transformations would 
rotate Schmidt basis, $|s_k\rangle  \rightarrow |\tilde s_k\rangle $. 
The rotated basis becomes a new `Schmidt', and this fact cannot be affected 
by unitary transformations of $\cal E$ -- by state rotations in the 
environment. But a state that has a different Schmidt decomposition from 
the original $|\psi_{\cal SE}\rangle $ is different. Hence, unitary 
transformation must be co-diagonal with Schmidt states to leave it envariant.
QED. 

\subsubsection{Born's rule from envariance}

When absolute values of some of the coefficients in Eq. (6.7) 
are equal, any orthonormal basis is `Schmidt' in the corresponding 
subspace of ${\cal H_S}$. This implies envariance of more general 
nature, e.g. under a {\it swap};
$$ u_{\cal S}(k \leftrightarrow j) = e^{i \phi_{kj}} |s_k\rangle  \langle  s_j|
+ h.c. \eqno (6.11)$$
Swap can be generated by a phase rotation, Eq. (6.8), but in a basis 
complementary to the one swapped. Its envariance does not contradict Lemma 6.2, 
as any orthonormal basis in this case is also `Schmidt'). So, when 
$|\alpha_k| = |\alpha_j|$, the effect of a swap on the system
can be undone by an obvious {\it counterswap} in the environment:
$$ u_{\cal S}(k \leftrightarrow j) = e^{- i ( \phi_{kj} + \phi_k - \phi_j
+ 2 \pi l_{kj})} |\varepsilon_k\rangle  \langle  \varepsilon_j| + h.c. 
\eqno (6.12)$$
Swap can be applied to states that do not have equal absolute values 
of the coefficients, but in that case it is no longer envariant. Partial 
swaps can be also generated e.g., by underrotating or by a 
$u_{\cal S}^{\{|r_i\rangle \}}$, Eq. (6.8), but with the eigenstates 
$\{ |r_i \rangle  \}$ intermediate between these of the swapped and the 
complementary (Hadamard) basis. Swap followed by a counterswap exchanges 
coefficients of the swapped states in the Schmidt expansion, Eq. (6.7). 

Classical correlated states can also exhibit something akin to
envariance under a classical version of ``swaps''. For instance, 
a correlated state of system and an apparatus described by:
$ \rho_{\cal SA} \sim |s_k\rangle  \langle  s_k||A_k \rangle  \langle  A_k| +
|s_j\rangle  \langle  s_j||A_j \rangle  \langle  A_j|$
can be swapped and counterswapped. The corresponding transformations would 
be still given by, in effect, Eqs. (6.11) - (6.12), but without phases, and
swaps could no longer be generated by the rotations around the complementary 
basis. This situation corresponds to the ``outsiders view'' of the measurement
process, Eq. (2.44c): Outsider can be aware of the correlation between 
the system and of the apparatus, but ignorant of their individual states. 
This connection between ignorance and envariance shall be exploited below.

Envariance based on ignorance may be found in the classical setting, but
envariance of pure states is purely quantum: Observers can know perfectly 
the quantum joint state of ${\cal SE}$, yet be provably ignorant of ${\cal S}$. 
Consider a measurement carried out on the state vector of ${\cal SE}$ from 
the point of view of envariance:
$$|A_0\rangle   \sum_{k=1}^N |s_k\rangle  
|\varepsilon_k\rangle  \rightarrow  \sum_{k=1}^N |A_k\rangle  |s_k\rangle  
|\varepsilon_k\rangle  \sim |\Phi_{{\cal SAE}}\rangle   \eqno(6.13)$$
Above, we have assumed that the absolute values of the coefficients
are equal (and omitted them for notational simplicity). We also have 
ignored phases (which need not be equal) since -- by the phase envariance 
theorem  -- they will not influence the state (and, hence, the probabilities) 
associated with ${\cal S}$.

Before the measurement observer with access to ${\cal S}$ cannot notice swaps 
in the state (such as Eq. (6.13)) with equal absolute values of the Schmidt 
coefficients.  This follows from envariance of the pre-measurement 
$|\psi_{\cal SE}\rangle  $ under swaps, Eq. (6.11). 

One could argue this point in more detail by comparing what happens for two 
very different input states; an entangled $|\psi_{\cal SE}\rangle $ with equal 
absolute values of Schmidt coefficients and a product state: 
$$|\varphi_{\cal SE}\rangle  = |s_J\rangle  |\varepsilon_J\rangle  \ . $$ 
When observer knows he is dealing with $\varphi_{\cal SE}$, he knows the state
and can predict outcome of the corresonding measurement on ${\cal S}$: 
Schr\"odinger equation or just the resulting truth table, Eq. (6.5), 
imply with certainty that his state -- the future state of his memory -- 
will be $|A_J\rangle $. Moreover, swaps involving $|s_J\rangle $ are 
{\it not} envariant for $\varphi_{\cal SE}$. They just swap the outcomes 
(i.e. when $u_{\cal S}(J\leftrightarrow L)$ precedes the measurement, 
memory will end up in $|A_L\rangle $).

By contrast, 
$$|\psi_{\cal SE} \rangle  \sim \sum_{k=1}^N e^{i\phi_k} |s_k\rangle  
|\varepsilon_k\rangle  $$
is envariant under swaps. This allows the observer (who knows the joint
state of ${\cal SE}$ exactly) to conclude that the probabilities of all
the envariantly swappable outcomes must be the same. {\it Observer cannot
predict his memory state after the measurement of ${\cal S}$ because he knows 
too much; the exact combined state of ${\cal SE}$}. 

For completeness, we note that when there are system states that are absent 
from the above sum -- i.e., states that `appear with zero amplitude' -- 
they cannot be envariantly swapped with the states present in the sum.
Of course, observer can predict with certainty he will not detect any of 
the corresponding zero - amplitude outcomes. 

This argument about 
{\it the ignorance of the observer concerning his future state -- concerning 
the outcome of the measurement he is about to perform -- is based on his 
perfect knowledge  of a joint state of ${\cal SE}$.}

Probabilities refer to the guess observer makes on the basis of his 
information {\it before the measurement} about the state of his memory -- 
``the future outcome'' -- after the measurement. As the left hand side 
of Eq. (6.13) is envariant under swaps of the system states,
probabilities of all the states must be equal. Thus, by normalisation;
$$ p_k = 1/N \ . \eqno(6.14)$$
Moreover, probability of $n$ mutually exclusive events that all appear 
in Eq. (6.13) with equal coefficients must be:
$$ p_{k_1 \vee  k_2 \vee ... \vee \ k_n} = n / N  \ . \eqno(6.15)$$
This concludes discussion of the equal probability case. Our case rests 
on the independence of the state of ${\cal S}$ entangled with ${\cal E}$ 
from the phases of the coefficients in the Schmidt representation -- the 
Theorem 6.1 -- which in the case of equal coefficients Eq. (6.13),
allows envariant swapping, and yields Eqs. (6.14)-(6.15).

After a measurement situation changes. In accord with our preceding 
discussion we interpret presence of the term $|A_k\rangle $ in Eq. (6.13) 
as evidence that an outcome $|s_k\rangle $ can be (or indeed has been -- 
the language here is somewhat dependent on the interpretation) recorded.
Conversely, absence of some $|A_{k'}\rangle $ in the sum above implies that 
the outcome $|s_{k'}\rangle $ cannot occur.  After a measurement memory 
of the observer who has detected $|s_k\rangle $ will contain the record 
$|A_k\rangle $. Futher measurements of the same observable on the same system
will confirm that ${\cal S}$ is in indeed in the state $|s_k\rangle $.

This post-measurement state is still envariant, but only under swaps that 
involve {\it jointly} the state of the system and the correlated state of 
the memory:
$$ u_{\cal AS}(k \leftrightarrow j) = e^{i \phi_{kj}} 
|s_k, A_k\rangle  \langle  s_j, A_j| + h.c. \eqno (6.16)$$
Thus, if another observer (`Wigner's friend') was getting ready to find out 
-- either by direct measurement of ${\cal S}$ or by communicating with 
observer ${\cal A}$ -- the outcome of his measurement, he would be (on the 
basis of envariance) provably ignorant of the outcome ${\cal A}$ has detected, 
but could be certain of the ${\cal AS}$ correlation. We shall employ this
joint envariance in the discussion of the case of unequal probabilities 
immediately below.

Note that our reasoning does not really appeal to the ``information lost 
in the environment'' in a sense in which this phrase is often used. Perfect 
knowledge of the combined state of the system and the environment is the basis 
of the argument for the ignorance od ${\cal S}$ alone: For entangled 
${\cal SE}$, perfect knowlegde of ${\cal SE}$ is incompatible with perfect 
knowledge of ${\cal S}$. This is really a consequence of indeterminacy -- 
joint observables with entangled eigenstates such as $\psi_{\cal SE}$ simply 
do not commute (as the reader is invited to verify) with the observables 
of the system alone. Hence, ignorance associated with envariance is 
ultimately mandated by Heisenberg indeterminacy.


The case of unequal coefficients can be reduced to the case of equal
coefficients. This can be done in several ways, of which we choose one 
that makes use of the preceding discussion of envariance of the 
post-measurement state. We start with:
$$|\Phi_{{\cal SAE}}\rangle  \sim \sum_{k=1}^N \alpha_k |A_k\rangle  |s_k\rangle  
|\varepsilon_k\rangle  \eqno(6.17)$$
where $\alpha_k \sim \sqrt m_k$ and $m_k$ is a natural number (and, by phase 
envariance theorem, we drop the phases). To get an envariant state we 
``increase the resolution'' of ${\cal A}$ by assuming that;
$$ |A_k\rangle  = \sum_{j_k=1}^{m_k} |a_{j_k}\rangle  / \sqrt m_k \eqno(6.18)$$
An increase of resolution is a standard trick, used in classical probability 
theory ``to even the odds''. Note that we assume that basis states such as 
$|A_k\rangle $ are normalised (as they must be in a Hilbert space). This leads 
to:
$$|\Phi_{{\cal SAE}}\rangle  \sim \sum_{k=1}^N \sqrt m_k
{ { \sum_{j_k=1}^{m_k} |a_{j_k}\rangle  } \over \sqrt m_k }
|s_k\rangle  |\varepsilon_k\rangle  \eqno(6.19)$$
We now assume that ${\cal A}$ and ${\cal E}$ interact (e.g., through a 
{\tt c-shift} of section 2, with a truth table $|a_{j_k}\rangle 
|\varepsilon_k\rangle  \rightarrow|a_{j_k}\rangle |e_{j_k}\rangle $ where 
$\{|e_{j_k}\rangle \}$ are all orthonormal), so that after simplifying 
and re-arranging terms we get a sum, over a new `fine-grained' index,
with the states of ${\cal S}$ that remain the same within `coarse - grained
cells', i.e., intervals measured by $m_k$: 
$$|\tilde \Phi_{{\cal SAE}}\rangle  \sim \sum_{k=1}^N |s_k\rangle  
( \sum_{j_k=1}^{m_k} |a_{j_k}\rangle  |e_{j_k}\rangle  )
= \sum_{j=1}^M |s_{k(j)}\rangle  |a_{j}\rangle  |e_{j}\rangle   
\eqno(6.20)$$
Above, $M=\sum_{k=1}^N m_k$, and $k(j)=1$ for $ j \leq m_1$, $k(j) = 2$ for
$j\leq m_1 + m_2$, etc. The above state is envariant under combined swaps;
$$u_{\cal SA}(j \leftrightarrow j') = \exp(i \phi_{jj'}) |s_{k(j)}, a_j \rangle  
\langle  a_{j'}, s_{k(j')}| + h.c. $$
Suppose that an additional observer measures ${\cal SA}$ in the obviously
swappable joint basis.  By our equal coefficients argument, Eq. (6.14), we get
$p(s_{k(j)}, a_j) = 1/M . $
But the observer can ignore states $a_j$. Then the probability of 
different Schmidt states of ${\cal S}$ is, by Eq. (6.15);
$$p(s_k) = m_k / M = |\alpha_k|^2 \ . \eqno(6.21)$$
This is Born's rule. 

The case with coefficients that do not lead to commensurate probabilities
can be treated by assuming continuity of probabilities as a function of 
the amplitudes and taking appropriate (and obvious) limits. This can 
be physically motivated: One would not expect probabilities to change
drastically depending on the infinitesimal changes of the state. One can also 
extend the strategy outlined above to deal with probabilities (and probability 
densities) in cases such as $|s(x)\rangle $, i.e. when the index of the 
state vector changes continuously. This can be accomplished by discretising 
it (so that the measurement of Eq. (6.17) correlates different apparatus
states with small intervals of $x$) and then repeating the strategy of
Eqs. (6.17) - (6.21). The wavefuction $s(x)$ should be sufficiently smooth 
for this strategy to succeed.
 
We note that the ``increase of resolution'' we have exploited, 
Eqs. (6.18)-(6.21), need not be physically implemented for the argument 
to go through: The very possibility of carrying out these steps within 
the quantum formalism forces one to adopt Born's rule. For example, 
if the apparatus did not have the requisite extra resolution, Eq. (6.18), 
interaction of the environment with a still different `counterweight' system
${\cal C}$ that yields
$$ |\Psi_{\cal SAEC}\rangle  = \sum_{k=1}^N \sqrt m_k |s_k\rangle  |A_k\rangle 
|\varepsilon_k\rangle  |C_k\rangle  \eqno(6.22)$$ 
would lead one to the Born's rule through steps similar to these we have 
invoked before, providing that $|C_k\rangle  \}$ has the requisite resolution, 
$ |C_k\rangle  = \sum_{j_k=1}^{m_k} |c_{j_k}\rangle /\sqrt m_k \ . $
Interaction resulting in a correlation, Eq. (6.22), can occur between ${\cal E}$
and ${\cal C}$, and happen far from the system of interest or from the 
apparatus. Thus, it will not influence probabilities of the outcomes of
measurements carried out on ${\cal S}$ or of the records made by ${\cal A}$.
Yet, the fact that it can happen leads us to the desired conclusion.

\subsubsection{Relative frequencies from envariance}

Relative frequency is a common theme studied with the aim of
elucidating the physical meaning of probabilities in quantum theory 
(Everett, 1957; Hartle, 1968; DeWitt, 1970; Graham, 1970; 
Farhi, Goldstone, and Gutmann, 1989; Aharonov and Resnik, 2002).
In particular, in the context of the ``no collapse'' MWI
relative frequency seem to offer the best hope of arriving at the 
Born's rule and elucidating its physical significance. Yet, 
it is generally acknowledged that the MWI derivations 
offered to date have failed to attain this goal (Kent, 1990). 

We postpone brief discussion of these efforts to the next section, 
and describe an approach to 
relative frequencies based on envariance. Consider an ensemble of many 
(${\cal N}$) distinguishable systems prepared in the same initial state:
$$|\sigma_{\cal S}\rangle  = \alpha |0\rangle  + \beta |1\rangle  \eqno(6.23)$$
We focus on the two state case to simplify the notation. We also assume that
$|\alpha|^2$ and $|\beta|^2$ are commensurate, so that the state vector of 
the whole ensemble of correlated triplets ${\cal SAE}$ after the requisite 
increases of resolution (see Eqs. (6.18)-(6.20) above) is given by: 
$$|\Phi^{\cal N}_{\cal SAE}\rangle  \sim  \bigl(\sum_{j=1}^m|0\rangle 
|a_j\rangle |e_j\rangle  + \sum_{j=m+1}^M|1\rangle |a_j\rangle 
|e_j\rangle \bigr)^{\otimes{\cal N}} \eqno(6.24)$$
save for the obvious normalisation.
This state is envariant under swaps of the joint states $|s,a_j\rangle $,
as they appear with the same (absolute value) of the amplitude in 
Eq. (6.24). (By Theorem 6.1 we can omitt phases.)

After the exponentiation is carried out, and the resulting product states are
sorted by the number of 0's and 1's in the records, we can calculate the number
of terms with exactly $n$ 0's, $ \nu_{\cal N} (n) = 
{ {\cal N} \choose n} m^n (M-m)^{{\cal N}-n} \ . $
To get probability, we normalise:
$$ p_{\cal N}(n) =  { {\cal N} \choose n} {{m^n (M-m)^{{\cal N}-n}} \over M^{\cal N}}  
=  { {\cal N} \choose n} |\alpha|^{2n}|\beta|^{2({\cal N}-n)} \ . \eqno(6.25)$$
This is the distribution one would expect from Born's rule. To establish the
connection with relative frequencies we appeal to the de Moivre - Laplace 
theorem (Gnedenko, 1982) which allows one to approximate above $p_{\cal N}(n)$ 
with a Gaussian:
$$ p_{\cal N}(n) \simeq { 1 \over { \sqrt{ 2 \pi {\cal N}} |\alpha \beta|}}
e^{- { 1 \over 2} \bigl( {{n - {\cal N}|\alpha|^2} \over 
\sqrt {{\cal N} |\alpha \beta|}} \bigr)^2} \eqno(6.26)$$
This last step requires large ${\cal N}$, but our previous discussion 
including Eq. (6.25) is valid for arbitrary ${\cal N}$. Indeed, equation 
(6.21) can be regarded as the ${\cal N} = 1$ case. 

Nevertheless, for large ${\cal N}$ relative frequency is sharply peaked 
around the expected $\langle  n \rangle  = {\cal N} |\alpha|^2$. 
Indeed, in the limit ${\cal N} \rightarrow \infty $ appropriately
rescaled $ p_{\cal N}(n)$ tends to a Dirac $\delta(\upsilon - |\alpha|^2)$ 
in relative frequency $\upsilon = n/{\cal N}$. This justifies the 
relative frequency interpretation of the squares of amplitudes 
as probabilities in the MWI context. `Maverick universes' with different 
relative frequencies exist, but have a vanishing {\it probability}
(and not just vanishing Hilbert space measure) for large ${\cal N}$. 

Our derivation of the physical significance of probabilities -- while
it led to the relative frequency argument -- was based on a very different
set of assumptions than previous derivations. The key idea behind it is
the connection between a symmetry (envariance) and ignorance
(impossibility of knowing something). The unusual feature of our
argument is that this ignorance (for an individual system ${\cal S}$) is
demonstrated by appealing to the perfect knowledge of the larger joint
system that includes ${\cal S}$ as a subsystem.

We emphasize that one could not carry out the basic step of our
argument -- the proof of the independence of the likelihoods from the
phases of the Schmidt expansion coefficients -- for an equal
amplitude pure state of a single, isolated system. The problem with:
$ |\psi\rangle  \ = \ N^{-{1 \over 2}} \sum_k^N \exp( i \phi_k) |k\rangle  $
is the accessibility of the phases. Consider, for instance;
$ |\psi\rangle  \sim  |0\rangle  + |1\rangle  - |2\rangle $ and
$ |\psi'\rangle  \sim |2\rangle  + |1\rangle  - |0\rangle $.
{\it In absence of decoherence} swapping of $k$'s is detectable:
Interference measurements (i.e., measurements of the observables
with phase-dependent eigenstates 
$|1\rangle +|2\rangle ; \ |1\rangle -|2\rangle ,$ etc.) would have
revealed the difference between $|\psi\rangle $ and $|\psi'\rangle $. 
Indeed, given an ensemble of identical {\it pure} states an observer
will simply find out what they are. Loss of phase coherence is essential 
to allow for the shuffling of the states and coefficients.

Note that in our derivation environment and einselection play an additional,
more subtle role: Once a measurement has taken place -- i.e., a correlation
with the apparatus or with the memory of the observer was established
-- one would hope that records will
retain validity over a long time, well beyond the decoherence timescale.
This is a pre-condition for the axiom (iv). Thus,
a ``collapse'' from a multitude of possibilities to a single reality
can be confirmed by subsequent
measurements only in the einselected pointer basis.

\subsubsection{Other approaches to probabilities}

Gnedenko (1982), in his classic textbook, lists three classical approaches 
to probability:
\begin{enumerate}
\item [a.] Definitions that appeal to {\it relative frequency}
of occurrence of events in a large number of trials.
\item [b.] Definitions of probability as a
{\it measure of certainty} of the observer.
\item [c.] Definitions that reduce probability to the more
primitive notion of {\it equal likelihood}.
\end{enumerate}

In the quantum setting, the relative frequency approach has been to date the
most popular, especially in the context of the ``no collapse'' MWI '
(Everett, 1957a\&b; Graham, 1970; DeWitt, 1970). Counting the number of
the ``clicks'' seems most directly tied to the experimental manifestations 
of probability. Yet, Everett interpretation versions were generally found 
lacking (Kent, 1990; Squires, 1990), as they relied on circular reasoning, 
invoking {\it without physical justification} an abstract measure of 
Hilbert space to obtain a physical measure (frequency). Some of the criticisms 
seem relevant also for the versions of this approach that allow for 
the measurement postulates (iii) and (iv) (Hartle, 1968; Farhi, Goldstone 
and Guttmann, 1989). Nevertheless, for the infinite 
ensembles considered in the above references, (where, in effect, the Hilbert 
space measure of the MWI branches that violate relative frequency predictions 
is zero) that the eigenvalues of the {\it frequency operator} acting on 
a large or infinite ensemble of identical states will be consistent with 
the (Born formula) prescription for probabilities. 

However, the infinite size of the ensemble necessary to prove this point is
troubling (and unphysical) and taking the limit starting from a finite case
is difficult to justify (Stein, 1984; Kent, 1990; Squires, 1990). Moreover, 
the frequency operator is a collective observable of the whole ensemble. 
It may be possible to relate observables defined for such an infinite 
ensemble supersystem to the states of individual subsystems, but 
the frequency operator does not do it. This is well illustrated by 
the gedankenexperiment envisaged by Farhi et al. (1989). To provide 
a physical implementation of the frequency operator they consider a version 
of the Stern-Gerlach experiment where all the spins are attached to 
a common lattice, so that -- during the passage through the inhomogeneity 
of the magnetic field -- the center of mass of the whole lattice
is deflected by an angle proportional to the projection of the net
magnetic moment associated with the spins on the direction defined by the
field gradient. The deflection is proportional to the eigenvalue of
the frequency operator that is then a {\it collective} observable --
states of individual spins remain in superpositions, uncorrelated with
anything outside. This difficulty can be addressed with the help of decoherence
(Zurek, 1998a), but using decoherence without justifying Born's formula
first is fraught with danger of circularity.

{\it Measure of certainty} seems to be a rather vague concept. Yet,
Cox (1946) has demonstrated that Boolean logic leads -- after addition of 
a few reasonable assumptions -- to the definition of probabilities that, 
in a sense, appear as an extension of the logical truth values. However, 
the rules of symbolic logic that underlie Cox's theorems are {\it classical}. 
One can adopt this approach (Zurek, 1998a) 
to probabilities in quantum physics only after decoherence ``intervenes'' 
restoring the validity of the distributive law, which is not valid in
quantum physics (Birkhoff and von Neumann, 1936).

One can carry out equal likelihood approach in the context of decoherence
(Zurek, 1998a). The problem is -- as pointed out before -- the use of
trace, and the dangers of circularity. An attempt to pursue a strategy akin 
to equal likelihood in the quantum setting at the level of pure states of 
individual systems has been also made by Deutsch in his (unpublished) 
``signalling'' approach to probabilities. The key idea is to consider 
a source of pure states, and to find out when the permutations of a set 
of basis states can be detected, and, therefore, used for communication. 
When permutations are undetectable, probabilities of the permuted set of 
states are declared equal. The problem with this idea (or with its more formal 
version described by DeWitt, 1998) is that it works only for superposition 
that have all the coefficients identical, {\it including their phases}. 
Thus, as we have already noted, for closed systems phases matter and there is 
no invariance under swapping. In a recent paper Deutsch (1999) has adopted 
a different approach based on decision theory. The basic argument focuses 
again on individual states of quantum systems, but -- as noted in the critical 
comment by Barnum et al. (2000) -- seems to make appeal to some of the aspects 
of decision theory that do depend on probabilities. In my view, it also
leaves the problem of the phase dependence of the coefficients unaddressed.

Among other approaches, recent work of Gottfried (2000) shows that
in a discrete quantum system coupled with a continuous quantum system
Born's formula follows from the demand that the continuous system
should follow classical mechanics in the appropriate limit. A somewhat
different strategy, with a focus on the coincidences of the expected 
magnitude of fluctuations was proposed by Aharonov and Resnik (2002).

In comparison with all of the above strategies, `probabilities from envariance'
is the most radically quantum, in that it ultimately relies on entanglement
(which is still sometimes regarded as `a paradox', and `to be explained': 
I have used it as an explanation). This may be the reason why it has not 
been discovered untill now. The insight offered by envariance into the nature 
ignorance and information sheds a new light on probabilities in physics. 
The (very quantum) ability to prove ignorance of a part of a system 
by appealing to perfect knowledge of the whole may resolve some of 
the difficulties of the classical approaches. 

\section{ENVIRONMENT AS A WITNESS}

Emergence of classicality can be viewed either as a consequence of
the widespread dissemination of the information about the pointer states 
through the environment, or as a result of the censorship imposed by 
decoherence.  So far I have focused on this second view, defining 
{\it existence} as persistence -- predictability in spite of the environmental
monitoring. Predictability sieve is a way of discovering states that
are classical in this sense (Zurek, 1993a\&b; Zurek, Habib and Paz, 1993;
Gallis, 1996).

A complementary approach focusses {\it not} on the system, but on
the records of its state spread throughout the environment.  Instead of
``the least perturbed states'' one can ask ``what states of the system are
easiest to discover by looking at the environment''. Thus, environment is 
no longer just a source of decoherence, but acquires a role of a communication
channel with a basis - dependent noise that is minimised by the preferred
pointer states.

This approach can be motivated by the old dilemma: On the one hand,
quantum states of isolated systems are purely ``epistemic'' (see e.g., Peres,
(1993); Fuchs and Peres (2000)). Quantum cryptography (Bennett and DiVincenzo,
2000; Nielsen and Chuang, 2000, and references therein) uses this 
impossibility of finding out what is an unknown state of an isolated 
quantum system. On the other hand classical reality seems to be made up 
of quantum building blocks: States of macroscopic systems exist objectively
--  they can be found out by many observers independently
without being destroyed or re-prepared.  So -- the question arises --
how can objective existence --- the ``reality'' of the classical
states -- emerge from ``purely epistemic'' wavefunctions?

There is not much one can do about this in case of a single state 
of an isolated quantum system. But open systems are subject to einselection and 
can bridge the chasm dividing their epistemic and ontic roles. The most direct 
way to see this arises from the recognition of the fact that we never 
{\it directly} observe any system. Rather, we discover states of macroscopic 
systems from the imprints they make on the environment: A small
fraction of the photon environment intercepted by our eyes is often all
that is needed. States that are recorded most redundantly in the rest of 
the Universe (Zurek, 1983; 1998a; 2000) are also easiest to discover. 
They can be found out indirectly, from multiple 
copies of the evidence imprinted in the environment, without a threat to 
their existence. Such states exist and are real -- they can be found 
out without being destroyed as if they were really classical.

Environmental monitoring creates an ensemble of ``witness states'' in
the subsystems of the environment, that allow one to invoke
some of the methods of the statistical interpretation (Ballentine, 1970)
while subverting its ideology  -- to work with an {\it ensemble} of objective
evidence of a state of a {\it single} system. From this ensemble of witness
states one can infer the state of the quantum system that has led to such
``advertising''. This can be done without disrupting the einselected states.

Predictability sieve selects states that entangle least with the environment. 
Question about predictability simultaneously lead to states that are most 
redundantly recorded in the environment. Indeed, this idea touches on the
``quantum Darwinism'' we have alluded to in the introduction: The einselected 
pointer states are not only best in surviving the environment, but, also, 
they broadcast the information about themselves -- spread out their ``copies''
-- throughout the rest of the Universe: Amplified information is easiest to
amplify. This leads to analogies with ``fitness'' in the Darwinian 
sense, and suggests looking at einselection as a sort of natural selection.

\subsection{Quantum Darwinism}

Consider the ``bit by byte'' example of Section IV.  Spin - system ${\cal S}$
is correlated with the environment:
\begin{eqnarray}
|\psi_{{\cal SE}}\rangle  & = & a| \uparrow \rangle  |00 \ldots 0\rangle  + 
b|\downarrow\rangle |11 \ldots 1\rangle 
\nonumber \\
& = & a| \uparrow \rangle  |{\cal E}_{\uparrow}\rangle  + b|\downarrow\rangle |{\cal 
E}_{\downarrow}\rangle 
\eqnum{7.1}
\end{eqnarray}
The basis $\{|\uparrow \rangle  , |\downarrow \rangle  \}$ of ${\cal S}$
is singled out by the redundancy of the record. By comparison, the same 
$|\psi_{\cal SE}\rangle $ is:
\begin{eqnarray}
|\psi_{{\cal SE}} \rangle  & = & |\odot \rangle  ( a| 00 \ldots 0 \rangle  \ + \ b |11 \ldots 1\rangle  )
\sqrt{2}
\nonumber \\
&  + & |\otimes \rangle  (a | 00 \ldots 0 \rangle  -  b|11 \ldots 1 \rangle  ) / \sqrt{2} \nonumber
\\
&  = &
(|\odot\rangle |{\cal E}_\odot\rangle +|\otimes \rangle |{\cal E}_\otimes\rangle )/\sqrt2
\eqnum{7.2}
\end{eqnarray}
in terms of the Hadamard-transformed $\{|\odot\rangle , |\otimes \rangle  \}$.

One can find out whether ${\cal S}$ is $|\uparrow \rangle $ or
$|\downarrow \rangle $ from a small subset of the environment bits.
By contrast, states $\{|\odot\rangle $, $|\otimes\rangle \}$ cannot 
be easily inferred from the environment. States 
$\{|{\cal E}_\odot\rangle ,|{\cal E}_\otimes \rangle \}$
are typically not even orthogonal,
$\langle {\cal E}_\odot | {\cal E}_\otimes \rangle  = |a|^2 - |b|^2.$
And even when $|a|^2 - |b|^2 = 0$, the record in the environment
is fragile: Only one relative phase distinguishes $|{\cal E}_\odot \rangle $
from $|{\cal E}_\otimes \rangle $ in that case, in contrast with multiple 
records of the pointer states in $|{\cal E}_\uparrow \rangle $ and 
$|{\cal E}_\downarrow \rangle $. Remarks that elaborate this 
observation follow. They correspond to several distinct 
measures of `fitness' of states.

\subsubsection{Consensus and algorithmic simplicity}

>From the state vector $|\psi_{{\cal SE}} \rangle $, Eqs. (7.1) and
(7.2), observer can find the state of the quantum system just by looking 
at the environment.  To accomplish this, the total $N$ of the environment 
bits can be divided into samples of $n$ bits each, with $1 \ll n \ll N$. 
These samples can be then measured using observables that are the same within 
each sample, but differ between samples. They may correspond, for example, 
to different antipodal points in the Bloch spheres of the environment bits.  
In the basis $\{|0\rangle , |1\rangle \}$
(or bases closely aligned with it) the record inferred from the bits of
information scattered in the environment will be easiest to come by.
Thus, starting from the environment part of $|\psi_{{\cal SE}}\rangle $,
Eq. (7.1), {\it the observer can find out, with no prior knowledge, the state
of the system}:  Redundancy of the record in the environment allows for
a trial-and-error `indirect' approach while leaving the system untouched.

In particular, measurement of $n$ environment bits in a Hadamard transform of 
the basis $\{|0\rangle  , |1\rangle  \}$ yields a random-looking sequence of 
outcomes (i.e., $\{| + \rangle  _1, | - \rangle _2, \ldots | - \rangle _n\}$).
This record is algorithmically random: Its algorithmic complexity
is of the order of its length (Li and Vit\`anyi, 1994):
$$
K(\langle {\cal E}_n |+,- \rangle ) \simeq n \eqno(7.3)
$$
By contrast, algorithmic complexity of the measurement outcomes in the
$\{|0\rangle , |1\rangle \}$ basis will be small:
$$
K(\langle {\cal E}_n|0,1 \rangle) \ll n \ , \eqno(7.4)
$$
since the outcomes will be either $00 \ldots 0$ or $11 \ldots 1$.
Observer seeking preferred states of the system by looking
at the environment should then search for the minimal record size and,
thus, for the maximum redundancy in the environmental record. States of
the system that are recorded redundantly in the environment must have
survived repeated instances of the environment monitoring, and are
obviously robust and predictable.

Predictability we have utilized before to devise a ``sieve'' to
select preferred states is used here again, but in a different
guise: Rather than search for predictable sets of states of
the system we are now looking for the records of the states of
the system in the environment. Sequences of states of environment
subsystems correlated with pointer states are mutually predictable
and, hence, collectively algorithmically simple. States that are
predictable in spite of the interactions with the environment
are also easiest to predict from their impact on  its state.

The state of the form of Eq. (7.1) can serve as an example of
amplification. Generation of redundancy through amplification brings about
objective existence of the otherwise subjective quantum states.
States $|\uparrow \rangle $ and $|\downarrow \rangle $ of the system 
can be found out reliably from a small fraction of the environment.
By contrast, to find out whether the system was in a state
$|\odot \rangle $ or $|\otimes \rangle $ one would need to detect {\it all}
of the environment.  Objectivity can be defined as
the ability of many observers to reach consensus independently.
Such consensus concerning states $|\uparrow\rangle $ and
$|\downarrow\rangle $ is easily established -- many ($\sim N/n$)
can independently measure fragments of the environment.

\subsubsection{Action distance}

One measure of robustness of the environmental records is the action distance 
(Zurek, 1998a). It is given by the total {\it action} necessary to undo
the distinction between the states of the environment corresponding
to different states of the system, subject to the constraints
arising from the fact that the environment consists of subsystems.
Thus, to obliterate the difference between  $|{\cal E}_\uparrow \rangle $
and $|{\cal E}_\downarrow \rangle $ in Eq. (7.1), one needs to ``flip''
one-by-one $N$ subsystems of the environment. That implies an action
-- i.e., the least total angle by which a states must be rotated, see 
Section IIB -- of:
$$
\Delta (|{\cal E}_\uparrow \rangle  , |{\cal E}_\downarrow \rangle  ) = N \left[
\frac{\pi}{2} \cdot \hbar
\right ] \ . \eqno (7.5)
$$
By contrast a ``flip'' of phase of just one bit will reverse
the correspondence between the states of the system and of
the environment superpositions that make up $|{\cal E}_\odot \rangle  $
and $|{\cal E}_\otimes \rangle  $ in Eq (7.2). Hence:
$$
\Delta (|{\cal E}_\odot \rangle  , |{\cal E}_\otimes\rangle  ) = 1 \left[ \frac{\pi}{2}
\cdot \hbar
\right ] \ . \eqno (7.6)
$$

Given a fixed division of the environment into subsystems the action distance
is a metric on the Hilbert space (Zurek, 1998a). That is;
$$ \Delta(|\psi \rangle  , | \psi \rangle ) = 0 \ , \eqno(7.7)$$
\vspace{-.25 in}
$$ \Delta(|\psi \rangle ,| \varphi \rangle ) = \Delta(|\varphi \rangle ,| \psi \rangle )\geq 0, 
\eqno(7.8)$$
and the triangle inequality:
$$
\Delta(|\psi \rangle  , | \varphi \rangle ) + \Delta(|\varphi \rangle , |\gamma \rangle  ) \geq \Delta
(|\psi
\rangle , |\gamma \rangle ) \eqno (7.9)
$$
are all satisfied.

In defining $\Delta$ it is essential to restrict rotations to the subspaces 
of the subsystems of the whole Hilbert space, and to insist that the unitary 
operations used in defining distance act on these subspaces. It is possible
to relax constraints on such unitary operations by allowing, for example, 
pairwise or even more complex interactions between subsystems. Clearly, 
in absence of any restrictions the action required to rotate any 
$|\psi \rangle $ into any $|\varphi \rangle $ would be no more than
$\frac{\pi}{2}  \hbar$.  Thus, constraints imposed by the natural
division of the Hilbert space of the environment into subsystems play
an essential role.  Preferred states of the system can be sought by
extremizing action distance between the corresponding
record states of the environment. In simple cases (e.g., see
``bit-by-byte'', Eq. (4.7) and below) the action distance
criterion for preferred states coincides with the predictability sieve
definition (Zurek, 1998a).

\subsubsection{Redundancy and mutual information}

The most direct measure of reliability of the environment as a witness
is the information-theoretic redundancy of einselection itself. When
environment monitors the system (see Fig. 4), the information 
about its state will spread to more and more subsystems of the environment.
This can be represented by the state vector $|\psi_{\cal {SE}} \rangle $, 
Eq. (7.1), with increasingly long sequences of 0's and 1's in the record states.
The record size -- the number $N$ of the subsystems of the environment 
involved -- does not affect the density matrix of the system ${\cal S}$. 
Yet, it obviously changes accessibility abd robustness of the information
analogues of the Darwinian ``fitness''. As an illustration, let us consider
{\tt c-shift}'s. One subsystem of the environment (say, ${\cal E}_1$)
with the dimension of the Hilbert space no less than that of the system
$$
Dim {\cal H_{E}}_1 \geq Dim {\cal H_S}
$$
suffices to eradicate off-diagonal elements of $\rho_{\cal S}$ in the
control basis. On the other hand, when $N$ subsystems of the environment
correlate with the same set of states of ${\cal S}$, the information
about these states is simultaneously accessible more widely.  While
$\rho_{\cal S}$ is no longer changing, spreading of the information makes
the existence of the pointer states of ${\cal S}$ more objective -- they are
easier to discover without being perturbed.

Information theoretic {\it redundancy} is defined as the difference
between the least number of bits needed to uniquely specify the message
and the actual size of the encoded message. Extra bits allow for detection
and correction of errors (Cover and Thomas, 1991). In our case, the message
is the state of the system, and the channel is the environment.
The information about the system will often spread over all of the Hilbert
space ${\cal H_E}$ that is enormous compared to ${\cal H_S}$. Redundancy
of the record of the pointer observables of selected systems can be also
huge.
Moreover, typical environments consist of obvious subsystems (i.e., photons,
atoms, etc.). It is then useful to define redundancy of the record by
the number of times the information about the system has been copied, or
by how many times it can be independently extracted from the environment.

In the simple example of Eq. (7.1) such {\it redundancy ratio}
${\cal R}$ for the $\{|\uparrow \rangle  , |\downarrow \rangle  \}$ basis will be given
by $N$, the number of environment bits perfectly correlated with
the obviously preferred basis of the system. More generally, but in
the same case of perfect correlation we obtain:
$$
{\cal R} = \frac{\lg (Dim {\cal H_E})}{\lg (Dim {\cal H_S})} =
\log_{Dim {\cal H_S}} (Dim {\cal H_E}) = N \eqno (7.10)
$$
where ${\cal H_E}$ is the Hilbert space of the environment perfectly
correlated with the pointer states of the system.

On the other hand, with respect to the $\{|\odot \rangle,|\otimes\rangle\}$
basis, the redundancy ratio for $|\psi_{\cal SE} \rangle $ of Eq. (7.2) is only
$\sim$1 (see also Zurek, 1983; 2000): Redundancy measures the number of errors 
that can obliterate the difference between two records, and in this basis 
one phase flip is clearly enough. This basis dependence of 
redundancy suggests an alternative strategy to seek preferred states.

To define ${\cal R}$ in general we can start with mutual information
between the subsystems of the environment ${\cal E}_k$ and the
system ${\cal S}$. As we have already seen in section IV,
definition of mutual information in quantum mechanics
is not straightforward. The basis-independent formula:
$$
{\cal I}_k = {\cal I(S:E}_k) = H({\cal S}) + H({\cal E}_k) - H({\cal S, E}_k)
\eqno (7.11)
$$
is simple to evaluate (although it does have some strange features; see
Eqs. (4.30) - (4.36)). In the present context it involves
joint density matrix:
$$
\rho_{{\cal SE}_k} = Tr_{{\cal E/E}_k} \rho_{{\cal SE}} \eqno
(7.12)
$$
where the trace is carried out over all of the environment except for
its singled out fragment ${\cal E}_k$.  In the example of Eq. (7.1) for
any of the environment bits
$$
\rho_{{\cal S E}_k} = |a|^2 |\uparrow \rangle \langle  \uparrow | \ | 0 \rangle \langle  0 | +
|b|^2 |\downarrow \rangle \langle  \downarrow | \ |1 \rangle \langle  1 | \ \ .
$$
Given the partitioning of the environment into subsystems, the redundancy
ratio can be defined as:
$$
{\cal R}_{{\cal I}{\left (\{\otimes {\cal H}_{{\cal E}_k}\} \right)}}
= \sum_{k} \ {\cal I(S:E}_k) / H({\cal S}) \ \ . \eqno(7.13)
$$
When ${\cal R}$ is maximized over all of the possible partitions,
$$
{\cal R}_{{\cal I}max} =
R_{\{\otimes {\cal
H_E}_k\}}
\eqno(7.14)
$$
obtains. Roughly speaking, and in the case when the number of the environment 
subsystems is large, ${\cal R_I}_{max}$ is the total number of copies of 
the information about (the optimal basis of) ${\cal S}$ that exist in 
${\cal E}$. Maximal redundancy ratio ${\cal R_I}_{max}$ is of course 
basis - independent.

The information defined through the symmetric ${\cal I}_k$, Eq. (7.11),
is in general inaccessible to observers who interrogate environment
one subsystem at a time (Zurek, 2002a). It makes therefore a lot of sense
to consider the basis-dependent locally accessible information
and define the corresponding redundancy ratio ${\cal R_J}$ using:
$$
{\cal J}_k = {\cal J}({\cal S : E}_k) =
H({\cal S}) + H({\cal E}_k) -(H({\cal S}) + H({\cal E}_k|{\cal S})) . 
\eqno (7.15)
$$
Conditional entropy must be computed in a specific basis of
the system (see Eq. (4.32)). All of the other steps that have led
to the definition of ${\cal R}_{{\cal I}max}$ can be now repeated
using ${\cal J}_k$. In the end, a basis - dependent:
$$
{\cal R}_{\cal J}(\{|s\rangle \}) = {\cal R}_{\cal J} (\otimes {\cal H_E}_k)
\eqno (7.16)
$$
obtains. ${\cal R_J}(\{|s\rangle \})$ quantifies the mutual information between
the collection of subsystems ${\cal H_E}_k$ of the environment and the basis
$\{|s\rangle \}$ of the system. We note that the condition of non-overlapping
partitions guarantees that all of the corresponding measurements commute,
and that the information can be indeed extracted independently from each
environment fragment ${\cal E}_k$.

Preferred basis of ${\cal S}$ can be now defined by maximizing ${\cal
R_J}(\{|s\rangle \})$ with respect to the selection of $\{|s\rangle \}$:
$$
{\cal R}_{{\cal J}max} = \max_{\{|s\rangle \};\{\otimes {\cal H_E}_k\}}
{\cal R_J} (\{|s\rangle \}) \eqno (7.17)
$$
This maximum can be sought either by varying the basis of the system only
(as it is indicated above) or by varying both the basis and the partition
of the environment.

It remains to be seen whether and under what circumstances pointer
basis ``stands out'' through its definition in terms of ${\cal R_J}$.
The criterion for a well - defined set of pointer states $\{|p\rangle \}$ would be:
$$
{\cal R}_{{\cal J}max} = {\cal R_J} (\{|p\rangle \}) \gg {\cal R_J}(\{|s\rangle \})
\eqno (7.18)
$$
where $\{|s\rangle \}$ are typical superpositions of states belonging
to different pointer eigenstates.

This definition of preferred states directly employs the notion
of multiplicity of records available in the environment.
Since ${\cal J} \leq {\cal I}$, it follows that:
$$
{\cal R}_{{\cal J}max} \leq {\cal R}_{{\cal I}max} \eqno (7.19)
$$
The important feature of either version of ${\cal R}$ that makes them useful 
for our purpose is their independence on $H({\cal S})$: The dependence on
$H({\cal S})$ is in effect ``normalized out'' of ${\cal R}$. ${\cal R}$ 
characterizes the ``fan-out'' of the information about the preferred basis 
throughout environment, without a reference to what is known about the system.
The usual redundancy (in bits) is then $\sim {\cal R \cdot H(S)}$,
although other implementations of this program (Ollivier, Poulin, and
Zurek, 2002) employ different measures of redundancy, which may be even
more specific than the redundancy ratio we have described above. Indeed, 
what is important here is the genreal idea of measuring classicality of 
quantum states through the number of copies they imprint 
throughout the Universe. This is a very Darwinian approach: We define
classicality related to einselection in ways reminiscent of `fitness' 
in natural selection: States that spawn most of the (information-theoretic) 
progeny are the most classical.

\subsubsection{Redundancy ratio rate}

The rate of change of redundancy is of interest as another 
measure of `fitness', perhaps closest to the definitions of fitness used in 
modeling natural selection. Redundancy can increase either as a result of 
interactions between the system and the environment, or because the environment 
already correlated with ${\cal S}$ is passing on the information to more 
distant environments. In this second case `genetic information' is passed 
on by the `progeny' of the original state. Even an observer
consulting the environment becomes a part of such
a more distant environment. Redundancy rate is defined as:
$$
\dot{{\cal R}} = \frac{d}{dt} {\cal R} \eqno(7.20)
$$
Either basis-dependent or basis-independent versions of
$\dot {\cal R}$ may be of interest.

In general, it may not be easy to compute either ${\cal R}$ or
$\dot{{\cal R}}$ exactly. This is nevertheless possible in models
(such as those leading to Eqs. (7.1) - (7.2)). The simplest illustrative
example corresponds to the {\tt c-not} model of decoherence in Fig. 4.
One can imagine that the consecutive record bits get correlated with
the two branches (corresponding to $|0\rangle $ and $|1\rangle $ in 
the ``control'') at discrete moments of time. ${\cal R}(t)$ would be then 
the total number of {\tt c-not}s that have acted over the time $t$, and
$\dot{{\cal R}}$ is the number of new {\tt c-not}s added per unit time.

Redundancy rate measures information flow from the system to
the environment. Note that, after the first {\tt c-not} in the example of
Eqs. (7.1) - (7.2), ${\cal R_I}$ will jump immediately from 0 to 2 bits,
while the basis-specific ${\cal R_J}$ will increase from 0 to 1.
In our model this initial discrepancy (which reflects quantum discord, 
Eq. (4.36), between ${\cal I}$ and ${\cal J}$) will disappear after
the second {\tt c-not}.


Finally, we note that ${\cal R}$ and, especially, $\dot{{\cal R}}$
can be used to introduce new predictability criteria:
The states (or the observables)
that are being recorded most redundantly are the obvious candidates
for the ``objective'', and therefore for the ``classical''.

\subsection{Observers and the Existential Interpretation}

Von Neumann (1932), London and Bauer (1939) and Wigner (1963) have all
appealed to the special role of the {\it conscious} observer. 
Consciousness was absolved from following unitary evolution, and, thus,
could collapse the wavepacket.
Quantum formalism has led us to a different view, that nevertheless
allows for a compatible conclusion. In essence, macroscopic systems
are open, and their evolution is almost never unitary. Records
maintained by the observers are subject to einselection. In a
binary alphabet decoherence will allow only for the two logical
states, and prohibit their superpositions (Zurek, 1991). For human
observers neurons conform to this binary convention and the decoherence
times are short (Tegmark, 2000). Thus, even if a cell of the observer
entangles through a premeasurement with a pure quantum state, the record will
become effectively classical almost instantly: As a result, it will be
impossible to ``read it off'' in any basis except for the einselected one. 
This censorship of records is the key
difference between the existential interpretation and the original Everett's
MWI. 

Decoherence treats observer as any other macroscopic
system. There is, however, one feature distinguishing observers from the
rest of the Universe: They are {\it aware} of the content of their memory.
Here we are using {\it aware} in a down - to - earth sense: Quite
simply, observers know what they know. Their information processing machinery
(that must underlie higher functions of the mind such as ``consciousness")
can readily consult the content of their memory.

The information stored in the memory comes with `strings attached'.
The physical state of the observer is described in part by the data
in his records. {\it There is no information without representation.}
The information observer has could be, in principle, deduced from his
physical state. Observer is -- in part -- information. Moreover,
this information encoded in states of macroscopic quantum systems
(neurons) is by no means secret: As a result of lack of isolation
the environment -- having redundant copies of the relevant data --
`knows' in detail everything observer knows. Configurations of neurons
in our brains, while at present undecipherable, are in principle
as objective and as widely accessible as the information about the states
of other macroscopic objects.

The observer {\it is} what he knows. In the unlikely case of a flagrantly
quantum input the physical state of the observers memory will decohere, 
resulting almost instantly in the einselected alternatives, each of them 
representing simultaneously both observer and his memory. The `advertising' 
of this state throughout the environment makes it effectively objective.

An observer perceiving the Universe from within is in a very different
position than an experimental physicist studying a state vector of
a quantum system. In a laboratory, Hilbert space of the investigated system
is typically tiny. Such systems can be isolated, so that often the
information loss to the environment can be prevented. Then the evolution
is unitary. The experimentalist can know everything there is to know about it.
Common criticisms of the approch advocated in this paper are based on an 
unjustified extrapolation of the above laboratory situation to the case 
of the observer who is a part of the Universe. Critics of decoherence often
note that the differences between the laboratory example above
and the case of the rest of the Universe are `merely quantitative'
-- the system under investigation is bigger, etc.
So why cannot one analyze -- they ask -- interactions of the observer
and the rest of the Universe as before, for a small
isolated quantum system?

In the context of the existential interpretation the analogy with the 
laboratory is, in effect, turned ``upside down": For, now the
observer (or the apparatus, or anything effectively classical)
is continuously monitored by the rest of the Universe.
Its state is repeatedly collapsed -- forced into the einselected states --
and very well (very redundantly) `known' to the rest of the Universe.

The `higher functions' of observers -- e.g., consciousness, etc. --
may be at present poorly understood, but it is safe to assume that 
they reflect physical processes in the information processing 
hardware of the brain. Hence, mental processes are in effect
objective, as they leave an indelible imprint on the environment:
The observer has no chance of perceiving either his memory, or any
other macroscopic part of the Universe in some arbitrary superposition.
Moreover, memory capacity of observers is miniscule compared to
information content of the Universe. So, while observers may know
exact state of laboratory systems, their records of the Universe will be
very fragmentary. By contrast, the Universe has enough memory capacity
to acquire and maintain detailed records of states of macroscopic systems 
and their histories.

\subsection{Events, Records, and Histories}

Suppose that instead of a monotonous record sequence
in the environment basis corresponding to the pointer states of
the system $\{|\uparrow\rangle , |\downarrow\rangle \}$ implied by Eq. (7.1)
the observer looking at the environment detects:
$$ 000...0111...1000...0111... $$
Given appropriate additional assumptions, such sequences consisting of
long stretches of record 0's and 1's justify inference of the history
of the system. Let us further assume that observer's records come from 
intercepting a small fragment of the environment. Other observers will be 
then able to consult their independently accessible environmental records, 
and will infer (more or less) the same history. Thus, in view of the 
``preponderance of evidence'' history defined as a sensible inference 
from the available records can be probed by many independently, and 
can be regarded as classical and objective.

The redundancy ratio of the records ${\cal R}$ is a measure of
this objectivity. Note that this {\it relatively objective existence}
(Zurek, 1998a) is an operational notion, quantified by the number
of times the state of the system can be found out independently,
and not some ``absolute objectivity''. However, and in a sense that
can be rigorously defined, relative objectivity tends to absolute
objectivity in the limit ${\cal R} \longrightarrow \infty$. For example,
cloning of unknown states becomes possible (Bruss, Ekert, and Macchiavello,
1999, Jozsa, 2002) in spite of the ``no cloning'' theorem  
(Wootters and Zurek, 1982, Dieks 1982). In that limit, and given same reasonable
constraints on the nature of the interactions and on the structure
of the environment which underlie the definition of ${\cal R}$,
it would take infinite resources such as action, Eqs. (7.5) - (7.9),
to hide or subvert evidence of such an objective history. 

There are differences and parallels between relatively objective
histories introduced here and consistent histories proposed
by Griffiths (1984, 1996), and investigated by Gell-Mann and Hartle
(1990; 1993; 1997), Omn\`es (1988; 1992; 1994), Halliwell (1999), and
others (Dowker and Kent, 1996; Kiefer, 1996).  Such histories are defined 
as time-ordered sequences of projection operators $P^1_{\alpha_1}(t_1), 
P^2_{\alpha_2}(t_2), \dots,  P^n_{\alpha_n}(t_n)$
and are abbreviated $[P_{\alpha}]$. Consistency is achieved when they
can be combined into coarse grained sets (where the projectors defining
coarse - grained set are given by sums of the projectors in the original set)
while obeying probability sum rules: Probability of a bundle of histories
should be a sum of the probabilities of the constituent histories.
The corresponding condition can be expressed in terms of the
{\it decoherence functional} (Gell-Mann and Hartle, 1990);
$$ D([P_{\alpha}],[P_{\beta}]) = Tr \left ( (P^n_{\alpha_n}(t_n) \dots
P^1_{\alpha_1}(t_1) \rho P^1_{\beta_1}(t_1) \dots P^n_{\beta_n}(t_n)
\right ) \eqno(7.21)$$
Above, the state of the system of interest is described by the density matrix
$\rho$.  Griffiths' condition is equivalent to the vanishing of
the real part of the expression above,
$ Re \{D([P_{\alpha}],[P_{\beta}]\} = p_{\alpha} \delta_{\alpha, \beta}. $
As Gell-Mann and Hartle (1990) emphasize, it is more convenient -- and
in the context of and emergent classicality more realistic -- to demand
instead $ D([P_{\alpha}],[P_{\beta}]) = p_{\alpha} \delta_{\alpha, \beta}$.
Both weaker and stronger conditions for the consistency of histories were
considered (Goldstein and Page, 1995; Gell-Mann and Hartle, 1997).
The problem with all of them is that the resulting histories are very
subjective: Given an initial density matrix of the Universe it is in general
quite easy to specify many different, mutually incompatible consistent
sets of histories. This subjectivity leads to serious interpretational
problems (d'Espagnat, 1989; 1995; Dowker and Kent, 1996).
Thus, a demand for exact consistency as one of the conditions
for classicality is both uncomfortable (overly restrictive)
and insufficient (as the resulting histories are very non-classical).
Moreover, coarse-grainings that help secure approximate consistency
have to be, in effect, guessed.

The attitudes adopted by the practitioners of the consistent histories
approach in view of its unsuitability for the role of the cornerstone 
of the emergent classicality differ. Initially -- before difficulties 
became apparent -- it was hoped that such approach would answer all of 
the interpretational questions, perhaps when supplemented by a subsidiary 
condition, i.e. some assumption about favored coarse-grainings. At present, 
some still aspire to the original goals of deriving classicality from 
consistency alone. Others may uphold the original aims of the program, but
they also generally rely on environment-induced decoherence, using in
calculations variants of models we have presented in this paper. This
strategy has been quite successful -- after all, decoherence leads 
to consistency. For instance, special role of the hydrodynamic observables
(Gell-Mann and Hartle, 1990; Dowker and Halliwell, 1992; Halliwell, 1999;
Brun and Hartle, 1999) can be traced to their predictability,
or to their approximate commutativity with the total Hamiltonian 
(see Eq. (4.41)). On the other hand, the original
goals of Griffiths (1984, 1996) have been more modest: Using consistent
histories, one can discuss sequences of events in an
evolving quantum system without logical contradictions.
The ``golden middle'' is advocated by Griffiths and Omn\`es (1999) who
regard consistent histories as a convenient language, rather than
as an explanation of classicality.

The origin of the effective classicality can be traced to decoherence
and einselection. As was noted by Gell-Mann and Hartle (1990), and elucidated
by Omn\`es (1992; 1994) decoherence suffices to ensure approximate
consistency. But consistency is both not enough and too much -- it is too
easy to accomplish, and does not necessarily lead to classicality
(Dowker and Kent, 1996). What is needed instead is the objectivity
of events and their time-ordered sequences -- histories. As we have seen
above, both can appear as a result of einselection.

\subsubsection{Relatively Objective Past}

We have already provided an operational definition of relatively objective
existence of quantum states. It is easy to apply it to events and histories:
When many observers can independently gather compatible evidence concerning
an event, we call it {\it relatively objective}. Relatively objective
history is then a time-ordered sequence of relatively objective events.

Monitoring of the system by the environment leads
to decoherence and einselection. It will also typically lead to redundancy 
and hence effectively objective classical existence in the sense of 
`quantum Darwinism'. Observers can independently access redundant records
of events and histories imprinted in the environmental degrees of freedom.
The number of observers who can examine evidence etched in the environment
can be of the order of, and is bounded from above by ${\cal R_J}$.
Redundancy is a measure of this objectivity and classicality.

As observers record their data, ${\cal R_J}$ changes: Consider an observer 
who measures the `right observable' of ${\cal E}$ (i.e., the one with the 
eigenstates  $|0\rangle ,|1\rangle $ in the example of Eq. (7.1)). Then 
his records and -- as his records decohere, their environment
-- become a part of evidence, and are correlated with the preferred 
basis of the system. Consequently, ${\cal R_J}$ computed from Eq. (7.14) 
increases. Every interaction that increases the number of the records 
also increases ${\cal R_J}$. This is obvious for 
the ``primary'' interactions with the system, but it is also true for 
the secondary, tertiary, etc. acts of replication of the information obtained 
from the observers who recorded primary state of the system, from 
the environment, from the environment of the environment, and so on.

A measurement reveals to the observer ``his'' branch of the universal
state vector. The correlations established alter observer's state
-- his records -- and ``attach'' him to this branch. He will share it
with other observers who examined the same set of observables,
and who have recorded compatible results.

It is also possible to imagine a stubborn observer who insists on measuring
either the relative phase between the two obvious branches of the environment
in Eq. (7.2), or the state of the environment in the Hadamard-transformed
basis $\{ |+\rangle ,|-\rangle  \}$. In either case the distinction between 
the two outcomes could determine the state of the spin in the 
$\{|\odot\rangle , |\otimes\rangle  \}$  basis. However, in that basis 
${\cal R_J}=1$. Hence, while in principle these measurements can be carried 
out and yield the correct result, the information concerning 
$\{|\odot\rangle , |\otimes\rangle  \}$ basis is not redundant, and, 
therefore, not objective: Only one stubborn observer can access it directly. 
As a result ${\cal R_J}$ will decrease. Whether the 
${\cal R_J} (\{|\odot\rangle , |\otimes\rangle  \})$ will become larger than
${\cal R_J} (\{|\uparrow\rangle , |\downarrow\rangle \})$ was before
the measurement of the stubborn observer will depend on detailed
comparison of the initial redundancy with the amplification involved, 
decoherence, etc.

There is a further significant difference between the two stubborn observers
considered above. When the observer measures the phase between
the two sequences of 0's and 1's in Eq. (7.2), correlations between
the bits of the environment remain. Thus, even after his measurement
one could find relatively objective evidence of the past event -- past state
of the spin -- and, in more complicated cases, of the history. On the other
hand, measurement of all the environment bits in the 
$\{|+\rangle ,|-\rangle \}$ basis will obliterate evidence of such a past.

Relatively objective existence of events is the strongest condition we have
considered here. It is a consequence of the existence of multiple records
of the same set of states of the system. It allows for such manifestations
of classicality as unimpeded cloning. It implies einselection of states most 
closely monitored by the environment. Decoherence is clearly weaker and 
easier to accomplish.

``The past exists only insofar as it is recorded in the present'' (dictum
often repeated by Wheeler) may the best summary of the above discussion. 
Relatively objective reality of few selected observables in our familiar 
Universe is measured by their fitness -- by the redundancy with which they are
recorded in the environment. This multiplicity of available copies of the
same information can be regarded as a consequence of amplification, and
as a cause of indelibility. Multiple records safeguard objectivity of our 
past. 

\section{DECOHERENCE IN THE LABORATORY}

The biggest obstacle in the experimental study of decoherence is 
-- paradoxically -- its effectiveness. In the macroscopic domain only 
the einselected states survive. Their superpositions are next to impossible 
to prepare. In the mesoscopic regime one may hope to adjust the size 
of the system, and, thus, interpolate between quantum and classical. The
strength of the coupling to the environment is the other parameter
one may employ to control the decoherence rate. 

One of the key consequences of monitoring by the environment is the inevitable
introduction of the Heisenberg uncertainty into the observable complementary
to the one that is monitored. One can {\it simulate} such uncertainty
without any monitoring environment by introducing classical noise.
In each specific run of the experiment -- for each realization of
time-dependent noise -- quantum system will evolve deterministically.
However, after averaging over different noise realizations
evolution of the density matrix describing an {\it ensemble} of systems
may approximate decoherence due to an entangling quantum environment. 
In particular, the master equation may be essentially the same as for 
true decoherence, although the interpretational implications are more
limited. Yet, using such strategies one can simulate much of the dynamics of
open quantum systems.

The strategy of simulating decoherence can be taken further: Not just
the effect of the environment, but also the dynamics of the quantum system
can be simulated by classical means. This can be accomplished
when classical wave phenomena follow equations of motion 
related to Schr\"odinger equation. We shall discuss
experiments that fall into all of the above categories.

Last not least, while decoherence -- through einselection -- helps solve
the measurement problem, it is also a major obstacle to quantum information
processing. We shall thus end this section briefly describing strategies
that may allow one to tame decoherence.

\subsection{Decoherence due to entangling interactions}

Several experiments fit this category, and more have been proposed. Decoherence due to emission or scattering of photons has been investigated by the MIT group
(Chapman et al., 1995; Kokorowski et al., 2001) using atomic interferometry. 
Emission or scattering deposits a record in the environment. It can store 
information about the path of the atom providing photon wavelength is shorter 
than the separation between the two of the atoms. In case of emission this 
record is not redundant, as the atom and photon are simply entangled
-- ${\cal R_J} \sim 1$ in any basis. Scattering may involve more photons,
and the recent careful experiment has confirmed the saturation of decoherence
rate at distances in excess of photon wavelength (Gallis and Fleming, 1990;
Anglin, Paz, and Zurek, 1997).

There is an intimate connection between interference and complementarity
in the two-slit experiment on one hand, and the entanglement on the other
(Wootters and Zurek, 1979). Consequently, appropriate measurements of
the photon allow one to restore interference fringes in the conditional
subensembles corresponding to a definite phase between the two photon
trajectories (see especially Chapman et al. (1995), as well as
Kwiat, Steinberg and Chiao (1993); Pfau et al., 1994;
Herzog, Kwiat, Weinfurter and Zeilinger, (1995) for implementations
of this ``quantum erasure'' trick due to Hillery and Scully (1983)).
Similar experiments have been also carried out using neutron interferometry
(see e.g. Rauch, (1998)).

In all of these experiments one is dealing with a very simplified situation
involving a single microsystem and a single ``unit'' of decoherence
(${\cal R_J} \sim 1$) caused by a single quantum of the environment.
Experiments on a mesoscopic system monitored by the environment are obviously
much harder to devise. Nevertheless, Serg\'e Haroche, Jean-Michel Raimond,
Michel Brune and their colleagues (Brune et al., 1996; Haroche 1998, Raimond,
Brune, and Haroche, 2001) have carried out a spectacular experiment
of this type, yielding solid evidence in support of the basic tenets of
the environment-induced transition from quantum to classical. Their system
is a microwave cavity. It starts in a coherent state with an amplitude
corresponding to a few photons.

``Schr\"odinger cat'' is created by introducing an atom in a superposition
of two Rydberg states, $|+\rangle = |0\rangle  + |1\rangle $: The atom 
passing through the cavity puts its refractive index in a superposition 
of two values. Hence, the phase of the coherent state shifts by the amount 
correlated with the state of the atom, creating an entangled state:
$$ |\rightarrow\rangle (|0\rangle  + |1\rangle ) \ \Longrightarrow \ |\nearrow\rangle |0\rangle  +
|\searrow\rangle |1\rangle  = |\vartheta\rangle \eqno(8.1)$$
Arrows indicate relative phase space locations of coherent states. States 
of the atom are $|0\rangle $ and $|1\rangle $. ``Schr\"odinger kitten'' 
is prepared from this entangled state by measuring the atom in the 
$\{|+\rangle ,|-\rangle \}$ basis:
$$ |\vartheta\rangle  =
(|\nearrow\rangle  + |\searrow\rangle )|+\rangle  + (|\nearrow\rangle  - |\searrow\rangle )|-\rangle \eqno(6.2)$$
Thus, atom in the state $|+\rangle $ implies preparation of a ``positive cat''
$|\uplus\rangle  = |\nearrow\rangle  + |\searrow\rangle $ in the cavity.
Such superpositions of coherent states could survive forever if
there was no decoherence. However, radiation leaks out of the cavity.
Hence, environment acquires information about the state inside.
Consequences are tested by passing another atom in the state
$|+\rangle  = |0\rangle  + |1\rangle $ through the cavity. In absence 
of decoherence the state would evolve as:
\begin{eqnarray}
|\uplus\rangle |+\rangle  \ = \ (|\nearrow\rangle  + |\searrow\rangle )(|0\rangle  + |1\rangle ) \Longrightarrow
\nonumber \\
(|\uparrow\rangle |0\rangle +|\rightarrow\rangle |1\rangle )\ +\  (|\rightarrow\rangle |0\rangle  + |\downarrow\rangle |1\rangle )
\nonumber \\
    = (|\uparrow\rangle |0\rangle  + |\downarrow\rangle |1\rangle ) + \sqrt 2 |\rightarrow\rangle |+\rangle 
\ . \eqnum{8.3}
\end{eqnarray}
Above we have omitted the overall normalization, but retained the (essential)
relative amplitude.

For the above state detection of $|+\rangle $ in the first (preparatory)
atom implies the conditional probability of detection of $|+\rangle $,
$p_{+|+} = {3 \over 4}$, for the second (test) atom. Decoherence
will suppress off-diagonal terms of the density matrix, so that,
some time after the preparation, $\rho_{cavity}$ that starts, say,
as $|\uplus\rangle \langle \uplus|$ becomes:
\begin{eqnarray}
\rho_{cavity}& = & (|\nearrow\rangle \langle \nearrow|+|\searrow\rangle \langle \searrow|)/2 \nonumber \\
& + & z(|\nearrow\rangle \langle \searrow| +  |\searrow\rangle \langle \nearrow|)/2 \ . \eqnum{8.4}
\end{eqnarray}
When $z=0$ the conditional probability is $p(+|+) = {1 \over 2}$.

In the intermediate cases intermediate values of this and other relevant
conditional probabilities are predicted. The rate of decoherence, and,
consequently, the time-dependent value of $z$ can be estimated from the
cavity quality factor $Q$, and from the data about the coherent state initially
present in the cavity. Decoherence rate is a function
of separation of the two components of the cat $|\uplus\rangle $.
Experimental results agree with predictions.

The discussion above depends on the special role of coherent states.
Coherent states are einselected in harmonic oscillators, and, hence,
in the underdamped bosonic fields (Anglin and Zurek, 1996).
Thus, they are the pointer states of the cavity.
Their special role is recognized implicitly above: If number eigenstates
were einselected, predictions would be obviously quite different.
Therefore, while the ENS experiment is focused on decoherence rate,
confirmation of the predicted special role of coherent states
in bosonic fields is its important (albeit implicit) corollary.

\subsection{Simulating decoherence with classical noise}

>From the fundamental point of view, the distinction between cases when
decoherence is caused by entangling interactions with the quantum state of
the environment and when it is simulated by a classical noise in
the observable complementary to the ``pointer'' is essential. However,
from the engineering point of view (adopted, e.g., by the practitioners
of quantum computation, see Nielsen and Chuang, 2000 for discussion)
this may not matter. For instance, quantum error correction techniques
(Shor, 1995; Steane, 1996; Preskill, 1999) are capable of dealing with either.
Moreover, experimental investigations of this subject often involve both.

The classic experiment in this category was carried out recently by
David Wineland, Chris Monroe, and their collaborators (Myatt et al, 2000;
Turchette et al., 2000). They use ion trap to study behavior of
individual ions in a ``Schr\"odinger cat'' state (Monroe et al., 1996)
under the influence of injected classical noise. They also embark
on a preliminary study of ``environment engineering''. 

Superpositions of two coherent states as well as of number eigenstates 
were subjected to simulated high temperature amplitude and phase
``reservoirs''.
This was done through time-dependent modulation of the self-Hamiltonian
of the system. For the amplitude noise these are in effect random
fluctuations of the location of the minimum of the harmonic trap. Phase
noise corresponds to random fluctuations of the trap frequency.

In either case, the resulting loss of coherence is well described by
the exponential decay with time, with the exponent that scales with
the square of the separation between the two components of the macroscopic
quantum superposition (e.g., Eq. (5.34)). The case of the amplitude noise
approximates decoherence in quantum Brownian motion in that the coordinate
is monitored by the environment, and, hence, the momentum is perturbed.
(Note that in the underdamped harmonic oscillator rotating wave approximation
blurs the distinction between $x$ and $p$, leading to einselection of
coherent states.) The phase noise would arise in an environment monitoring
the number operator, thus leading to uncertainty in phase. Consequently,
number eigenstates are einselected.

The applied noise is classical, and the environment does not acquire any
information about the ion (${\cal R_I} = 0$). Thus, following a particular
realization of the noise the state of the system is still pure. Nevertheless,
an ensemble average over many noise realizations is represented by
the density matrix that follows an appropriate master equation. Thus,
as Wineland, Monroe, and their colleagues note, decoherence simulated
by the classical noise could be in each individual case -- for each 
realisation -- reversed by simply measuring
the corresponding time-dependent noise run either beforehand or afterwards,
and then applying the appropriate unitary transformation to the state
of the system.
By contrast, in the case of entangling interactions, two measurements --
one preparing the environment before the interaction with the environment,
the other following it -- would be the least required for a chance of undoing
the effect of decoherence.

The same two papers study decay of a superposition of number
eigenstates $|0\rangle $ and $|2\rangle $ due to an indirect coupling with the vacuum.
This proceeds through entanglement with the first order environment (that,
in effect, consists of the other states of the harmonic oscillator) and
a slower transfer of information to the distant environment. Dynamics
involving the system and its first - order environment leads to non-monotonic
behavior of the off-diagonal terms representing coherence. Further studies
of decoherence in the ion trap setting are likely to follow, as this is
an attractive implementation of the quantum computer (Cirac and Zoller, 1995).

\subsubsection{Decoherence, Noise, and Quantum Chaos}

Following a proposal of Graham, Schlautmann, and Zoller (1993) Mark Raizen
and his group (Moore et al., 1994) used a one-dimensional optical lattice
to implement a variety of 1-D chaotic systems including the ``standard map''.
Various aspects of the behavior expected
from a quantized version of a classically chaotic system were subsequently
found, including, in particular, dynamical localization (Casati and Chirikov,
1995a; Reichl, 1992).

Dynamical localization establishes, in a class of driven quantum chaotic
systems, a saturation of momentum dispersion, and leads to a characteristic
exponential form of its distribution (Casati and Chirikov, 1995a). 
Localization is obviously a challenge to the quantum-classical correspondence, 
since in these very same systems classical prediction has the momentum 
dispersion growing unbounded, more or less with the square root of time. 
However, it sets in after $t_L \sim \hbar^{-\alpha}$, where $\alpha \sim 1$
(rather than on the much shorter $t_{\hbar} \sim  \ln \hbar^{-1}$
we have discussed in Section III) so it can be ignored for macroscopic 
systems. On the other hand, its signature is easy to detect.

Demonstration of dynamical Anderson localization in the optical lattice
implementation of the $\delta$-kicked rotor and related studies of quantum
chaos are a significant success (Moore et al., 1994).
More recently, attention of both Raizen and his group in Texas as well
as of Christensen and his group in New Zealand has shifted towards 
the effect of decoherence on quantum chaotic evolution
(Ammann et al, 1998; Klappauf et al., 1998).

In all of the above studies the state of the chaotic system ($\delta$-kicked
rotor) was perturbed by spontaneous emission from the trapped atoms, that was
induced by decreasing detuning of the lasers used to set up the optical
lattice. In addition, noise was occasionally introduced into the potential.
Both groups find that, as a result of spontaneous emission, localization
disappears, although the two studies differ in some of the details. More
experiments, including some that allow gentler forms of monitoring by
the environment (rather than spontaneous emission noise) appear to be
within reach.

In all of the above cases one deals, in effect, with a large ensemble of
identical atoms. While each atom suffers repeated disruptions of its
evolution due to spontaneous emission, the ensemble evolves
smoothly and in accord with the appropriate master equation. The situation
is reminiscent of the ``decoherence simulated by noise''.

Experiments that probe the effect of classical noise on chaotic systems
have been carried out earlier (Koch, 1995). They were, however,
analyzed from the point of view that does not readily shed light on
decoherence.

A novel experimental approach to decoherence and to
irreversibility in open complex quantum systems is
pursued by Levstein, Pastawski, and their colleagues (Levstein, Usaj,
and Pastawski, 1998; Levstein et al., 2000).
Using NMR techniques the investigate reversibility
of dynamics by implementing a version of spin echo. This promising
``Loschmidt echo'' approach has led to renewed interest in the issues
that touch on quantum chaos, decoherence, and related subjects (see 
e.g., Gorin and Seligman, 2001; Prosen, 2001; Jalabert and 
Pastawski, 2001; Jacquod, Silvestrov and Beenakker, 2001).

\subsection{Analogue of decoherence in a classical system}

Both the system and the environment are effectively classical in the last
category of experiments, represented by the work of Cheng and Raymer (1999).
They have investigated
behavior of transverse spatial coherence during the propagation of the
optical beam through a dense, random dielectric medium. This problem
can be modelled by a Boltzmann-like transport equation for the Wigner function
of the wave field, and exhibits characteristic increase of decoherence
rate with the square of the spatial separation, followed by a saturation
at sufficiently large distances. This saturation contrasts with the simple
models of decoherence in quantum Brownian motion that are based on dipole
approximation. However, it is in good accord with the more sophisticated
discussions which recognize that, for separations of the order of the
prevalent wavelength in the environment, dipole approximation fails
and other more complicated behaviors can set in
(Gallis and Fleming, 1990; Anglin, Paz and Zurek, 1997; Paz and Zurek, 1999).
Similar result in a completely quantum case was obtained using atomic
interferometry by Kokorowski et al. (2001).

\subsection{Taming decoherence}

In many of the applications of quantum mechanics the quantum nature of the
information stored or processed needs to be protected. Thus, decoherence
is an enemy. Quantum computation is an example of this situation.
Quantum computer can be thought of as a sophisticated interference device
that works by performing in parallel a coherent superposition of a
multitude of classical computations. Loss of coherence would disrupt
this quantum parallelism essential for the expected speedup. 

In the absence of the ideal -- completely isolated absolutely perfect
quantum computer, something easy for a theorist to imagine but impossible 
to attain in the laboratory -- one must deal with imperfect hardware 
``leaking'' some of its information
to the environment. And maintaining isolation while simultaneously achieving
a reasonable ``clock time'' for the quantum computer is likely to be difficult
since both are in general controlled by the same interaction (although there
are exceptions -- for example, in the ion trap proposal of Cirac and Zoller
(1995) interaction is in a sense ``on demand'', and is turned on by the
laser coupling internal states of ions with the vibrational degree
of freedom of the ion chain).

The need for error correction in quantum computation was realized early on
(Zurek, 1984b) but methods for accomplishing this goal have evolved
dramatically from the Zeno effect suggested then to very sophisticated
(and much more effective) strategies in the recent years. This is fortunate:
Without error correction even fairly modest quantum computations (such as
factoring number 15 in an ion trap with imperfect control of the duration
of laser pulses) go astray rapidly as a consequence of relatively small
imperfections (Miquel, Paz, and Zurek, 1997).

Three different, somewhat overlapping approaches that aim to control and
tame decoherence, or to correct errors caused by decoherence or by the other
imperfections of the hardware have been proposed. We summarize
them very briefly, spelling out main ideas and pointing out references
that discuss them in greater detail.

\subsubsection{Pointer states and noiseless subsystems}

The most straightforward strategy to suppress decoherence is to isolate the
system of interest (e.g., quantum computer). Failing that, one may try to
isolate some of its observables with degenerate pointer subspaces, which
then constitute niches in the Hilbert space of the information processing
system that do not get disrupted in spite of its the coupling
to the environment. Decoherence free subspaces (DFS's for brevity)
are thus identical in conception with the pointer subspaces introduced
some time ago (Zurek, 1982), and satisfy (exactly or approximately)
the same Eqs. (4.22) and (4.41) or their equivalents (given, e.g., in terms
of ``Krauss operators'' (Krauss, 1983)) that represent non-unitary consequences
of the interaction with the environment in the Lindblad (1976) form of
the master equation). DFS's were (re)discovered in the context of quantum
information processing. They appear as a consequence of an exact or
approximate symmetries of the Hamiltonians that govern the evolution of
the system and its interaction with the environment
(Zanardi and Rasetti, 1997; Duan and Guo, 1998; Lidar et al., 1998;
Zanardi, 1998; 2000).

An active extension of this approach aimed at finding quiet corners
of the Hilbert space is known as dynamical decoupling. There the
effectively decoupled subspaces are induced by time-dependent modifications
of the evolution of the system deliberately introduced from the outside
by time-dependent evolution and / or measurements (see e.g. Viola
and Lloyd, 1998; Zanardi, 2000). A further generalization
and unification of various techniques leads to the concept of
noiseless quantum subsystems (Knill, Laflamme, and Viola, 1999;
Zanardi, 2000), which may be regarded as a non-abelian
(and quite non-trivial) generalization of pointer subspaces.

A sophisticated and elegant strategy that can be regarded as a version of the
decoherence free approach was devised independently by Kitaev (1997a,~b).
He has advocated using states that are topologically stable, and, thus,
that can successfully resist arbitrary interactions with the environment. 
The focus here (in contrast to much of the DFS work) is on devising
a system with self-Hamiltonian that -- as a consequence of the structure
of the gap its energy spectrum related to the ``cost'' of topologically
non-trivial excitations -- acquires a subspace {\it de facto} isolated
from the environment. This approach has been further developed
by Bravyi and Kitaev (1998) and by Freedman and Meyer (1998).

\subsubsection{Environment engineering}

This strategy involves altering the (effective) interaction Hamiltonian between 
the system and the environment or influencing the state of the environment
to selectively suppress decoherence. There are many ways
to implement it, and we shall describe under this label variety of proposed
techniques (some of which are not all that different from the strategies
we have just discussed) that aim to protect quantum information stored
in selected subspaces of the Hilbert space of the system, or even to exploit
pointer states induced or redefined in this fashion.

The basic question that started this line of research --
can one influence the choice of the preferred pointer states? --
arose in the context of ion trap quantum computer proposed by
Cirac and Zoller (1995). The answer given by the theory is, of course, that
the choice of the einselected basis is predicated on the details of the
situation, and, in particular, on the nature of the interaction between
the system and the environment (Zurek, 1981; 1982; 1993a). Yet, Poyatos,
Cirac and Zoller (1996) have suggested a scheme suitable for implementation
in an ion trap, where interaction with the environment -- and, in accord
with Eq. (4.41), the pointer basis itself -- can be adjusted. The key idea
is to recognize that the effective coupling between the vibrational degrees
of freedom of an ion (the system) and the laser light (which plays the
role of the environment) is given by:
$$ H_{int} = {{\Omega} \over 2}
(\sigma_+e^{-i\omega_Lt} +\sigma_-e^{i\omega_Lt})
\sin[\kappa(a + a^+) + \phi]\ . \eqno(8.5)$$
Above, $\Omega$ is the Rabi frequency, $\omega_L$ the laser frequency, $\phi$
is related to the relative position of the center of the trap with respect
to the laser standing wave, $\kappa$ is the Lamb-Dicke parameter of the
transition, while $\sigma_-$ ($\sigma_+$) and $a$ ($a^+$) are
the annihilation (creation) operators of the atomic transition
and of the harmonic oscillator (trap).

By adjusting $\phi$ and $\omega_L$ and adopting appropriate set of
approximations (that include elimination of the internal degrees
of freedom of the atom) one is led to the master equation for the system
-- i.e, for the density matrix of the vibrational degree of freedom:
$$ \dot \rho \ = \ \gamma (2f\rho f^+ - f_+f\rho - \rho f_+f) \eqno(8.6)$$
Above, $f$ is the operator with a form that depends on the adjustable
parameters $\phi$ and $\omega_L$ in $H_{int}$, while $\gamma$ is a constant
that depends also on $\Omega$ and $\eta$. As Poyatos et al. show, one can
alter the effective interaction between the slow degree of freedom
(the oscillator) and the environment (laser light) by adjusting parameters
of the actual $H_{int}$.

First steps towards realization of this ``environment engineering''
proposals were taken by the NIST group (Myatt et al., 2000;
Turchette et al., 2000). Similar techniques can be employed to protect
deliberately selected states from decoherence (Carvalho et al., 2000).

Other ideas aimed at channeling decoherence have been also explored
in contexts that range from quantum information processing (Beige et al., 2000)
to preservation of Schr\"odinger cats in Bose-Einstein condensates
(Dalvit, Dziarmaga, and Zurek, 2000).

\subsubsection{Error correction and resilient computing}

This strategy is perhaps most sophisticated and comprehensive, and capable
of dealing with the greatest variety of errors in a most hardware - independent
manner. It is a direct descendant of the error correction techniques employed
in dealing with the classical information, based on redundancy:
Multiple copies of the information are made, and the errors are found and
corrected by sophisticated ``majority voting' techniques.

One might have thought that implementing error correction in the 
quantum setting
will be difficult for two reasons. To begin with, quantum states -- and, hence,
quantum information -- cannot be ``cloned'' (Dieks, 1982; Wootters
and Zurek, 1982). Moreover, quantum information is very ``private'', and
a measurement that is involved in majority voting would infringe on this
privacy and destroy quantum coherence, making quantum information classical.
Fortunately, both of these difficulties can be simultaneously overcome by
encoding quantum information in entangled states of several qubits.
Cloning turns out not to be necessary. And measurements can be carried out
in a way that identifies errors while keeping quantum
information untouched. Moreover, error correction is {\it discrete} --
measurements that reveal error syndromes have ``yes - no'' outcomes.
Thus, even though information stored in a qubit represents a continuum of
possible quantum states (e.g., corresponding to a surface of the Bloch sphere)
error correction is discrete, allaying one of the earliest worries concerning
the feasibility of quantum computation  -- the unchecked ``drift''
of the quantum state representing information (Landauer, 1995).

This strategy (discovered by Shor (1995) and Steane (1996)) has been since
investigated by many (Ekert and Macchiavello, 1996; Bennett et al., 1996;
Laflamme et al., 1996) and codified into a mathematically appealing
formalism (Gottesman, 1996; Knill and Laflamme, 1997). Moreover, first
examples of successful implementation (see e.g. Cory et al., 1999) are already
at hand.

Error correction allows one, at least in principle, to compute forever,
providing that the errors are suitably small ($\sim 10^{-4}$ per computational
step seems to be the error probability threshold sufficient for most error
correction schemes). Strategies that accomplish this encode qubits in already
encoded qubits (Aharonov and Ben-Or 1997; Kitaev 1997c; Knill, Laflamme and
Zurek, 1996; 1998a,~b; Preskill 1998). The number of layers of such 
concatenation necessary
to achieve fault tolerance -- the ability to carry out arbitrarily long
computations -- depends on the size (and the character) of the errors,
and on the duration of the computation, but when error probability is smaller
than the threshold, that number of layers is finite. Overviews of fault
tolerant computation are already at hand (Preskill, 1999; Nielsen and
Chuang, 2000, and references therein).

An interesting subject related to the above discussion is quantum process
tomography, anticipated by Jones (1994), and described in the context of
quantum information processing by Chuang and Nielsen (1997) and by
Poyatos, Cirac and Zoller (1997). The aim here is to characterize completely
a process -- such as a quantum logical gate -- and not just a state. 
First deliberate implementation of this procedure (Nielsen, Knill,
and Laflamme, 1998) has also demonstrated experimentally that einselection
is indeed equivalent to an unread measurement of the pointer basis by 
the environment, and can be regarded as such from the standpoint of 
applications (i.e., NMR teleportation in the example above).

\section{CONCLUDING REMARKS}

Decoherence, einselection, pointer states, and even predictability sieve have
become familiar concepts to many in the past decade. The first goal of this
paper was to review these advances and to survey, and -- where
possible, to address -- the remaining difficulties. The second related aim was
to ``preview'' the future developments. This has led to considerations
involving information, as well as to the operational, physically motivated 
discussions of seemingly esoteric concepts such as objectivity. Some of the 
material presented (including the `Darwinian' view of the emergence of 
objectivity through redundancy, as well as the discussion of envariance
and probabilities) are rather new, and a subject of research,
hence the word ``preview'' applies here.

New paradigms often take a long time to gain ground. Atomic theory
of matter (which, until early XX century, was `just an interpretation') is
the case in point. Some of the most tangible applications and consequences
of new ideas are difficult to recognize immediately. In the case of atomic
theory, Brownian motion is a good example: Even when the evidence is
out there, it is often difficult to decode its significance.

Decoherence and einselection are no exception. They have been investigated
for about two decades. They are the only explanation of classicality that
does not require modifications of quantum theory, as do the alternatives
(Bohm, 1952; Leggett, 1980, 1988, 2002; Penrose, 1986, 1989; Holland, 1993;
Goldstein, 1998; Pearle, 1976;~1993; Ghirardi, Rimini, and Weber, 1986;~1987,
Gisin and Percival, 1992; 1993a-c). Ideas based on the immersion of the
system in the environment have recently gained enough support to be described
(by sceptics!) as ``the new orthodoxy'' (Bub, 1997). This is a dangerous
characterization, as it suggests that the interpretation based
on the recognition of the role of the environment is both complete
and widely accepted. Neither is certainly the case.

Many conceptual and technical issues
(such as what constitutes `a system') are still open.
As for the breadth of acceptance, ``the new orthodoxy'' seems to be
an optimistic (mis-)characterization of decoherence
and einselection, especially since this explanation
of the transition from quantum to classical has (with very few
exceptions) not made it to the textbooks. This is intriguing, 
and may be as much a comment on the way in which quantum physics has been 
taught, especially on the undergraduate level, as on the status of the
theory we have reviewed and its level of acceptance among the physicists.

Quantum mechanics has been to date, by and large, presented in a manner that
reflects its historical development. That is, Bohr's planetary model of atom
is still often the point of departure, Hamilton-Jacobi equations are
used to ``derive'' Schr\"odinger equation, and an oversimplified version
of the quantum - classical relationship (attributed to Bohr, but generally
not doing justice to his much more sophisticated views) with
the correspondence principle, kinship of commutators and Poisson brackets,
Ehrenfest theorem, some version of the Copenhagen interpretation, and other
evidence that quantum theory is really not all that different from classical
-- especially when systems of interest become macroscopic, and all one cares
about are averages -- is presented.

The message seems to be that the there is really no problem and that
quantum mechanics can be ``tamed'' and confined to the microscopic domain.
Indeterminacy and double slit experiment are of course discussed, but
to prove peaceful coexistence within the elbow room assured by the
Heisenberg's principle and complementarity. Entanglement is rarely explored.
This is quite consistent with the aim of the introductory quantum mechanics
courses, which has been (only slightly unfairly) summed up by the memorable
phrase ``shut up and calculate''.  Discussion of measurement is
either dealt with through models based on the CI ``old orthodoxy'',
or not at all. An implicit (and sometime explicit) message is: Those 
who ask questions that do not lend themselves to an answer through 
laborious, preferably perturbative calculation are ``philosophers'', 
and should be avoided.

The above description is of course a caricature. But given that
the calculational techniques of quantum theory needed in atomic, nuclear,
particle, or condensed matter physics are indeed difficult to master,
and given that -- to date -- most of the applications had nothing
to do with the nature of quantum states, entanglement, and such, the
attitude of avoiding the most flagrantly quantum
aspects of quantum theory is easy to understand.

Novel applications force one to consider questions about the information
content, the nature of ``the quantum'', and the emergence of the classical
much more directly, with a focus on states and correlations, rather than on
the spectra, cross sections and the expectation values. 
Hence, problems that are usually bypassed will come to the fore:
It is hard to brand Schr\"odinger cats and entanglement as ``exotic''
and make them a centerpiece of a marketable device. I believe that
as a result decoherence will become a part of the textbook lore.
Indeed, at the graduate level there are already some notable exceptions
among monographs (Peres, 1993) and specialized texts (Walls and
Milburn, 1994; Nielsen and Chuang, 2000). 


Moreover, the range of subjects already influenced by decoherence and 
einselection -- by the ideas originally motivated by the quantum theory 
of measurements -- is beginning to extend way beyond its original domain.
In addition to the atomic physics, quantum optics, and quantum information
processing (which were all mentioned throughout this review) it stretches
from material sciences (Karlsson, 1998; Dreismann, 2000), surface science,
where it seems to be an essential ingredient explaining emission of electrons
(Brodie, 1995; Durakiewicz et al., 2001) through heavy ion collisions 
(Krzywicki, 1993) to quantum gravity and cosmology
(Zeh, 1986,~1988,~1992;  Kiefer, 1987; Kiefer and Zeh, 1995; Halliwell, 1989;
Brandenberger, Laflamme and Mijic, 1990; Barvinsky and Kamenshchik, 
1990,~1995; Paz and Sinha, 1991,~1992; Castagnino et al., 1993, Mensky and
Novikov, 1996).
Given the limitations of space we have not done justice to most of these
subjects, focusing instead on issues of principle. In some areas reviews
already exist. Thus, Giulini et al. (1996) is a valuable collection of essays,
where, for example, decoherence in field theories is addressed. Dissertation
of Wallace (2002) offers a good (if somewhat philosophical) summary of the
role of decoherence with a rathe different emphasis on similar field-theoretic
issues. Conference
proceedings edited by Blanchard et al. (2000) and, especially, an extensive
historical overview of the foundation of quantum theory from the modern
perspective by Auletta (2000) are also recommended. More specific technical
issues with implications for decoherence and einselection have also been
reviewed. For example, on the subject of master equations there are several
reviews with very different emphasis including Alicki and Lendi (1987),
Grabert, Schramm, and Ingold (1988), Namiki and Pascazio (1993), as well as 
-- more recently --
Paz and Zurek (2001). In some areas -- such as atomic BEC's -- the study
of decoherence has only started  (Anglin, 1997; Dalvit, Dziarmaga,
and Zurek, 2001). In many situations (e.g, quantum optics) a useful
supplement to the decoherence view of the quantum - classical interface
is afforded by `quantum trajectories' -- a study of the state of the system
inferred from the intercepted state of the environment (see Carmichael, 1993;
Wiseman and Milburn, 1993; Gisin and Percival, 1993a-c). This approach
``unravels'' evolving density matrices of open systems into trajectories
conditioned upon the measurement carried out on the environment, and may
have -- especially in quantum optics -- intriguing connections with the
``environment as a witness'' point of view (see Dalvit, Dziarmaga,
and Zurek, 2001). In other areas -- such condensed matter
-- decoherence phenomena have so many variations and are so pervasive that
a separate ``decoherent review'' may be in order, especially as intriguing
experimental puzzles seem to challenge the theory (Mohanty and Webb, 1997;
Kravtsov and Altshuler, 1999).

Physics of information and computation is a special case. Decoherence
is obviously a key obstacle in the implementation of information
processing hardware that takes advantage of the superposition principle.
While we have not focused on quantum information processing, the discussion
was often couched in the language inspired by the information theory. This
is no accident: It is the belief of this author that many of the remaining
gaps in our understanding of quantum physics and its relation 
to the classical domain -- such as the definition of systems,
or the still mysterious details of the ``collapse'' -- shall follow
pattern of the ``predictability sieve'' and shall be expanded into
new areas investigation by considerations that simultaneously elucidate
nature of ``the quantum'' and of ``the information''.

\section{ACKNOWLEDGMENTS}

John Archibald Wheeler has -- quarter century ago -- taught a course
on the subject of quantum measurements at the University of Texas
in Austin. The questions raised then have since evolved into ideas
presented here, partly in collaboration with Juan Pablo Paz, and through
interactions with many colleagues, including Andreas Albrecht, Jim Anglin,
Charles Bennett, Robin Blume-Kohout, Carlton Caves, Ike Chuang, Diego Dalvit, 
David Deutsch, David Divincenzo, Jacek Dziarmaga, Richard Feynman, 
Murray Gell-Mann, Daniel Gottesmann, Robert Griffiths, Salman Habib,
Jonathan Halliwell, Serge Haroche, James Hartle, Chris Jarzynski, Erich Joos,
Manny Knill, Raymond Laflamme, Anthony Leggett, Seth Lloyd, Gerard Milburn,
Michael Nielsen, Harold Ollivier, Asher Peres, David Poulin, R\"udiger Schack,
Ben Schumacher, Kosuke Shizume, Bill Unruh, David Wallace, Eugene Wigner,
Bill Wootters, Dieter Zeh, Anton Zeilenger, and Peter Zoller. Moreover,
Serge Haroche, Mike Nielsen, Harold Ollivier, Juan Pablo Paz and David Wallace
have provoided me with extensive written comments on earlier versions of
the manuscript. Its preparation was assisted by authors' participation
in two ITP programs on decoherence - related subjects, and was in part
supported by a grant from the NSA. Last not least, this paper has evolved
in course of over a dozen years, along with the field, under a watchful eye
of a sequence of increasingly impatient editors of RMP, their feelings
shared by my family. Perseverance of all afflicted was very much appreciated
by the author (if thoroughly tested by the process).


\noindent Figure Captions


\bigskip

\noindent Fig. 1. Snapshots of the quantum ($\hbar = 0.16$) Wigner function 
(a-c) and the classical probability distribution in phase space (d) for 
the chaotic evolution generated from the same initial Gaussian by the 
Hamiltonian:
$$ H \ =  \ p^2/2m \ - \ \kappa \cos (x - l \sin t) \ + \ a x^2/2$$
For $m=1$, $\kappa = 0.36, \ l=3 $ and $ a = 0 - 0.01$ 
it exhibits chaos with Lyapunov exponent $\Lambda = 0.2$.
Quantum (a) and classical (d) are obtained
at the same instant $t=20$. They exhibit some similarities (i.e., the 
shape of the regions of significant probability density, ``ridges'' in
the topographical maps of (a) and (d)), but the difference -- the presence
of the interference patterns with $W(x,p)$ assuming negative values
(marked with blue) is striking. Saturation of the size of the smallest 
patches is anticipated already at this early time, as the ridges of the
classical probability density are narrower than in the corresponding quantum
features. Saturation is even more visible in (c) taken at $t=60$ and (d), 
$t=100$ (note change of scale). Sharpness of the classical features makes 
simulations going beyond $t=20$ unreliable, but quantum simulations can be 
effectivelly carried out much further, as the necessary resolution can be 
anticipated in advance from Eqs. (3.15) - (3.16).

\medskip

\noindent Fig. 2. Difference between the classical and quantum average of the 
dispersion of momentum $\Delta^2=\langle p^2 \rangle - \langle p \rangle^2$
is plotted in (a) for the same initial condition, but three different values
of $\hbar$ in the model defined in Fig. 1, but with the parameter $a=0$. The
instant when the departure between the classical and quantum averages becomes 
significant varies with $\hbar$ in a manner anticipated from Eqs. (3.5) and
(3.6) for the Ehrenfest time, as is seen in the inset. Figure (b) shows the
behaviors for the same value of $\hbar$, but for four different initial
conditions. Inset appears to indicate that the typical variance difference
$\delta$ varies only logarithmically with $\hbar$, although the large error
bars (tied to the large systematic changes of behavior for different initial
conditions) preclude one from arriving at a firmer conclusion. (See 
Karkuszewski, Zakrzewski, and Zurek, 2002, for further details and discussion).

\medskip

\noindent Fig. 3.  Schematic representation of the effect of
decoherence on the Bloch sphere. When the interaction with the
environment singles out pointer states located on the poles of the
Bloch sphere, pure states (which lie on its surface) will
evolve towards the vertical axis. This classical core is a set of all
the mixtures of the pointer states.

\medskip

\noindent Fig. 4. Information transfer in a {\tt c-not} caricature of
measurement, decoherence, and decoherence with
noise. Bit-by-bit measurement is shown on the top. It is the
fundamental logical circuit used to represent decoherence affecting
the
apparatus. Note that the direction of the information flow in
decoherence -- from the decohering apparatus and to the
environment -- differs from the information flow associated with
noise. In short, as a result of decoherence environment is
perturbed by the state of the system. Noise is -- by contrast --
perturbation inflicted by the environment. Preferred
pointer states are selected so as to minimize the effect of the
environment -- to minimize the number of {\tt c-not}s pointing
from the environment at the expense of these pointing towards it.

\medskip

\noindent Fig. 5. Time-dependent coefficients of the perturbative
master equation for quantum Brownian motion. The parameters
used in these plots (where the time is measure in units of
$\Omega^{-1}$) are $\gamma/\Omega = 0.05, \ \Gamma/\Omega = 100, \
k_BT/\hbar \Omega = 10, 1, 0.1.$ Plots on the right show the initial
portion of the plots on the left -- the initial
transient -- illustrating its independence of temperature (although
higher temperatures produce higher final values of the coefficients).
Plots on the right
show that the final values of the coefficients strongly depend on
temperature, and that the anomalous diffusion is of importance
only for very low temperatures.

\medskip

\noindent Fig. 6. Evolution of the Wigner function of the decohering
harmonic oscillator. Note the difference between the rate
at which the interference term disappears for the initial
superposition of two minimal uncertainty Gaussians in position and in
momenta.

\medskip

\noindent Fig. 7. Predictability sieve in action. The plot shows
purity $Tr \rho^2$ for mixtures that have evolved from initial
minimum uncertainty wavepackets with different squeeze parameters $s$
in an underdamped harmonic oscillator with $\gamma / \omega =
10^{-4}$. Coherent states -- which
have the same spread in position as in momentum, $s=1$ -- are clearly
most predictable.

\medskip

\noindent Fig. 8.  Snapshots of a chaotic system with a double-well potential:
$ H \ = \ p^2/2m \ + \ A x^4 \ - \ B x^2 \ + \ C x \cos(ft) $.
In the example discussed here $m=1$, $A=0.5$, $B=10$, $f=6.07$ and $C=10$ 
yielding the Lyapunov exponent $\Lambda \approx 0.45$ (see Habib, Shizume,
and Zurek, 1998). 
All figures were obtained after approximately eight periods of the driving
force. The evolution started from the same minimum uncertainty
Gaussian, and proceeded according to the quantum Moyal bracket (a),
the Poisson bracket (b), and (c) the
Moyal bracket with decoherence (constant $D=0.025$ in Eq. (5.64)). In
the quantum cases $\hbar = 0.1$, which corresponds to the
area of the rectangle in the image of the Wigner function above.
Interference fringes are clearly visible in (a), and the Wigner
function shown there is only vaguely reminiscent of the classical
probability distribution in (b). Even modest
decoherence ($D=0.25$ used to get (c) corresponds to coherence length
$\ell_c = 0.3$) dramatically improves the correspondence
between the quantum and the classical. The remaining interference
fringes appear on relatively large scales, which implies small
scale quantum coherence.

\medskip

\noindent Fig. 9.  Classical and quantum expectation values of position 
$\langle x\rangle $ as a function of time for an example of Fig. 8. Evolution started 
from a minimum uncertainty Gaussian. Noticeable discrepancy between
the quantum and classical averages appears on a timescale consistent with the
Ehrenfest time $t_{\hbar}$. Decoherence -- even in modest doses --
dramatically decreases differences between the expectation values.

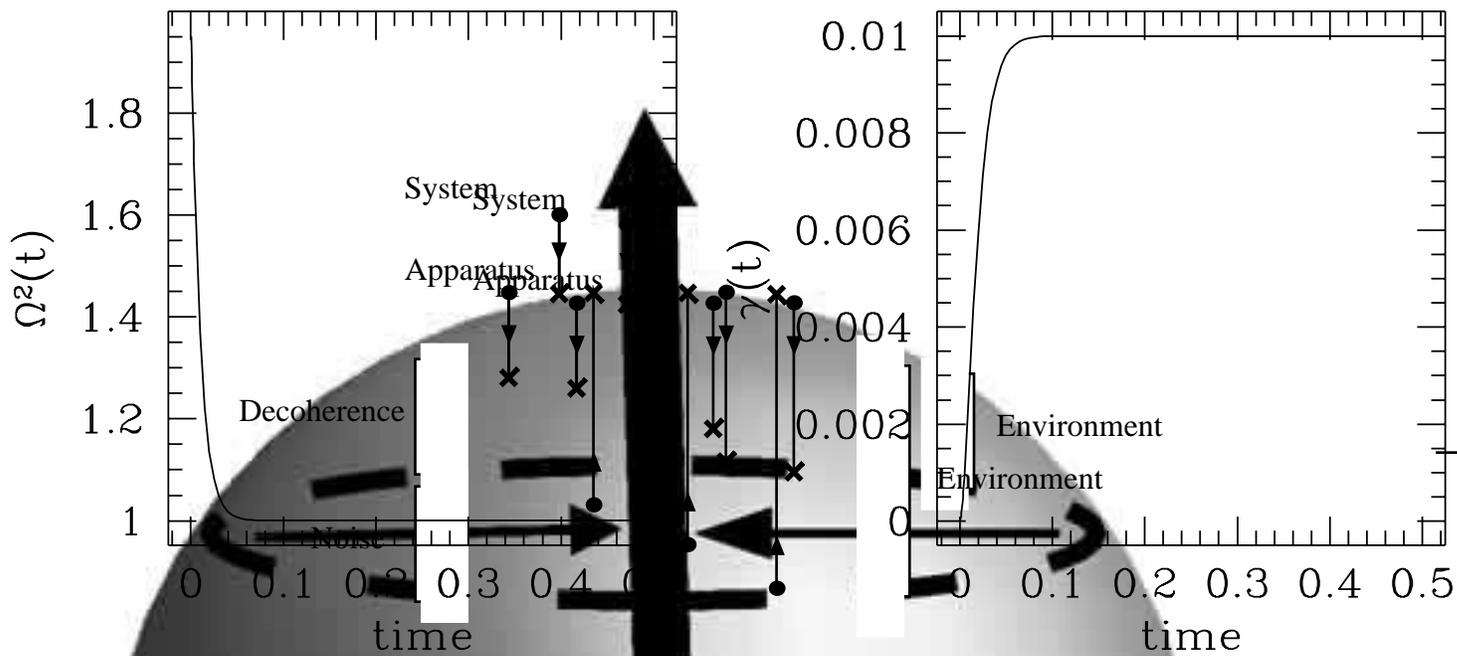
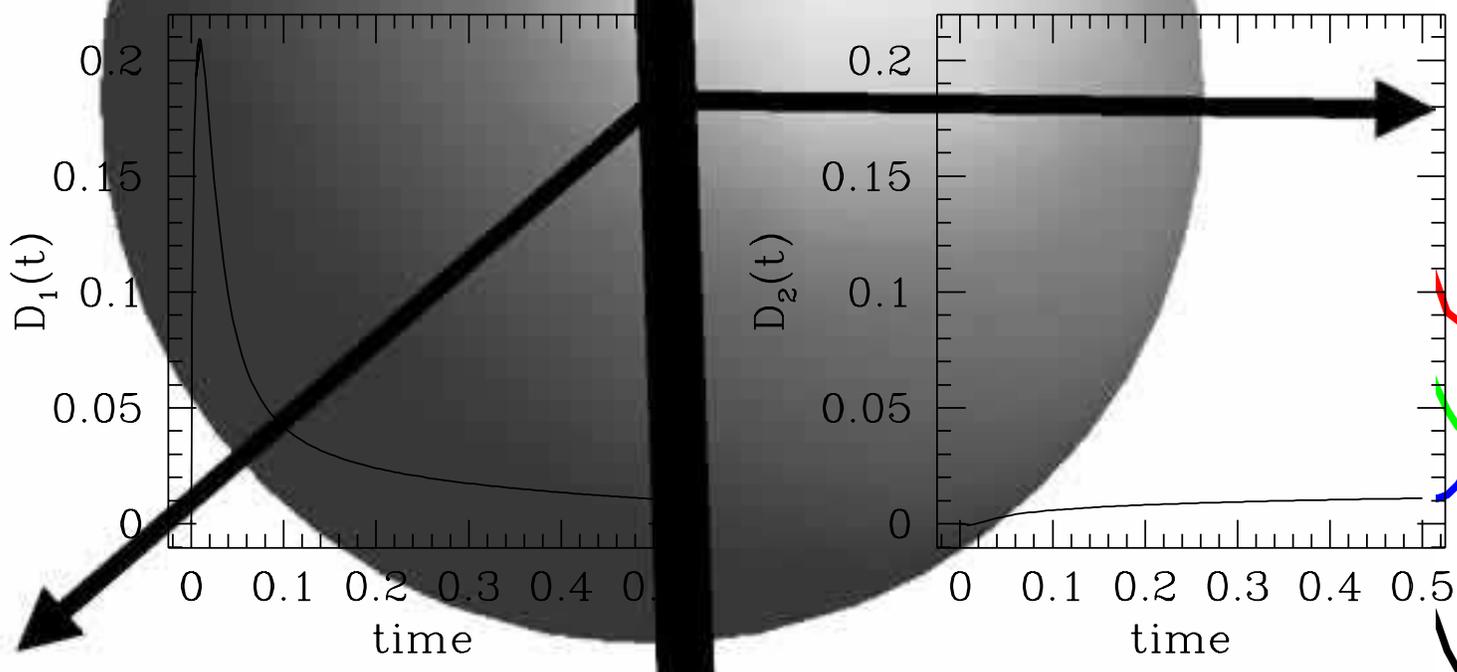
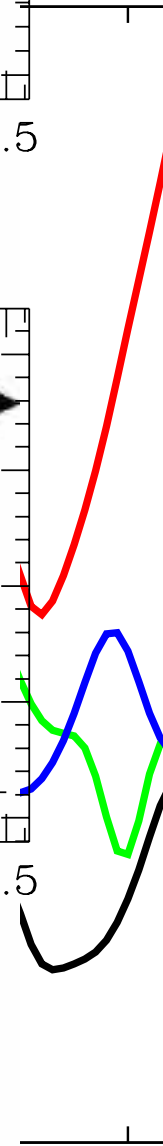
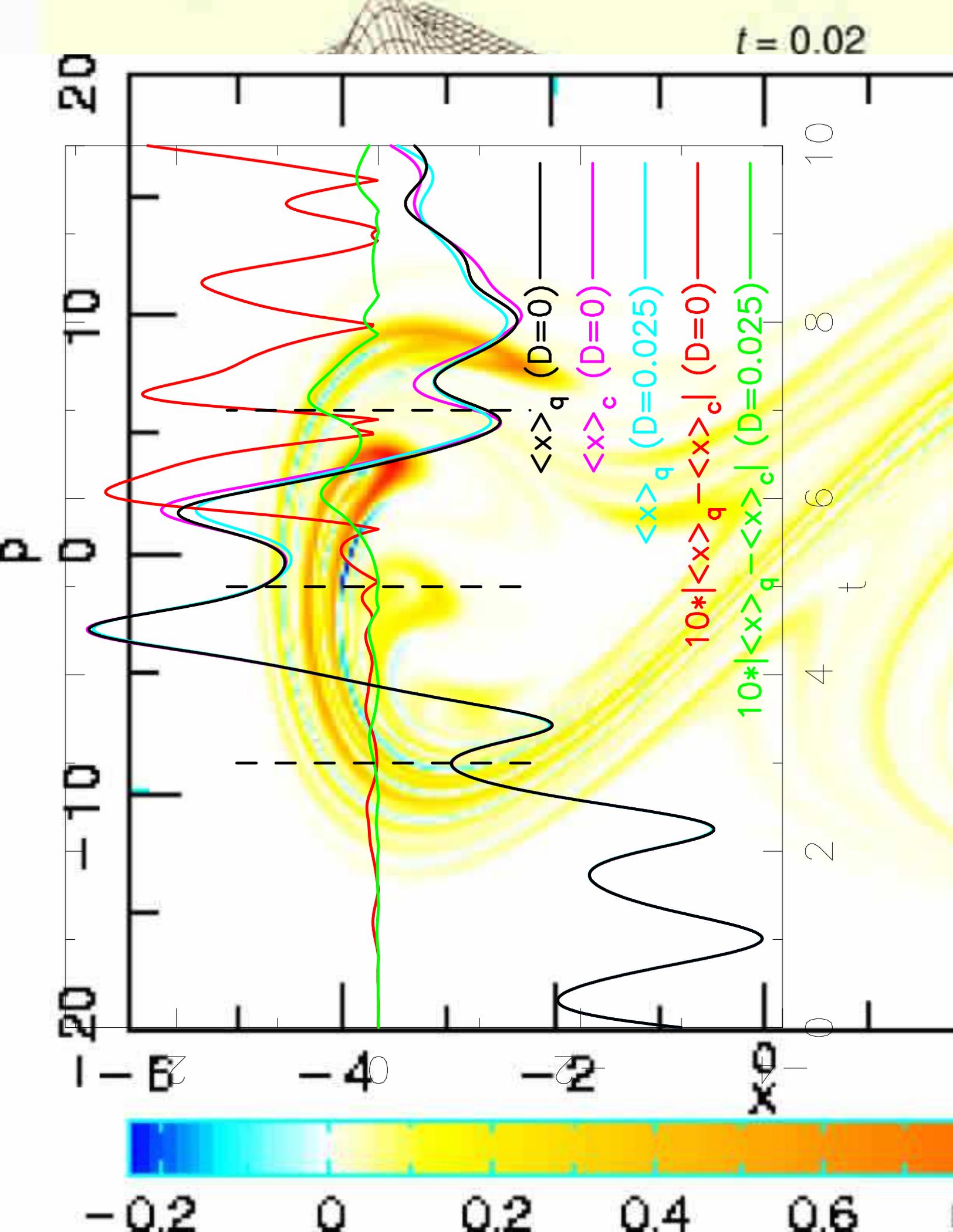
\end{document}